\documentclass[a4paper,11pt]{article}
\pdfoutput=1 
\usepackage{jheppub}

\usepackage[utf8]{inputenc}

\usepackage{amssymb,amsmath,amsfonts}
\usepackage{mathtools}
\usepackage{mathrsfs}
\usepackage{bbm}
\usepackage{slashed}
\usepackage{nicefrac}
\usepackage{graphicx}
\usepackage{caption}
\usepackage{subcaption}
\usepackage{esint}

\usepackage{graphicx}
\usepackage[dvipsnames]{xcolor}
\usepackage{array}

\usepackage{simplewick}

\usepackage{hyperref}
\usepackage{xparse}
\usepackage{xspace}

\usepackage{tikz}
\usetikzlibrary{decorations.pathmorphing}
\usetikzlibrary{automata,positioning}

\usepackage{cancel}
\usepackage[normalem]{ulem}

\usepackage{xifthen}
\usepackage{dsfont}
\usepackage[titletoc]{appendix}
\usepackage{booktabs}
\usepackage{units}

\newcommand{\gettitle}{}
\hypersetup{linkcolor=black
	colorlinks,
	linkcolor={red!75!black},
	citecolor={blue!75!black},
	urlcolor={blue!75!black},
	pdftitle={\gettitle},
	pdfauthor={Boussarie, Mehtar-Tani},
	pdfkeywords={Perturbative QCD} {Small x}{DIS},
	bookmarksopen=true,
	bookmarksopenlevel=2,
	bookmarksnumbered=true
}
\makeatletter
\newcommand\makebig[2]{%
  \@xp\newcommand\@xp*\csname#1\endcsname{\bBigg@{#2}}%
  \@xp\newcommand\@xp*\csname#1l\endcsname{\@xp\mathopen\csname#1\endcsname}%
  \@xp\newcommand\@xp*\csname#1r\endcsname{\@xp\mathclose\csname#1\endcsname}%
}
\makeatother
\makebig{biggg} {1.3}

\def\del{\partial}

\newcommand{\eqn}[1]{Eq.~\eqref{#1}}

\long\def\comment#1{ }

\newcommand{\nn}{\nonumber\\ }

\def\be{\begin{eqnarray*}}
\def\ee{\end{eqnarray*}}
\def\beq{\begin{eqnarray}}
\def\eeq{\end{eqnarray}}

\def\0{{\boldsymbol 0}}
\def\k{{\boldsymbol k}}

\def\r{{\boldsymbol r}}
\def\x{{\boldsymbol x}}
\def\y{{\boldsymbol y}}
\def\z{{\boldsymbol z}}

\def\bell{{\boldsymbol \ell}}
\def\A{{\boldsymbol A}}

\def\X{{\boldsymbol X}}

\def\A{{\boldsymbol A}}

\def\rme{{\rm e}}
\def\rmd{{\rm d}}

\def\and{ \quad\text{and}\quad}

\def\scal{\text{scal}}

\newcommand{\tr}{{\rm tr}}

\def\cP{{\cal P}}

\def\scal{{\rm scal}}

\begin{document}

\title{Gluon-mediated inclusive Deep Inelastic Scattering from Regge to Bjorken kinematics}

\author[a]{Renaud Boussarie}
\emailAdd{renaud.boussarie@polytechnique.edu}
\affiliation[b]{Centre de Physique Th\'{e}orique, \'{E}cole polytechnique, CNRS, I.P. Paris, F-91128 Palaiseau, France}

\author[b]{Yacine Mehtar-Tani,}
\emailAdd{mehtartani@bnl.gov}
\affiliation[b]{Physics Department and RIKEN BNL Research Center, Brookhaven National Laboratory, Upton, NY 11973, USA}

\date{\today}

\abstract{We revisit high energy factorization for gluon mediated inclusive Deep Inelastic Scattering (DIS) for which we propose a new semi-classical approach that accounts systematically for the longitudinal extent of the target in contrast with the shockwave limit. In this framework, based on a partial twist expansion, we derive a factorization formula that involves a new gauge invariant unintegrated gluon distribution which depends explicitly on the Feynman $x$ variable. It is shown that both the Regge and Bjorken limits are recovered in this approach.  We reproduce in particular the full one loop inclusive DIS cross-section in the leading twist approximation and the all-twist dipole factorization formula in the strict $x=0$ limit. Although quantum evolution is not discussed explicitly in this work, we argue that the proper treatment of the $x$ dependence of the gluon distribution encompasses the kinematic constraint that must be imposed on the phase-space of gluon fluctuations in the target to ensure stability of small-$x$ evolution.}


\date{\today}
\maketitle
\flushbottom

\section{Introduction}\label{sec:intro}

In the range of validity of perturbative QCD, two main kinematic regimes need to be distinguished when factorizing out the so-called hard sub-processes for which perturbation theory can be applied. For a given observable with the hard scale $Q$ at a given squared center-of-mass energy $s$, the Bjorken regime applies when $Q^2 \sim s$. It is then characterized by a single large momentum scale which allows for an expansion of the cross-section in powers of $Q^2 \gg \Lambda_{\rm QCD}^2$ as well as the resummation to all order of logarithmically enhanced contributions of the form $(\alpha_s \ln Q^2)^n$. On the other hand, the Regge limit, in which $\sqrt{s}\gg Q$, involves an expansion in powers of $x_{\rm Bj}\sim  Q^2/s$ and the resummation of $(\alpha_s \ln x_{\rm Bj})^n$.  In the latter, although no constraint on $Q^2$ is imposed at the outset, an energy dependent hard scale $Q_s^2(x)$, the so-called saturation scale, is expected to emerge as a result of non-linear gluon dynamics that causes the saturation and unitarization of the cross-section \cite{glr}. 

Twist expansion has been successful in the study of hadronic structure at large $Q^2$  but moderate $x_{\rm Bj}$ where the partonic interpretation is manifest. However, this approach is expected to break down at sufficiently small $x_{\rm Bj}$ or $Q^2\sim Q_s^2(x)$, where gluon saturation effects, neglected in the Bjorken limit, play an important role in taming the rapid rise of the gluon density when $x\to 0$. In the saturation regime, it turns out that the relevant degrees of freedom are strong classical gauge field rather than point-like quarks and gluons. These strong classical fields are the building blocks of the semi-classical approaches to small-$x$ factorization such as the Color Glass Condensate (CGC) effective theory~\cite{CGC-1,CGC-2}. Such approaches fully account for the relevant multiple gluon scattering effects, and gluon recombination effects are embedded into the Balitsky, Jalilian-Marian, Iancu, McLerran, Weigert, Leonidov, Kovner (B-JIMWLK) evolution equation~\cite{balitsky, jimwlk-1,jimwlk-2} as well as in the Balitsky-Kovchegov equation~\cite{balitsky, kovchegov-1,kovchegov-2} in the mean field approximation.

These frameworks for high energy scattering  suffer, however, from instability issues at Next-to-Leading Logarithmic (NLL) accuracy because of insufficient resummation of logarithms of $Q$, leading up to negative cross sections~\cite{negativeXS-1,negativeXS-2}. This problem is in fact already a significant feature of the linear component of small-$x$ evolution and was first diagnosed in the BFKL equation at Next-to-Leading-Logarithmic (NLL) accuracy in the late 1990's where the kernel appeared to be more singular than its Leading-Logarithmic (LL) part in the collinear and anti-collinear regions of the gluon emission kernel~\cite{nlobfkl-1,nlobfkl-2}. Several \textit{ad hoc} bottom-up corrections of the BFKL and BK evolution equations have been suggested to postpone the instabilities to higher perturbative corrections and to higher values of $s$~\cite{collinear-logs-Beuf,collinear-logs-Edmond-1,collinear-logs-Edmond-2, collinear-logs-Edmond-3,collinear-logs-Edmond-4,collinear-logs-Edmond-5,collinear-logs-zhongbo}, but none has eradicated the issue in a first principle top-down approach which would be more amenable to systematic higher order calculations.

In this article, we set to address this question at the level of the cross-section and discuss the gluon distribution whose derivation was outlined in a prior publication \cite{Boussarie:2020fpb} alongside the perturbative impact factor. Our approach amounts to revisiting high energy factorization which will allow for a systematic treatment of the collinear regions.  We focus on the inclusive Deep Inelastic Scattering (DIS) cross section where $Q^2$ and $s$ are the only available scales and we propose a framework with a built-in expansion in powers of $x_{\rm Bj}/Q^2$ which amounts to performing a partial twist expansion that systematically resums to all orders the higher twists that are enhanced at small $x_{\rm Bj}$ in particular when saturation sets in. The leading term we obtain spans by construction the leading term in both the Bjorken regime and the Regge limit, by virtue of it containing the leading power of $x_{\rm Bj}$ and of $1/Q$, respectively. This extrapolation enables a top-down understanding of how the $Q \rightarrow \infty$ limit of the Regge result compares to the $x_{\rm Bj}\rightarrow 0$ limit of the Bjorken result. We show how these two limits actually do not commute in the semi-classical descriptions of the Regge regime, and how this can lead to an insufficient account of the collinear corner of phase space, which explains the aforementioned instabilities which were only noticed numerically at NLL accuracy for reasons we will explore.

The article is structured as follows. In Section~\ref{sec:setup}, we lay the basis of our semi-classical framework inspired by the small-$x$ description of high energy scattering in the so-called shock wave approximation, and we detail how our approach goes beyond this approximation. In Section~\ref{sec:DIS}, we apply our framework to the gluon mediated inclusive Deep Inelastic Scattering cross section in full generality. In Section~\ref{sec:uPDF}, we factorize the cross section further within $x_{\rm Bj}/Q^2$ accuracy and we find and discuss a new expression for an unintegrated gluon distribution with explicit dependence on $x$. In Section~\ref{sec:Bjorken}, we expand our approximated expression in powers of the hard scale and recover the full known leading twist expression for gluon-induced DIS in the Bjorken limit. In Section~\ref{sec:Regge}, we take the $x=0$ limit of our approximated expression and we recover the full known eikonal expression for DIS in the Regge limit.  We argue in particular that these two limits only commute when one makes a strong assumption on the behavior of parton distributions at small $x$.  The consequences of this non-commutation and how our approach addresses the issue is then discussed in Section~\ref{sec:log-discussion}.

\section{Background field method for high energy scattering}\label{sec:setup}

\subsection{The classical target gauge field}
At high energy, the scattering of a dilute projectile such as a virtual photon on a hadronic target is dominated by the exchange of soft gluons that can be described by classical fields which solve the Yang-Mills equation of emotion, as prescribed by the color glass condensate (CGC) theory~\cite{CGC-1,CGC-2}. In this context, it is customary to use the light cone variables (aka Sudakov variables) defined as follows. 

For any 4-momentum $k$, we may write:
\begin{equation}
	k^{-}\equiv n_{1}\cdot k = \frac{k_0-k_3}{\sqrt{2}},\quad k^{+}\equiv n_{2}\cdot k = \frac{k_0+k_3}{\sqrt{2}}\label{eq:light-cone-variables}\,,
\end{equation}
where the light cone vectors  $n_1$ and $n_2$ are defined as
\begin{equation}
	n_{1}\equiv\frac{1}{\sqrt{2}}(1,0,0,1),\quad n_{2}\equiv\frac{1}{\sqrt{2}}(1,0,0,-1)\label{eq:light-cone-vectors}.
\end{equation}

Working in $D\equiv d+2\equiv 4+2\epsilon$ dimensions, the remaining $d$ transverse components will be denoted by bold characters in Euclidean space and with a $\perp$ subscript in Minkowski space. This way, for any $(k,\ell)$ we have:
\begin{align}
	k\cdot\ell \equiv 	k_\mu \ell^\mu & =k^{+}\ell^{-}+k^{-}\ell^{+}+(k_{\perp}\cdot\ell_{\perp})\nonumber \\
	& =k^{+}\ell^{-}+k^{-}\ell^{+}-(\boldsymbol{k}\cdot\boldsymbol{\ell})\label{eq:scalar-product}.
\end{align}
The hadronic target at very high energies moves close to the light cone, i.e., $x^{-}=(t-z)/\sqrt{2}\sim0$, and can be described by a classical current $J^\mu \equiv J^\mu_a\,t^a$~\cite{CGC-1,CGC-2}
\begin{equation}
	 J^{-}(x)\approx J^{-}(x^{+},\boldsymbol{x}),\quad\text{and}\quad J^{+}\approx J^{i}\approx0\,,\label{eq:current}
\end{equation}
that generates a gauge field which only depends on light cone time
$x^{+}$ and the transverse coordinate $\boldsymbol{x}$. In such
a framework, it turns out that both covariant $\partial\cdot A=0$
and light cone $A^{+}=0$ gauges share a common solution. Indeed, it
immediately follows from $A^{+}=A^{i}=0$ and the independence on
$x^{-}$ that $\partial\cdot A=\partial^{+}A^{-}=0.$ The equation
of motion for the field reads 
\begin{equation}
	\left[D_{\mu},F^{\mu-}\right]=-\partial^{i}F^{i-}=-\boldsymbol{\partial}^{2}A^{-}=J^{-}.\label{eq:YM}
\end{equation}
where 
\begin{eqnarray}
	F^{\mu\nu}\equiv\partial^{\mu}A^{\nu}-\partial^{\nu}A^{\mu}-ig[A^{\mu},A^{\nu}]\quad\text{and}\quad D^{\mu}\equiv\partial^{\mu}-igA^{\mu}\,,\label{eq:field-stensor}
\end{eqnarray}
denote the field strength tensor and the covariant derivative.  Here, it is understood that $F \equiv F_a t_a$ and  $A \equiv A_a t_a$, where $t_a$ are generators of SU($N_c$) with $a=1...N_c^2-1$. With
the above choice of gauges the current is covariantly conserved since
$D^{+}J^{-}=\partial^{+}J^{-}=0$. Furthermore, note that although
$A^{-}$ obeys a Poisson equation, it is an exact solution of the
Yang-Mills equations.
The background field method for high energy scattering then relies on effective Feynman rules in the presence of this classical target field.
The possibility to include non-zero transverse gluon fields in this framework is worth acknowledging, despite the obvious fact that such gluons are pure gauges. Indeed, the existence of a residual gauge freedom in $A^{+}=0$ light cone gauge allows us to perform any gauge transformations with the form 
\begin{align}
	A^{-}(x^{+},\boldsymbol{x}) & \rightarrow\Omega_{\boldsymbol{x}}(x^{+})A^{-}(x^{+},\boldsymbol{x})\Omega_{\boldsymbol{x}}^{-1}(x^{-})-\frac{1}{ig}\Omega_{\boldsymbol{x}}(x^{+})\partial^{-}\Omega_{\boldsymbol{x}}^{-1}(x^{+})\nonumber \\
	A^{i}(x^{+},\boldsymbol{x}) & \rightarrow-\frac{1}{ig}\Omega_{\boldsymbol{x}}(x^{+})\partial^{i}\Omega_{\boldsymbol{x}}^{-1}(x^{+}),\label{eq:gt-cov}
\end{align}
where $\Omega_{\boldsymbol{x}}(x^{+})$ is an element of the gauge
group SU(3) that preserves the condition $A^{+}=0$. Although it would provide a useful tool for double checking, it is not necessary for our purposes to include transverse gluons which are not pure gauges. Indeed, transverse gluons contribute to inclusive DIS in two ways: they will form transverse gauge links in the distribution, and they will provide the $\partial^- A^i$ and non-Abelian parts of the field strength tensor $F^{-i}$. The former can easily be accounted for via path independent transverse gauge links evaluated at constant light cone time $x^+$, 
 as shown in Ref.~\cite{Boussarie:2020vzf}, with the help of the identity  
\beq
 \Omega_{\boldsymbol{x}}  \Omega^{-1}_{\boldsymbol{y}} =[\x,\y]_{x^+} \equiv \cP_\lambda \exp\left[ -ig \int_{\y}^{\x} \rmd \z(\lambda) \cdot \A(x^+,\z(\lambda))\right]
\eeq
where $\z(\lambda)\equiv (z^1,z^2)$ defines a trajectory in the transverse plane, that starts at $\y$ and ends at $\x$, and parametrized by the real number $0<\lambda<1$. The latter are discussed in Appendix~\ref{sec:transverse-gluons}.

Throughout this article, we will stick to a gauge and sub-gauge choice where only $A^-$ is non-zero.

\subsection{Quark propagator in the target field: the shock wave approximation}\label{sec:shock-waves}
In order to clarify the difference between our framework and the usual CGC effective theory, in this section we will detail the derivation of the effective quark propagator in the external target field.

Let us first analyze the propagator order by order in the background field. At the 0th order, the quark propagator $D_F$ is the standard free propagator $D_0$:
\begin{equation}
	D_{F}^{(0)}(x,y)=D_{0}(x-y)\equiv\frac{\Gamma\left(\frac{D}{2}\right)}{2\pi^{\frac{D}{2}}}\frac{i(\slashed{x}-\slashed{y})}{[-(x-y)^{2}+i0]^{\frac{D}{2}}}.\label{eq:D0}
\end{equation}
Including one scattering, one has:
\begin{equation}
	D_{F}^{(1)}(x,y)=\int\!{\rm d}^{D}z_{1}\,D_{0}(x-z_{1})ig\slashed{A}(z_{1})D_{0}(z_{1}-y).\label{eq:D0-1}
\end{equation}
Again, $D=4+2\epsilon$ stands for the number of dimensions, not to be confused with the Dirac propagator denoted $D_F$ and $D_0$.

In semi-classical descriptions of QCD in the Regge limit, i.e. $s\to \infty$, the key assumption is known as the \textit{shock wave approximation}. In $A^+=0$ light cone gauge, the classical target gluon field $A^{\mu}(z_1)=A^{-}(z_1^{+},\boldsymbol{z}_1)\, n_{2}^{\mu}$ is assumed to be very peaked around $z_1^+=0$ as a result of Lorentz contraction. Thus, we can neglect any dependence on $z_{1}^{+}$ in the propagators,
and factorize 
\begin{equation}
	[D_{0}(x-z_{1})\gamma^{+}D_{0}(z_{1}-y)]_{z_{1}^{+}=0}\label{eq:d0d0}
\end{equation}
from the quantity
\begin{equation}
	U_{\boldsymbol{z}_{1}}^{(1)}\equiv\int\!{\rm d}u^{+}igA^{-}(u^{+},\boldsymbol{z}_{1}),\label{eq:u1}
\end{equation}
where we relabeled the internal variable $z_{1}^{+}$ as $u^{+}$. Writing that
\begin{equation} [D_{0}(x-z_{1})\gamma^{+}D_{0}(z_{1}-y)]_{z_{1}^{+}=0}=\int\!{\rm d}z_{1}^{+}\delta(z_{1}^{+})D_{0}(x-z_{1})\gamma^{+}D_{0}(z_{1}-y),
\end{equation}
we obtain
\begin{equation}
	D_{F}^{(1)}(x,y)=\int\!{\rm d}^Dz_{1}\,\delta(z_{1}^{+})D_{0}(x-z_{1})\gamma^{+}U_{\boldsymbol{z}_{1}}^{(1)}D_{0}(z_{1}-y).\label{eq:D0-1-1}
\end{equation}
The reason for the $U_{\boldsymbol{z}_{1}}^{(1)}$ notation will become clear shortly.

Let us finally consider the effective propagator of a fermion with
two scatterings with the external field:
\begin{align}
	D_{F}^{(2)}(x,y) & =\int\!{\rm d}^{D}z_{2}\,{\rm d}^{D}z_{1}D_{0}(x-z_{2})ig\slashed{A}(z_{2})D_{0}(z_{2}-z_{1})ig\slashed{A}(z_{1})D_{0}(z_{1}-y)\\
	& =\int\!{\rm d}^{D}z_{2}\,{\rm d}^{D}z_{1}\!\int\!\frac{{\rm d}^{D}p_{2}}{(2\pi)^{D}}\frac{{\rm d}^{D}p_{1}}{(2\pi)^{D}}\frac{{\rm d}^{D}p_{0}}{(2\pi)^{D}}\,{\rm e}^{-ip_{2}\cdot(x-z_{2})-ip_{1}\cdot(z_{2}-z_{1})-ip_{0}\cdot(z_{1}-y)}\nonumber \\
	& \times D_{0}(p_{2})ig\slashed{A}(z_{2})D_{0}(p_{1})ig\slashed{A}(z_{1})D_{0}(p_{0}).
\end{align}
Recall that the external field reads $A^{\mu}(z)=A^{-}(z^{+},\boldsymbol{z})\, n_{2}^{\mu}$, hence, 
\begin{align}
	ig\slashed{A}(z_{2})D_{0}(p_{1})ig\slashed{A}(z_{1}) & =ig\slashed{A}(z_{2})\frac{i\slashed{p}_{1}}{p_{1}^{2}+i0}ig\slashed{A}(z_{1})\nonumber \\
	& =igA^{-}(z_{2})\frac{i\gamma^{+}\slashed{p}_{1}\gamma^{+}}{2p_{1}^{+}\left(p_{1}^{-}-\frac{\boldsymbol{p}_{1}^{2}}{2p_{1}^{+}}+i0p_{1}^{+}\right)}igA^{-}(z_{1})\\
	& =igA^{-}(z_{2})\frac{i\gamma^{+}}{p_{1}^{-}-\frac{\boldsymbol{p}_{1}^{2}}{2p_{1}^{+}}+i0p_{1}^{+}}igA^{-}(z_{1}),\nonumber 
\end{align}
where we used that $\gamma^{+}\slashed{p}_{1}\gamma^{+}=2p_{1}^{+}\gamma^{+}$.
This observation makes it clear that inserting more scatterings does
not change the Dirac structure, that is
\begin{equation}
	\slashed{p}_{2}\gamma^{+}\slashed{p}_{0}
\end{equation}
to all orders in $gA$. Hence, the scalar propagation factorizes
from the Dirac structure in the fermionic propagator.

Let us now take the $p_{1}^{-}$ integral, using the standard Cauchy pole integral
\begin{equation}
	\int{\rm d}p_{1}^{-}\frac{{\rm e}^{-ip_{1}^{-}(z_{2}^{+}-z_{1}^{+})}}{p_{1}^{-}-\frac{\boldsymbol{p}_{1}^{2}}{2p_{1}^{+}}+i0p_{1}^{+}}=-2\pi i\left[\theta(p_{1}^{+})\theta(z_{2}^{+}-z_{1}^{+})-\theta(-p_{1}^{+})\theta(z_{1}^{+}-z_{2}^{+})\right]{\rm e}^{-i\frac{\boldsymbol{p}_{1}^{2}-i0}{2p_{1}^{+}}(z_{2}^{+}-z_{1}^{+})}.\label{eq:Cauchy-int}
\end{equation}
The $\boldsymbol{p}_{1}$ integral is then reduced to a Gaussian,
with the phase
\begin{equation}
	-i\frac{\boldsymbol{p}_{1}^{2}-i0}{2p_{1}^{+}}(z_{2}^{+}-z_{1}^{+})+i\boldsymbol{p}_{1}\cdot(\boldsymbol{z}_{2}-\boldsymbol{z}_{1}).\label{eq:phase}
\end{equation}
First, note that the first term in that phase is suppressed
in the Regge limit, where $1/p_{1}^{+}\sim 1/\sqrt{s}\rightarrow 0$.
In this limit, it is then reasonable to approximate the Gaussian by
a delta function by neglecting the quadratic term. Integrating w.r.t.
$\boldsymbol{p}_{1}$ will yield $(2\pi)^{d}\delta^{(d)}(\boldsymbol{z}_{2}-\boldsymbol{z}_{1})$.
This is easy to generalize to all orders in $gA$: each intermediate
propagator yields a $\delta$ function, which means that all gluon fields
are evaluated at exactly the same transverse position $\boldsymbol{z}_{1}$ .

Similar considerations can be made for the two external propagators.
Because of the $\gamma^{+}$ matrix, the numerator
of the $p_{2}$ and $p_{0}$ propagators do not depend on $p_{2}^{-}$
and $p_{0}^{-}$. This means that the same integral as in Eq.~(\ref{eq:Cauchy-int})
can be taken for $p_{2}^{-}$ and $p_{0}^{-}$ as well. Furthermore, since the gluon fields do not depend on $-$ positions, the $+$ momentum
is conserved throughout the scattering, so $p_{2}^{+}=p_{1}^{+}=p_{0}^{+}\equiv p^{+}$, as expected in the eikonal approximation.
Performing the $p_{2}^{-}$ and $p_{0}^{-}$ integrations with Eq.~(\ref{eq:Cauchy-int}),
along with the result of the $p_{1}^{-}$ integral, will yield two
possible cases: either $x^{+}>z_{2}^{+}>z_{1}^{+}>y^{+}$ and $p^{+}>0$,
or $x^{+}<z_{2}^{+}<z_{1}^{+}<y^{+}$ and $p^{+}<0$. We will focus
on the former case by restricting ourselves to studying the propagator
for $x^{+}>y^{+}$.

To sum up the progress so far, at $(gA)^2$ order, the Dirac propagator reads
\begin{align}\label{eq:order2DF}
	\left.D_{F}^{(2)}(x,y)\right|_{x^{+}>y^{+}} & =\int\!{\rm d}^{d}\boldsymbol{z}_{1}\int_{y^{+}}^{x^{+}}\!\!{\rm d}z_{2}^{+}\int_{y^{+}}^{z_{2}^{+}}\!\!{\rm d}z_{1}^{+}igA^{-}(z_{2}^{+},\boldsymbol{z}_{1})igA^{-}(z_{1}^{+},\boldsymbol{z}_{1})\nonumber \\
	& \times\int\!\frac{{\rm d}^{D}p_{2}}{(2\pi)^{D}}\frac{{\rm d}^{D}p_{0}}{(2\pi)^{D}}(2\pi)\delta(p_{2}^{+}-p_{0}^{+})D_{0}(p_{2})\gamma^{+}D_{0}(p_{0})\,\theta(p_{1}^{+})\\
	& \times{\rm e}^{-ip_{2}^{+}x^{-}+ip_{2}^{-}(x^{+}-z_{2}^{+})+i\boldsymbol{p}_{2}\cdot(\boldsymbol{x}-\boldsymbol{z}_{1})+ip_{0}^{+}y^{-}-ip_{0}^{-}(z_{1}^{+}-y^{+})+i\boldsymbol{p}_{0}\cdot(\boldsymbol{z}_{1}-\boldsymbol{y})}.\nonumber
\end{align}
The last two steps are based on the observation that in the shock
wave approximation, the target gluon fields as seen from the projectile
are very peaked around light cone time $0^{+}$ because of a large
separation between typical life times of quantum fluctuations and scattering times. This enables us to write
\begin{equation}
	x^{+}-z_{2}^{+}\sim x^{+},\quad z_{1}^{+}-y^{+}\sim-y^{+}\label{eq:times}
\end{equation}
in the phases, and thus completely factorize out the following
quantity from the first line in \eqn{eq:order2DF}:
\begin{equation}
	[x^{+},y^{+}]_{\boldsymbol{z}_{1}}^{(2)}\equiv\int_{y^{+}}^{x^{+}}\!\!{\rm d}z_{2}^{+}\int_{y^{+}}^{z_{2}^{+}}\!\!{\rm d}z_{1}^{+}\,igA^{-}(z_{2}^{+},\boldsymbol{z}_{2})igA^{-}(z_{1}^{+},\boldsymbol{z}_{1}).
\end{equation}
This quantity is exactly the second power in the $(gA)$-expansion
of the Wilson line 
\begin{equation}
	[x^{+},y^{+}]_{\boldsymbol{z}_{1}}\equiv{\cal P}_+\exp\left[ig\int_{y^{+}}^{x^{+}}\!\!{\rm d}z^{+}A^{-}(z^{+},\boldsymbol{z}_{1})\right].
\end{equation}
A recursion on the number of scatterings can be used in order to prove that the effective fermionic propagator in its entirety can be cast into:
\begin{align}
	\left.D_{F}(x,y)\right|_{x^{+}>y^{+}} & =\int\!{\rm d}^{d}\boldsymbol{z}_{1}\,\,{\rm e}^{-ip_{2}^{+}x^{-}+ip_{2}^{-}x^{+}+i\boldsymbol{p}_{2}\cdot(\boldsymbol{x}-\boldsymbol{z}_{1})+ip_{0}^{+}y^{-}+ip_{0}^{-}y^{+}+i\boldsymbol{p}_{0}\cdot(\boldsymbol{z}_{1}-\boldsymbol{y})} \nonumber \\
	& \times[x^{+},y^{+}]_{\boldsymbol{z}_{1}} \, \, \int\!\frac{{\rm d}^{D}p_{2}}{(2\pi)^{D}}\frac{{\rm d}^{D}p_{0}}{(2\pi)^{D}}(2\pi)\delta(p_{2}^{+}-p_{0}^{+})D_{0}(p_{2})\gamma^{+}D_{0}(p_{0})\theta(p_{1}^{+}). 
\end{align}
Finally, rewriting the $\delta$ function as
\begin{equation}
	(2\pi)\delta(p_{2}^{+}-p_{0}^{+})=\int\!{\rm d}z_{1}^{-}{\rm e}^{i(p_{2}^{+}-p_{0}^{+})z_{1}^{-}}\,,
\end{equation}
and using
\begin{equation}
	{\rm e}^{ip_{2}^{-}x^{+}+ip_{0}^{-}y^{+}}=\int\!{\rm d}z_{1}^{+}\delta(z_{1}^{+})\,{\rm e}^{ip_{2}^{-}(x^{+}-z_{1}^{+})+ip_{0}^{-}(y^{+}-z_{1}^{+})},
\end{equation}
we recognize the free propagators in the coordinate space and we obtain 
\begin{align}
	\left.D_{F}(x,y)\right|_{x^{+}>y^{+}} & =\int{\rm d}^{D}z_{1}\delta(z_{1}^{+})D_{0}(x-z_{1})[x^{+},y^{+}]_{\boldsymbol{z}_{1}}D_{0}(z_{1}-y).
\end{align}
It is not customary to stop at this point. The last step is to use
once more the fact that gluon fields are peaked around light cone
time $0^{+}$. Indeed, say the gluon fields have an effective support
$\epsilon^{+}$ in light cone time. If $x^{+}>\epsilon^{+},$ we can
extend $[x^{+},y^{+}]_{\boldsymbol{z}_{1}}\rightarrow[\infty^{+},y^{+}]_{\boldsymbol{z}_{1}}$
with no cost because all fields beyond $x^{+}$ are null. A similar
extension to $-\infty^{+}$ can be performed provided that $y^{+}<-\epsilon^{+}$.
The large separation of light cone times one assumes when using semi-classical
small $x_{{\rm Bj}}$ physics provides the jusification for which
$|x^{+}|,|y^{+}|\gg\epsilon^{+}$. Using this extension yields the
usual infinite length Wilson line operators
\begin{equation}
	[x^{+},y^{+}]_{\boldsymbol{z}_{1}}\rightarrow U_{\boldsymbol{z}_{1}}\equiv[\infty^{+},-\infty^{+}]_{\boldsymbol{z}_{1}},\label{eq:infinite-lines}
\end{equation}
and the standard effective propagator
\begin{align}
	\left.D_{F}(x,y)\right|_{x^{+}>0>y^{+}} & =\int{\rm d}^{D}z_{1}\delta(z_{1}^{+})D_{0}(x-z_{1})\gamma^{+}U_{\boldsymbol{z}_{1}}D_{0}(z_{1}-y).
\end{align}
Let us summarize the derivation above. The effective propagator in the shock
wave approximation is obtained following three essential assumptions:
i) that we can neglect the quantum phases to approximate the intermediate propagators
between scatterings by $\delta$ functions (\ref{eq:phase}), ii) the existence of a large enough separation of light cone times to neglect
all scattering times w.r.t. any interaction time in the projectile
wave functions (\ref{eq:times}), iii) and that we can extend finite length Wilson lines into infinite ones without loss of precision (\ref{eq:infinite-lines}). All three
of these assumptions rely on the same observation: that the classical
target fields $A^{-}(x^{+},\boldsymbol{x})$ are very peaked around
$x^{+}=0$. The purpose of this article is to go beyond this approximation, by keeping a generic dependence on $x^{+}$ in the target fields.

\subsection{Quark propagator in the target field: beyond shock waves}\label{sec:beyond-shock-waves}
In this article, we will use a very similar framework to the one described in the previous subsection and we will thus decompose gluon fields between classical and quantum fields in the space of $+$ momenta. Indeed, in the Bjorken limit the non-perturbative matrix elements describe partons being emitted collinearly to the target, hence with small $+$ momentum components, and in the Regge limit the projectile partons have such high $+$ momentum components that $+$ momentum transfer from the target can be neglected. Either way, $+$ momentum is strongly ordered and it is thus adequate to split the gluon fields according to their momenta along this direction. This motivates us to build our framework in view of factorizing observables in rapidity space, similarly to what was successfully performed in~\cite{ian-andrey}.\\
The difference with standard semi-classical small $x_{\rm Bj}$ physics lies in the fact that we will get rid of the assumption that the target field is peaked around $x^+=0$.
This assumption was relaxed in several studies of semi-classical small-$x$ schemes beyond the shock wave approximation, although non-zero transverse fields as well as non-trivial dependence on $x^-$ have since been incorporated in these schemes ~\cite{subeikonal-tolga-1,subeikonal-tolga-2,subeikonal-tolga-3,subeikonal-tolga-4,subeikonal-tolga-5,subeikonal-tolga-6,subeikonal-giovanni-1,subeikonal-giovanni-2}. The purpose of this article is to focus on corrections to this assumption with a different approach. In~\cite{subeikonal-tolga-1,subeikonal-tolga-2,subeikonal-tolga-3,subeikonal-tolga-4,subeikonal-tolga-5,subeikonal-tolga-6}, a fixed variable $L^+$ for the so-called longitudinal extent of the target was introduced, and corrections were considered via an expansion in powers of $L^+$ of all effective Feynman rules. Similarly in~\cite{subeikonal-giovanni-1,subeikonal-giovanni-2} a systematic expansion of effective propagators in powers of $1/\sqrt{s}$ is performed. Our approach differs in several ways. Firstly, we will not perform an expansion around fields peaked at $x^+=0$. Instead of expanding universal building blocks like the quark propagator around the shock wave approximation, we will expand a given observable in terms of a physical variable involved: we will keep the gluon fields to be fully general functions of $x^+$ for the first part of this computation and in the second part we will then expand our results in a process-dependent fashion. This makes our approach more similar to that of~\cite{subeikonal-yuri-1,subeikonal-yuri-2,subeikonal-yuri-3,subeikonal-yuri-4,subeikonal-yuri-5,subeikonal-yuri-6,subeikonal-yuri-7} and~\cite{subeikonal-yoshitaka} where specific observables or distributions are studied rather than universal effective quantities, but more general in its first step and less reliant on the implicit hypothesis discussed in Section~\ref{sec:log-discussion}.
For the first part of this article, we will stick to the following two hypothesis in light cone gauge: the gluon field does not depend on $x^-$, and it can be written is such a way that its only non-zero component is $A^-$.
Because of these two approximations, we can perform most of the same steps as in the previous derivation and write in momentum space:
\beq\label{eq:dirac-p-mom}
D_F(\ell',\ell)= D_0(\ell)(2\pi)^D \delta^D(\ell-\ell')+i \frac{\slashed{\ell}' \gamma^+ \slashed{\ell}}{2\ell^+} [G_\scal(\ell',\ell)-G_0(\ell)(2\pi)^D \delta^D(\ell-\ell')]. 
\eeq
\begin{figure}
\begin{center}
\includegraphics[width=0.8\textwidth]{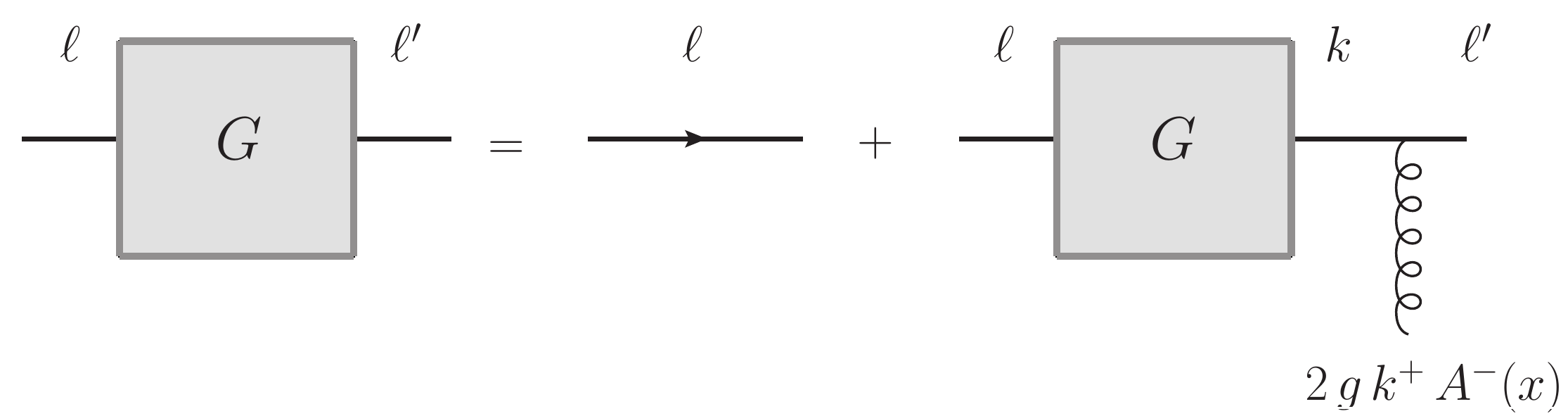} 
\end{center}
\caption{Diagrammatic illustration of the  scalar propagator in a background field. }
\label{fig:long-DIS-fig-1}
\end{figure}
Here, the first term corresponds to the free propagator (in the absence of scatterings) and in the second term we factorized out the Dirac matrix structure like before to isolate a scalar propagator $G_{\rm scal}$. This term contains at least one scattering, hence the subtraction of the free scalar propagator $G_0(\ell)\equiv1/(\ell^2+i0)$.
The scalar propagator is defined via a recursive relation in the number of gluon insertions (see Figure \ref{fig:long-DIS-fig-1}):
\begin{align}
	& G_{{\rm scal}}(\ell^{\prime},\ell)-G_{0}(\ell^{\prime})(2\pi)^D\delta^{D}(\ell^{\prime}-\ell)\label{eq:Gscal-rec}\\
	& =2g\int\!{\rm d}^{D}x\int\!\frac{{\rm d}^{D}k}{(2\pi)^{D}}{\rm e}^{i(\ell^{\prime}-k)\cdot x}G_{0}(\ell^{\prime})\,(k\cdot A)(x)\,G_{{\rm scal}}(k,\ell)\nonumber.
\end{align}
It can be rewritten in a more compact form in coordinate space where
\begin{equation}
	G_{{\rm scal}}(x,x_{0})\equiv\int\frac{{\rm d}^{D}\ell}{(2\pi)^{D}}\frac{{\rm d}^{D}\ell_{0}}{(2\pi)^{D}}{\rm e}^{-i(\ell^{\prime}\cdot x)+i(\ell\cdot x_{0})}G_{{\rm scal}}(\ell^{\prime},\ell)\label{eq:Gscal-fourier}
\end{equation}
is the solution of the Klein-Gordon equation in a potential 
\beq
\left[-\Box_x+2 ig  A(x)\cdot\del_x\right] G_\scal(x,x_0)=\delta^D(x-x_0) \, , \label{eq:KG}
\eeq
or similarly
\beq
G_\scal(x,x_0)\left[-\overleftarrow{\Box}_{x_0}-2 ig  \overleftarrow{\del}_{x_0}\cdot A(x_0)\right] =\delta^D(x-x_0) \,. \label{eq:KG-2}
\eeq
With the help of the simple relation
\begin{equation}
	i\frac{\slashed{\ell}\gamma^{+}\slashed{\ell}}{2\ell^{+}}G_{0}(\ell)=D_{0}(\ell)-i\frac{\gamma^{+}}{2\ell^{+}}\label{eq:G0-simp},
\end{equation}
it is finally possible to cast Eq.~(\ref{eq:dirac-p-mom}) into
\beq\label{eq:dirac-p-mom-simp}
D_F(\ell',\ell)= i\frac{\gamma^+}{2\ell^+}(2\pi)^D \delta^D(\ell^\prime-\ell)  +i \frac{\slashed{\ell}' \gamma^+ \slashed{\ell}}{2\ell^+} G_\scal(\ell',\ell). 
\eeq
The first term in this equation is reminiscent of the so-called instantaneous term, or Coulomb term, in light front perturbation theory (LFPT). It usually plays the role of a gauge invariance restoring counterterm.

It can be convenient to make use of basic properties of the scalar propagator to introduce another object. Because the target field does not depend on $x^-$, the propagator depends only  on the difference of $x^-$ coordinates. As a result, we can write
\begin{equation}
	G_{\mathrm{scal}}(x,x_{0})\equiv\int\!\frac{{\rm d}p^{+}}{2\pi}\frac{{\rm e}^{-ip^{+}(x^{-}-x_{0}^{-})}}{2ip^{+}} ( \boldsymbol{x} | \mathcal{G}_{p^{+}}(x^{+},x_{0}^{+}) |\boldsymbol{x}_{0}).\label{eq:cG-def}
\end{equation}
For more insight about the mathematical properties of ${\cal G}$ in another physical context, the reader is referred to~\cite{Blaizot:2015lma}. As a direct consequence of Eq.~(\ref{eq:KG}), the new object on the r.h.s satisfies the Schr\"{o}dinger equation 
\begin{equation}
	\left[i\frac{\partial}{\partial x^{+}}+\frac{\boldsymbol{\partial}_{x}^{2}}{2p^{+}}+gA^{-}(x)\right](\boldsymbol{x}|\mathcal{G}_{p^{+}}(x^{+},x_{0}^{+})|\boldsymbol{x}_{0})=i\delta(x^{+}-x_{0}^{+})\delta^{d}(\boldsymbol{x}-\boldsymbol{x}_{0})\label{eq:Schro},
\end{equation}
and a similar relation for $x_0$. This relation is particularly useful to recover the shock wave limit: by neglecting the $\frac{\boldsymbol{\partial}_{x}^{2}}{2p^{+}}$ term, one would be left with the equation which defines a finite length Wilson line between $x^+$ and $x_0^+$ at the fixed transverse coordinate $\boldsymbol{x}=\boldsymbol{x}_0$:
\begin{equation}
	(\boldsymbol{x}|\mathcal{G}_{p^{+}}(x^{+},x_{0}^{+})|\boldsymbol{x}_{0}) \to \delta^{d}(\x-\x_0) \, \theta(x^+-x^+_0)\, [x^+,x^+_0]_\x\, \label{eq:calG-sw},
\end{equation}
for $p^+>0$ and a similar relation for $p^+<0$.

Prolonging this line into an infinite one as described in the previous section, we would get the well-known propagator in the shock wave limit.

\section{Gluon mediated Deep Inelastic scattering}\label{sec:DIS}

\subsection{Kinematics }
Let us consider the DIS subprocess $\gamma^\ast(q) +{ \rm proton}\,(P)  \to  X$. We will use the standard DIS variables:
\beq
 s =(P+q)^2\,, \qquad Q^2 =-q^2\, , \qquad x_{\rm Bj}=\frac{Q^2}{2P\cdot q}\,.
\eeq
Owing to the optical theorem, the total cross-section is related to the forward scattering amplitude $\gamma^\ast(q) +{ \rm proton}\,(P)  \to  \gamma^\ast(q) +{ \rm proton}\,(P)$:
\begin{align}
	\sigma_{T,L}^{\gamma^{\ast}}(x_{{\rm Bj}},Q^{2}) & =\frac{1}{4P\cdot q}2{\rm Im}\left[\frac{{\cal A}_{T,L}^{\gamma^{\ast}(q)p(P)\rightarrow\gamma^{\ast}(q^{\prime})p(P^{\prime})}}{i(2\pi)^{D}\delta^{D}(q^{\prime}+P^{\prime}-q-P)}\right]_{q^\prime=q,\, P^\prime=P}\label{eq:opt-theo},
\end{align}
where ${\cal A}_T$, resp. ${\cal A}_L$ is the amplitude for the transition from a transverse (resp. longitudinal) photon to a transverse (resp. longitudinal) photon. Note that in the forward limit there is no transverse-to-longitudinal or longitudinal-to-transverse transition. With a slight abuse of notations, we will write:
\begin{align}
	\sigma_{T,L}^{\gamma^{\ast}}(x_{{\rm Bj}},Q^{2}) & =\frac{x_{{\rm Bj}}}{Q^{2}}{\rm Im}\frac{\,{\cal A}_{T,L}^{\gamma^{\ast}(q)p(P)\rightarrow\gamma^{\ast}(q)p(P)}}{i(2\pi)^{D}\delta^{D}(0)}\label{eq:opt-theo-1}.
\end{align}
We will focus on computing the imaginary part of the amplitude as requested in the relation above. For readability, the $\gamma^{\ast}(q)p(P)\rightarrow\gamma^{\ast}(q)p(P)$ superscript will be omitted from now on. The DIS structure functions will be recovered via their relation to the cross section of the subprocess and thus to the forward amplitude, see e.g.~\cite{Beuf:2011xd}:
\begin{equation}
	F_{T,L}(x_{{\rm Bj}},Q^{2})=\frac{x_{{\rm Bj}}}{\pi e^{2}}{\rm Im}\frac{\, {\cal A}_{T,L}}{i(2\pi)^{D}\delta^{D}(0)}\label{eq:FLT}.
\end{equation}
Finally, we will define the hadronic tensor to be the real part of the forward amplitude with open photon indices, i.e. by
\begin{equation}
	{\cal A}_L=\varepsilon_{L}^{\mu}W_{\mu\nu}\varepsilon_{L}^{\nu\ast},\quad	{\cal A}_T=\frac{1}{d}\sum_{\lambda=1...d}\varepsilon_{\lambda}^{\mu}W_{\mu\nu}\varepsilon_{\lambda}^{\nu\ast}.\label{eq:Wdef}
\end{equation}
In our $D=d+2$ dimensional regularization scheme where the incoming photon has $d$ transverse polarizations, the $1/d$ prefactor to the transverse contribution comes from averaging over photon helicities. We choose the frame in which the photon and proton momenta are alined with the $z$ axis, that is, 
\beq
q= q^+ n_1 -\frac{Q^2}{2q^+} n_2  \quad \text{and} \quad   P= P^- n_2 +\frac{m^2}{2 P^-} n_1\,,
\eeq
respectively. From here on, we shall neglect the proton mass $m$.

In Landau gauge for QED and in the considered frame, the longitudinal polarization vector may be chosen to be
\beq
 \varepsilon_L =\frac{1}{Q} \left( q^+ n_1 +\frac{Q^2}{2q^+}n_2\right)\,,
\eeq
while the transverse polarizations form a basis of the $d$-dimensional transverse Minkowski subspace. They satisfy the orthogonality and completeness relations:
\beq 
\varepsilon_\lambda \cdot \varepsilon^\ast_{\lambda^\prime} =\delta_{\lambda\lambda^\prime}\, \quad \text{ and}  \quad \sum_{\lambda=1...d} \varepsilon^\mu_\lambda \cdot \varepsilon^{\ast \nu}_\lambda = - g_\perp^{\mu\nu}\,.\label{eq:pol-comp}
\eeq

\subsection{The photon wave functions}\label{sec:wf}

\begin{figure}
\begin{center}
\includegraphics[width=0.5\textwidth]{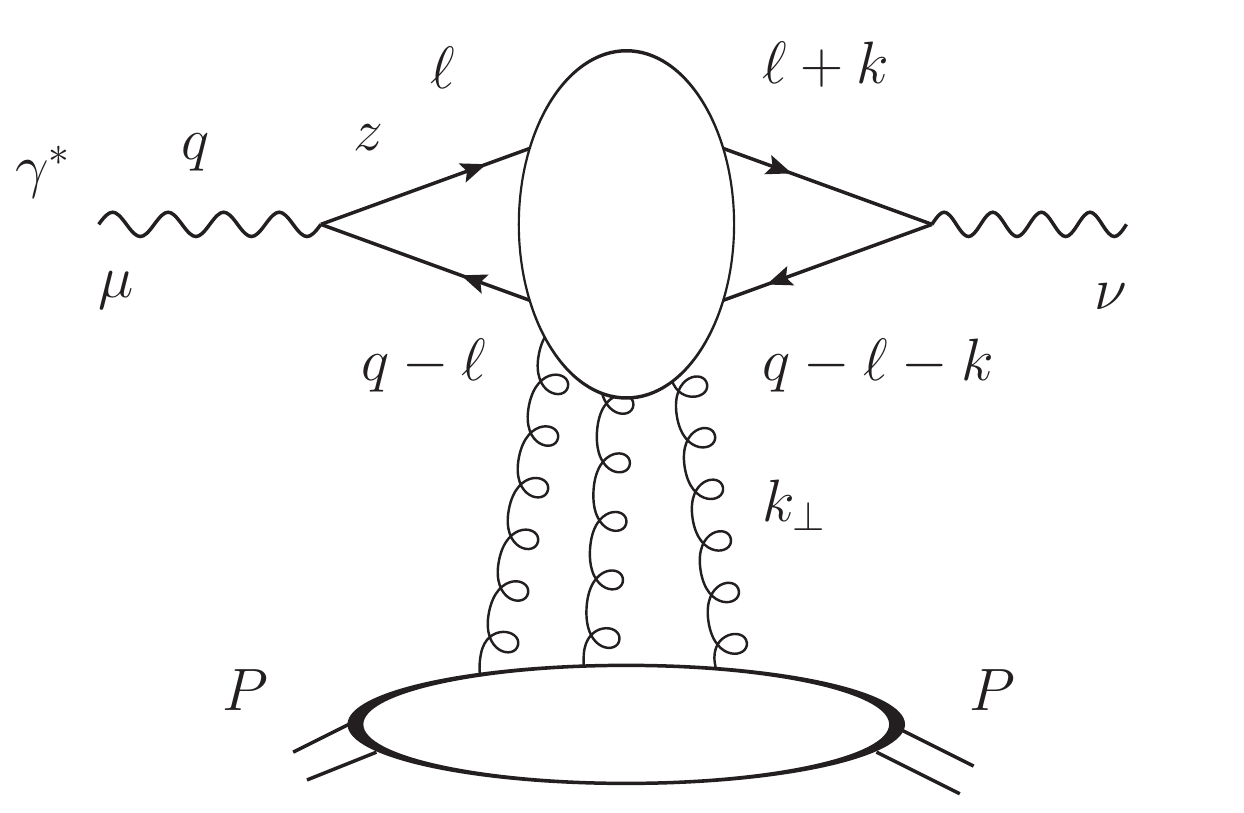} 
\end{center}
\caption{Depiction of the hadronic tensor in gluon mediated DIS. }
\label{fig:long-DIS-W}
\end{figure}
The hadronic tensor for gluon mediated DIS reads (cf.~Figure \ref{fig:long-DIS-W})
\beq\label{eq:h-tensor}
W^{\mu\nu} =\frac{e^2}{\mu^{d-2}} \sum_f q_f^2 \int \! \frac{\rmd^{D} \ell}{(2\pi)^D} \int \!  \frac{\rmd^{D} k }{(2\pi)^D}\,\langle P| \tr \left[ \gamma^\nu D_F(\ell+k,\ell) \gamma^\mu  D_F(-q+\ell, -q+\ell+k )\right] |P\rangle\,, \nn
\eeq
where $\mu$ is the dimensional regularization parameter. For readability, we will not write it explicitly until the end of the computation. $D_F(\ell+k,\ell)$ is the quark propagator in the background field $A^-(x^+,\x)$. This propagator, as written explicitly in Eq.~(\ref{eq:dirac-p-mom}), each contain two terms: one with $G_{\rm scal}$ and one with a $\delta$ function. In Appendix~\ref{sec:QEDinv} and in Appendix~\ref{sec:monopoles}, QED gauge invariance and the cancellation of the instantaneous terms are proven. Eventually, we can write $W^{\mu \nu}$ in such a way that no instantaneous term contributes, and that the QED Ward-Takahasi identity $q_\mu W^{\mu \nu}= W^{\mu \nu}q_\nu=0$ is satisfied.

In terms of the scalar propagator, \eqn{eq:h-tensor} finally reads:
\beq\label{eq:h-tensor-2}
&& W^{\mu\nu} = \, e^2\,\sum_f q_f^2 \, \int \frac{\rmd^{D} \ell}{(2\pi)^D} \,\int  \frac{\rmd^{D} k }{(2\pi)^D} \tr_{\rm c}\left[G_\scal(-q+\ell,-q+\ell+k)  G_\scal(\ell+k,\ell) \right] \nn
&& \times   \frac{1}{4 \ell^+ (q-\ell)^+} \, \tr_{\rm s} \left[( \slashed{q}-\slashed{\ell}) \gamma^+ (\slashed{q}-\slashed{\ell}-\slashed{k}) \gamma^\nu (\slashed{\ell}+\slashed{k}) \gamma^+ \slashed{\ell} \gamma^\mu \right ] \, .
\eeq
It is most convenient to replace $\varepsilon^\mu$ by $\varepsilon^\mu - q^\mu \varepsilon^+/q^+$ when computing contractions, or equivalently to substitute $ \gamma^\mu \rightarrow \gamma^\mu - n_2^\mu \slashed{q}/q^+$ in $W^{\mu \nu}$. The same trick can be performed for the second photon Lorentz index $\nu$. This operation is free thanks to the Ward-Takahashi identity but allows for a more efficient computation of the contributions from longitudinal photons.

The Dirac trace contains structures that should be familiar to readers who are experienced in LFPT. Indeed, structures of the form $\gamma^+ (\slashed{b}-\slashed{a}) \slashed{\varepsilon} \slashed{a} \gamma^+ $ are exactly what generates the numerators in LFPT light cone wave functions, in a specific light cone gauge and with the appropriate arbitrary conventions for helicities and spinors. We can separate the Dirac trace in \eqn{eq:h-tensor} into two such structures sandwiched between $\gamma^-$ matrices, by simply writing $\gamma^+ = \gamma^+\gamma^-\gamma^+/2 $ for both $\gamma^+$ matrices.

Introducing 
\beq 
z \equiv \frac{\ell^+}{q^+}  \equiv 1-\bar{z}\,,
\eeq
 and making use of $+$ momentum conservation during the scattering with the external field, which implies that $k^+=0$, we can finally introduce the following objects:
\beq\label{eq:phiLTdef}
\phi_{L/\lambda}(z,\boldsymbol{\ell}) \equiv \gamma^+ \slashed{\ell} \left(\slashed{\varepsilon}_{L/\lambda}- \frac{\slashed{q}}{q^+} \varepsilon_{L/\lambda}^+\right) (\slashed{q}-\slashed{\ell}) \gamma^+ \,.
\eeq
For the longitudinal contribution, the following relation makes the computation of the explicit expression for $\phi_L$ straightforward:
\begin{equation}
	\left(\slashed{\varepsilon}_{L}-\frac{\slashed{q}}{q^{+}}\varepsilon_{L}^{+}\right)=\left(\varepsilon_{L}^{-}-\frac{q^{-}}{q^{+}}\varepsilon_{L}^{+}\right)\gamma^{+}=\frac{Q}{q^{+}}\gamma^{+}\label{eq:eL-qL},
\end{equation}
Consequently, all one needs to use is the trivial relation
\begin{equation}
	\gamma^{+}\slashed{u}\gamma^{+}=2u^{+}\gamma^{+},
\end{equation}
for any 4-vector $u$. We find:
\beq
	\phi_{L}(z,\boldsymbol{\ell})=4z\bar{z}q^{+}Q\gamma^{+}\label{eq:PhiLexp}.
\eeq
In the transverse case, one has
\begin{equation}
	\phi_{T}=\gamma^{+}(zq^{+}\gamma^{-}+\slashed{\ell}_{\perp})\slashed{\varepsilon}_{T}(\bar{z}q^{+}\gamma^{-}-\slashed{\ell}_{\perp})\gamma^{+}.
\end{equation}
Using the fact that for any transverse vector $u$, 
\begin{equation}
	\gamma^{+}\slashed{u}_{\perp}\gamma^{+}=\gamma^{-}\slashed{u}_{\perp}\gamma^{-}=0,
\end{equation}
we get:
\begin{equation}
	\phi_{\lambda}=2\bar{z}q^{+}\gamma^{+}\slashed{\ell}_{\perp}\slashed{\varepsilon}_{\lambda}-2zq^{+}\slashed{\varepsilon}_{\lambda}\slashed{\ell}_{\perp}\gamma^{+}\label{eq:PhiT2}.
\end{equation}
It can be rewritten in a way that can be more convenient in some contexts, and closer to the light cone wave functions found in LFPT, using the following: for any $u$ and $v$,
\begin{equation}
	\slashed{u}\slashed{v}=(u\cdot v)+\frac{1}{2}[\slashed{u},\slashed{v}].
\end{equation}
Then,
\begin{align}
	\phi_{\lambda} & =q^{+}\left\{ [\slashed{\ell}_{\perp},\slashed{\varepsilon}_{\lambda}]-2(z-\bar{z})(\ell_{\perp}\cdot\varepsilon_{\lambda})\right\} \gamma^{+}\label{eq:PhiTfin}.
\end{align}
We are now left with the following traces:  
\beq
	{\cal T}_{L}(z,\boldsymbol{\ell},\boldsymbol{k})\equiv{\rm tr}_{s}\left[\frac{\gamma^{-}}{2}\phi_{L}(z,\boldsymbol{\ell})\frac{\gamma^{-}}{2}\phi_{L}^{\ast}(z,\boldsymbol{\ell}+\boldsymbol{k})\right]\label{eq:LLtracedef},
\eeq
and
\beq
	{\cal T}_{T}(z,\boldsymbol{\ell},\boldsymbol{k})\equiv\frac{1}{d}\sum_\lambda{\rm tr}_{s}\left[\frac{\gamma^{-}}{2}\phi_{\lambda}(z,\boldsymbol{\ell})\frac{\gamma^{-}}{2}\phi_{\lambda}^{\ast}(z,\boldsymbol{\ell}+\boldsymbol{k})\right]\label{eq:TTtracedef}\,,
\eeq
while the amplitude reads
\beq\label{eq:h-tensor-2}
 {\cal A}_{T,L} & = & \, e^2\,\sum_f q_f^2 \, \int\! \frac{\rmd^{D} \ell}{(2\pi)^D} \,\int \! \frac{\rmd^{D} k }{(2\pi)^D} \frac{1}{4 \ell^+ (q^+-\ell^+)} \, {\cal T}_{T,L}(z,\boldsymbol{\ell},\boldsymbol{k}) \nn
&& \times   \tr_{\rm c}\left[G_\scal(-q+\ell,-q+\ell+k)  G_\scal(\ell+k,\ell) \right]  \, .
\eeq
Trivial algebra leads to:
\beq
	{\cal T}_{L}(z,\boldsymbol{\ell},\boldsymbol{k})=32z^{2}\bar{z}^{2}(q^{+})^{2}Q^{2}\label{eq:traceLL}.
\eeq
For the transverse case, using \eqn{eq:pol-comp} contracted with $g_{\perp \mu \nu}$ leads to:
\beq
	{\cal T}_{T}(z,\boldsymbol{\ell},\boldsymbol{k})=8(q^{+})^{2}\left(1-\frac{4}{d}z\bar{z}\right)\boldsymbol{\ell}\cdot(\boldsymbol{\ell}+\boldsymbol{k})\label{eq:traceTT}.
\eeq
We thus recovered the numerators one finds in photon wave functions in the LFPT description of DIS. However, the so-called energy denominators are not explicit in the current form of our expressions. In LFPT, \eqn{eq:traceLL} and \eqn{eq:traceTT} would involve the following additional denominators:
\beq
\boldsymbol{\ell}^2+z\bar{z}Q^2, \quad (\boldsymbol{\ell}+\boldsymbol{k})^2+z\bar{z}Q^2,
\eeq
which can be related to differences in light cone energies: 
\begin{equation}
	\boldsymbol{\ell}^{2}+z\bar{z}Q^{2}=2z\bar{z}q^{+}\left(E_{\ell}^--E^-_{\ell-q}-E^-_{q}\right),
\end{equation}
with $E_\ell^- \equiv \boldsymbol{\ell}^2/(2\ell^+)$, $E_{\ell-q}^- \equiv (\boldsymbol{\ell}-\boldsymbol{q})^2/(2(\ell-q)^+)$, and $E_q^- \equiv (q^2+\boldsymbol{q}^2)/(2q^+)$. In the shock wave approximation, these differences of energies appear from light cone time integrals from $-\infty$ to $0$:
\beq
\frac{1}{E_\ell^- -E^-_{\ell-q}-E_q^--i0}=-i\int_{-\infty}^0\rmd x^+ \rme^{ix^+(E_\ell^- -E_{\ell-q}^--E_q^--i0)}.\label{eq:denomint}
\eeq
This is allowed provided that the propagation time between the photon splitting and the scatterings with the target is much longer than the scattering time, see Eq.~(\ref{eq:times}). In the present framework, this light cone time separation is not assumed, thus the emergence of energy denominators is non trivial. This question is answered through a related observation we will prove in the end of this Section: the dipole size in the wave functions which appear in the amplitude is the size of the quark-antiquark dipole at the light cone time of the first scattering with the target, and the light cone denominator corresponds to this size at this time. This goes beyond the shock wave approximation, where all scatterings occur at the same time, by enforcing an ordering in $-$ momenta which is otherwise absent. In other words, any gluon fluctuation in the target is bound to happen at shorter times than the photon splitting time.

\subsection{Operator algebra and energy denominators}\label{sec:denom}

We will now prove the emergence of light cone energy denominators by extracting free propagators from the scattering operators $G_{\rm scal}$. The main idea is to identify the first $A^-$ insertion that will set the first interaction time with the target which will allow to integrate the prior free propagators as is done in the shock wave framework, only instead of integrating the photon time down to the origin the integral is bounded by the first interaction time $z^+_1$ and the final one $z^+_2$. 
It is convenient to work temporarily in position space  
\beq\label{eq:G-mom}
&& \, G_\scal(\ell_2,\ell_1)= \int \rmd^D x_1 \int \rmd^D x_2  \, \rme^{i \ell_2\cdot x_2 - i \ell_1\cdot  x_1}  \,  G_\scal (x_2,x_1) \,.
\eeq
There, 
\beq\label{eq:GG-mom}
&& \, G_\scal(\ell_2,\ell_1) \, G_\scal(-q+\ell_1,-q+\ell_2)  = \int \rmd^D x_1 \int \rmd^D x_2  \, \rme^{i \ell_2\cdot x_2 - i \ell_1\cdot  x_1}  \nn
&&\times \int \rmd^D y_1 \int \rmd^D y_2  \, \rme^{- i (-q+\ell_2)\cdot y_2+i (-q+\ell_1)\cdot  y_1}  \, G_\scal (x_2,x_1)\,  G_\scal (y_1,y_2)   \,.
\eeq
In the hadronic tensor, the hard wave functions do not depend on the $-$ components of the momenta $\ell_1 = \ell$ and $\ell_2 = \ell + k $, which means the $\ell_1^-$ and $\ell_2^-$ integrals will set $x_1^+=y_1^+$ and $x_2^+=y_2^+$. In other words, the propagators are evaluated between the same light cone times in reverse order. It thus means that one propagator will be retarded and the other one will be advanced.

The interactions with the target field may occur at any light cone time between $x_1^+$ and $x_2^+$, which correspond to the times of the photon splitting into the quark-antiquark pair and the merging of the latter back into the photon, respectively. Starting from \eqn{eq:GG-mom}, our goal is to integrate over these two times in a similar fashion to \eqn{eq:denomint} but with an upper bound related to the time of the first scattering instead of $0$. To do so, we may identify the first and final dipole-target interactions.

Let us focus first on the first interaction. We begin by writing the retarded quark and the advanced anti-quark propagators as
\beq \label{eq:G-position}
G^R_\scal (x_2,x_1) =  G^R_0 (x_2-x_1)  - 2ig  \int \rmd^D x_3 \, G^R_\scal (x_2,x_3) A^-(x_3) \del_3^+ G^R_0(x_3,x_1) 
\eeq
and 
\beq \label{eq:G-anti-position}
G^A_\scal (y_1,y_2) =  G^A_0 (y_1-y_2)  - 2ig  \int \rmd^D y_3 \, G^A_0(y_1,y_3) A^-(y_3) \del_3^+ G_\scal^A(y_3,y_2) \,.
\eeq
Here, it is sufficient to define the retarded (resp. advanced) propagators $G^{A,R}(x_2,x_1)$ by imposing $x_2^+>x_1^+$ (resp. $x_2^+<x_1^+$). These relations enable us to extract the first interaction on the quark (retarded propagator $G^R$) and antiquark (advanced propagator $G^A$). It is now necessary to distinguish two cases: either the quark or the antiquark scatters first. Considering only the product of the second terms in the r.h.s. of \eqn{eq:G-position} and \eqn{eq:G-anti-position}, the distinction can be made by writing $1=\theta(x_3^+ -y_3^+)+\theta(y_3^+-x_3^+)$ in the integrand, and by simplifying both terms separately. Let us consider the $\theta(y_3^+-x_3^+)$ contribution for a while.

\begin{figure}
\begin{center}
\includegraphics[width=0.7\textwidth]{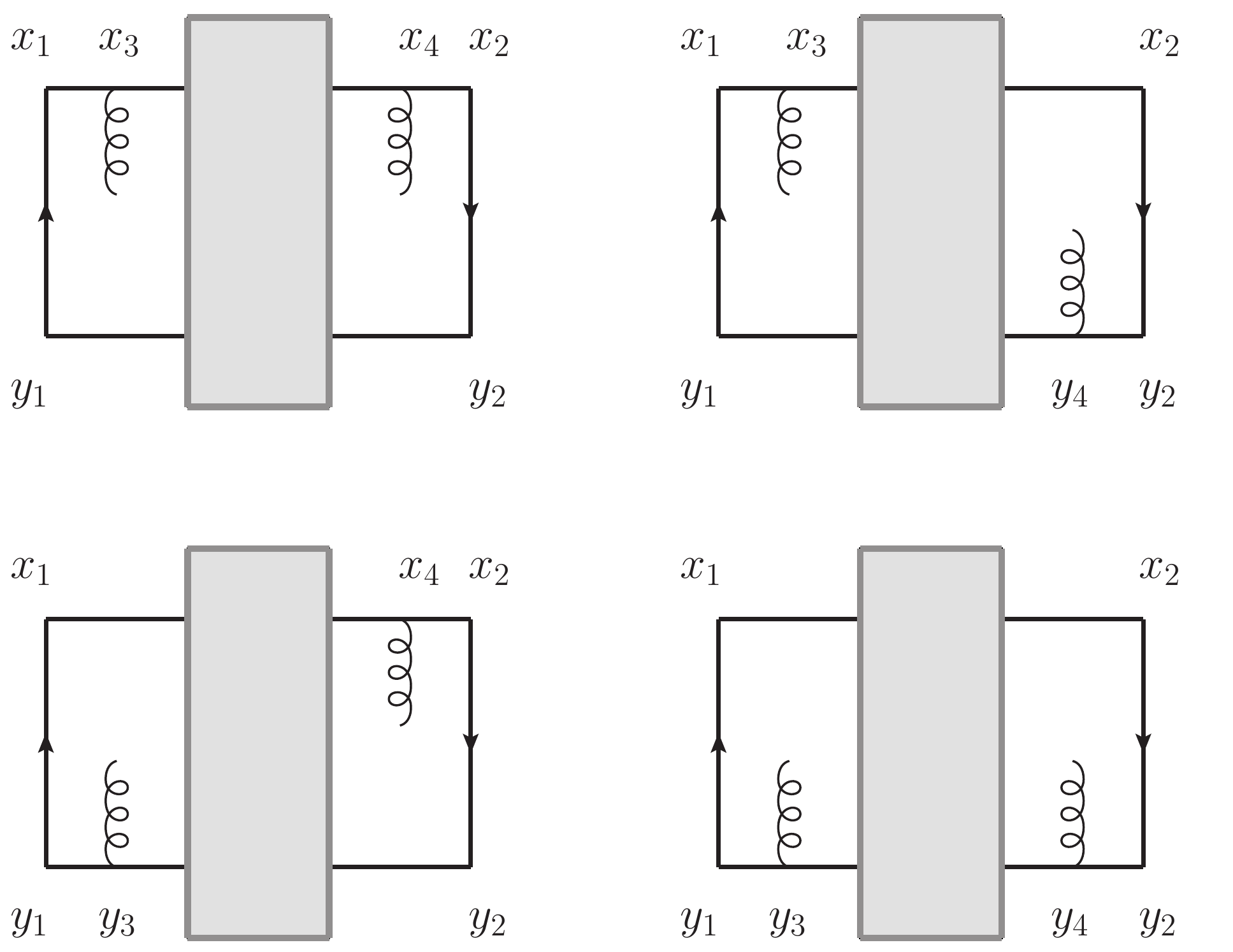} 
\end{center}
\caption{Illustration of the 4 topologies contributing to \eqn{eq:GRGAfin} where the first and last interactions with the background field are extracted. The gray boxes represent higher order gluon exchanges with the target as shown in Figure~\ref{fig:long-DIS-fig-1}. }
\label{fig:long-DIS-GG-evol-2}
\end{figure}

For any $x_3^+$ such that ${\rm max}\left(\,y_3^+,y_1^+\right)>x_3^+>{\rm min}\left(\,y_3^+,y_1^+\right)$ one can use the following relation:
\beq \label{eq:conv-prop}
G_0(y_1,y_3)= {\rm sgn}(y_1^+-y_3^+) \int \rmd^D u\, \delta(u^+-x_3^+) \, G_0(y_1,u)\,  2\del_u^+ \, G_0(u,y_3)\,.
\eeq
It is generalized to the full scalar propagator in Appendix~\ref{sec:convolution}, but we only need the free case for this computation.
As stated before, we will focus for the moment on the product between the second terms in the r.h.s. of \eqn{eq:G-position} and \eqn{eq:G-anti-position} with a $\theta(x_3^+-y_3^+)$ function. Following the discussion at the beginning of this section, we have $x_1^+=y_1^+$ so $y_3^+<x_3^+<y_1^+$. The contribution to this term from the antiquark propagator can be rewritten using \eqn{eq:conv-prop}
\begin{align}\label{eq:1st-inter}
	& - 2ig  \int \rmd^D y_3 \, G^A_0(y_1,y_3) A^-(y_3) \del_3^+ G_\scal^A(y_3,y_2)\theta(x_3^+-y_3^+) \nonumber \\
	 = & 4ig  \int \rmd^Du \,\delta(u^+-x_3^+)  \int \rmd^D y_3 \, G^A_0(y_1,u) \partial_u^+G^A_0(u,y_3) A^-(y_3) (\del_3^+  G_\scal^A)(y_3,y_2). \nn
\end{align}
Using \eqn{eq:G-anti-position} with $u$ instead of $y_1$ \eqn{eq:1st-inter} yields
\begin{align}
	& - 2ig  \int \rmd^D y_3 \, G^A_0(y_1,y_3) A^-(y_3) \del_3^+ G_\scal^A(y_3,y_2)\theta(x_3^+-y_3^+) \\
	= & -2   \int \rmd^Du \, \delta(u^+-x_3^+)  G^A_0(y_1,u) \partial_u^+
	[G_{\rm scal}^A(u,y_2)-G_0^A(u,y_2)] \nonumber.
\end{align}
Then, using again \eqn{eq:conv-prop} allows to cast the second term in the brackets into a counterterm to the first term in \eqn{eq:G-anti-position}. Also integrating the partial derivative by parts, we get:
\begin{align}
	& - 2ig  \int \rmd^D y_3 \, G^A_0(y_1,y_3) A^-(y_3) \del_3^+ G_\scal^A(y_3,y_2)\theta(x_3^+-y_3^+) \\
	= & -G_0^A(y_1,y_2)   +2   \int \rmd^Dy_3 \, \delta(y_3^+-x_3^+) (\partial_3^+G^A_0)(y_1,y_3) 
	G_{\rm scal}^A(y_3,y_2), \nonumber
\end{align}
where we renamed the local variable $u$ back to $y_3$ for more readable notations.
Combining the above result with the $\theta(y_3^+-x_3^+)$ contribution that can be  obtained in a similar fashion yields
\beq
&& \,\tr_c \,G^R_\scal (x_2,x_1)\, G^A_\scal (y_1,y_2) =\,N_c  \,G^R_0 (x_2,x_1)\, G^A_0 (y_1,y_2)  -4ig \,\int \rmd^D x_3\, \int \rmd^D y_3 \delta(x_3^+-y_3^+) \, \,  \nn
&& \times \tr_c \Big[   \, G^R_\scal (x_2,x_3) \left(A^-(x_3)-A^-(y_3)\right)\,G^A_\scal(y_3,y_2)   \Big]  \, (\del_3^+ G^R_0)(x_3,x_1) (\del_{3}^+ G^A_0)(y_1,y_3). \label{eq:initial-times}
\eeq 
We now isolated the first scattering on the target, which will occur at light cone time $x_3^+=y_3^+$. Note that the first term in the r.h.s. is a disconnected contribution where the photon never interacts with the target and it will thus be subtracted.
Using the cyclicity of the trace, we can now perform the same steps as from \eqn{eq:G-position} to \eqn{eq:initial-times}, this time on $G_{\rm scal}^A(y_3,y_2)G_{\rm scal}^R(x_2,x_3)$, in order to make the last scattering appear as well. The result reads
\begin{eqnarray}
	&  & G_{{\rm scal}}^{A}(y_{3},y_{2})G_{{\rm scal}}^{R}(x_{2},x_{3})=G_{0}^{A}(y_{3},y_{2})G_{0}^{R}(x_{2},x_{3})+4ig\int{\rm d}^{D}x_{4}\int{\rm d}^{D}y_{4}\delta(x_{4}^{+}-y_{4}^{+})\nonumber \\
	&  & \times G_{{\rm scal}}^{A}(y_{3},y_{4})\left[A^{-}(y_{4})-A^{-}(x_{4})\right]G_{{\rm scal}}^A(x_{4},x_{3})(\partial_{y_{4}}^{+}G_{0}^{A})(y_{4},y_{2})(\partial_{x_{4}}^{+}G_{0}^{R})(x_{2},x_{4})\label{eq:GAGR}.
\end{eqnarray}
Plugging \eqn{eq:GAGR} into \eqn{eq:initial-times} and using the fact that ${\rm tr}_c (A^-) = A_a^- {\rm tr}_c (t^a)=0$, we conclude:
\begin{align}
	& {\rm tr}G_{{\rm scal}}^{R}(x_{2},x_{1})G_{{\rm scal}}^{A}(y_{1},y_{2})\nonumber \\
	& =16g^{2}\int{\rm d}^{D}x_{3}\int{\rm d}^{D}x_{4}\int{\rm d}^{D}y_{3}\int{\rm d}^{D}y_{4}\delta(y_{3}^{+}-x_{3}^{+})\delta(x_{4}^{+}-y_{4}^{+})\label{eq:GRGAfin} \nonumber \\
	& \times(\partial_{x_{3}}^{+}G_{0}^{R})(x_{3},x_{1})(\partial_{x_{4}}^{+}G_{0}^{R})(x_2,x_{4})(\partial_{y_{3}}^{+}G_{0}^{A})(y_{1},y_{3})(\partial_{y_{4}}^{+}G_{0}^{A})(y_{4},y_{2}) \\
	& \times{\rm tr}\left\{ \left[A^{-}(y_{3})-A^{-}(x_{3})\right]G_{{\rm scal}}^{A}(y_{3},y_{4})\left[A^{-}(y_{4})-A^{-}(x_{4})\right]G_{{\rm scal}}^{R}(x_{4},x_{3})\right\}, \nonumber 
\end{align} 
where we subtracted the disconnected contribution. We finally found the form we set out to derive: the first and last interactionswith the target, respectively occurring at light cone times $x_3^+=y_3^+$ and $x_4^+=y_4^+$, have been isolated. There are 4 contributions, which are depicted in Figure \ref{fig:long-DIS-GG-evol-2}, given that each of them could occur either on the quark or on the antiquark, hence the $[A^-(y_3)-A^-(x_3)][A^-(y_4)-A^-(x_4)]$ structure. To the small $x$ expert, this is reminiscent of the structures encountered when performing the so-called dilute perturbative expansion of the non-perturbative Wilson line operators into Reggeons which appear in the shock wave approximation, see e.g.~\cite{Caron-Huot:2013fea}.

With Reggeons defined in terms of the infinite length Wilson line in the adjoint representation as
\begin{equation}
	R^{a}(\boldsymbol{x})=\frac{f^{abc}}{gC_{A}}\left(\ln U_{\boldsymbol{x}}^{{\rm adj}}\right)^{bc}\label{eq:reggeons},
\end{equation}
the proper perturbative expansion of the dipole operator reads:
\begin{align}
	1 - \frac{1}{N_c} {\rm tr}(U_{\boldsymbol{x}_{1}}U_{\boldsymbol{x}_{2}}^{\dagger}) & =\frac{\pi\alpha_s}{N_c}\left[R^{a}(\boldsymbol{x}_{1})-R^{a}(\boldsymbol{x}_{2})\right]\left[R^{a}(\boldsymbol{x}_{1})-R^{a}(\boldsymbol{x}_{2})\right]+O(g^3)\label{eq:Reggeons}.
\end{align}
Treating the target perturbatively in \eqn{eq:GRGAfin} would amount to replacing $G_{\rm scal} \rightarrow G_0$ and readily finding this exact structure. Note that it already appears in a more complicated form in full generality in this equation.
Keeping in mind that the target fields do not depend on $x^-$ in the present framework and that light cone times are identical ($x_{3,4}^+=y_{3,4}^+$), the coordinates at which the fields are evaluated in the differences only differ by their transverse components.
In a way, this difference of target fields with only a transverse separation, along with the propagators ending at the light cone times of those fields, is a generalization of the following property for straight Wilson lines on the light cone:
\begin{equation}
	\frac{\partial}{\partial x^{+}}[\,y^{+},x^{+}]_{\boldsymbol{x}_{1}}[\,x^{+},z^{+}]_{\boldsymbol{x}_{2}}=-ig[\,y^{+},x^{+}]_{\boldsymbol{x}_{1}}\left[A^{-}(x^{+},\boldsymbol{x}_{1})-A^{-}(x^{+},\boldsymbol{x}_{2})\right][\,x^{+},z^{+}]_{\boldsymbol{x}_{2}}\label{eq:WLder},
\end{equation}
where instead of the Wilson lines we have non-trivial propagators. In fact, this relation will be used in the Regge limit to relate the amplitude to its shock wave formulation in Section~\ref{sec:Regge}.
Because of our additional assumption that transverse target fields are gauged away, as is standard in small $x$ physics, we also have the following relation:
\beq \label{eq:FST}
A^-(x^+,\x)-A^-(x^+,\y) &=&  \int_0^1 \rmd s \frac{\rmd }{\rmd s}A^-(x^+,\y+s\r) \nn 
&=&  - \r^i\,  \int_0^1 \rmd s  \,   \del^i A^-(x^+,\y+s\r) \label{eq:AtoF} \\
&=&  - \int_{\y}^{\x}\rmd\z^i(s) \, F^{i-}(x^+,\z(s)) \nn 
\eeq
where 
\beq
\z(s) =  \x+s \r, \qquad \r=\x-\y\,
\eeq
and
\beq
F^{i-}(x^+,\z) = \del^i A^-(x^+,\z)
\eeq

\begin{figure}
\begin{center}
\includegraphics[width=0.7\textwidth]{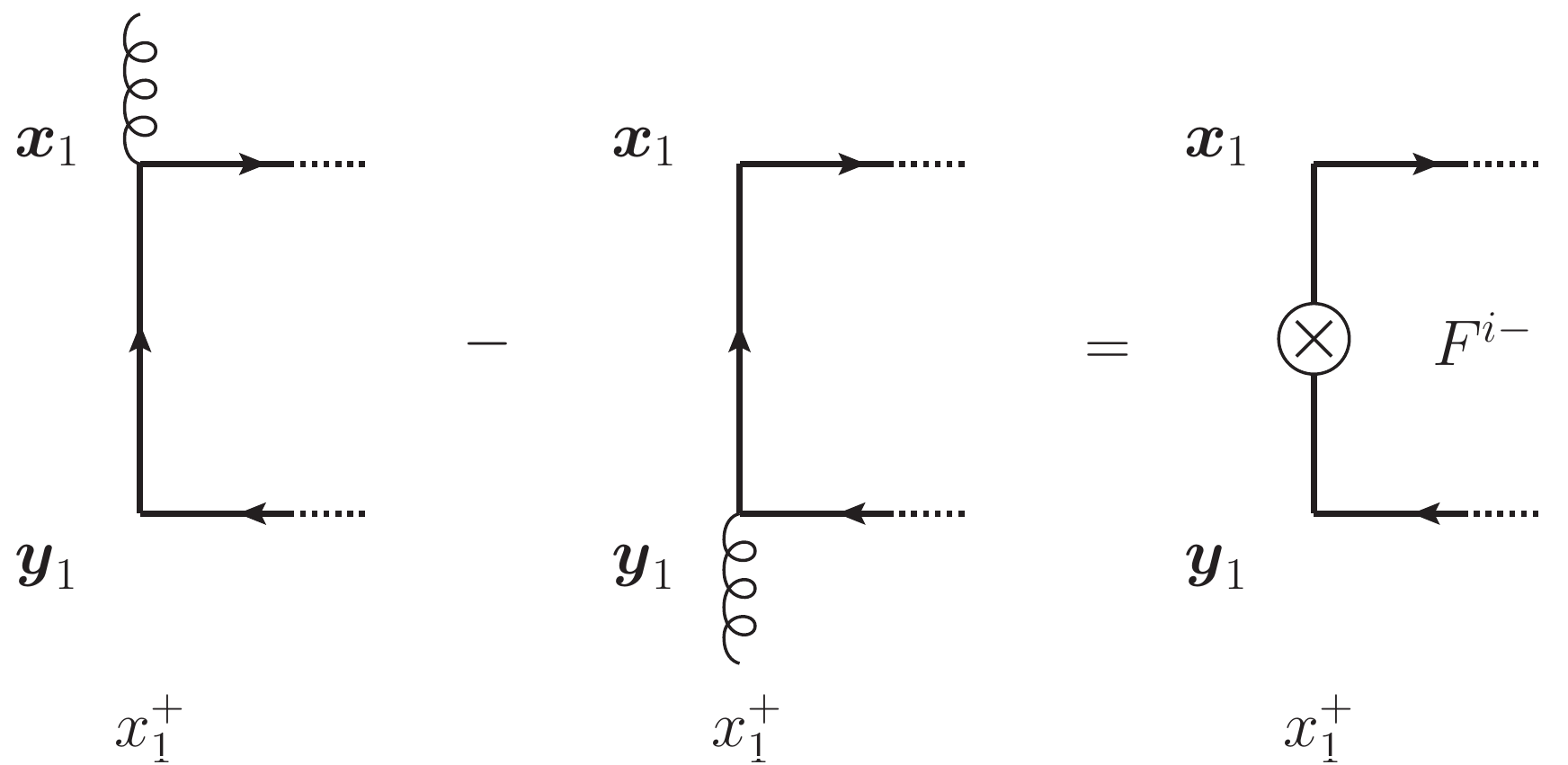} 
\end{center}
\caption{Diagrammatic representation of \eqn{eq:FST}. }
\label{fig:long-DIS-AA-F}
\end{figure}
%
Now that we have introduced the scattering coordinates $(x_3,x_4,y_3,y_4)$, we can finally take the Fourier transform over the initial coordinates $(x_1, x_2, y_1, y_2)$ in \eqn{eq:GG-mom}. Let us now show how it will lead to a more physical version of \eqn{eq:denomint} where the upper bound is the light cone time of the first scattering, which will result in the energy denominators along with a non-zero phase.
For the $x_1$ and $y_1$ integrals, we will need the following:
\begin{align}
	& \int{\rm d}^{D}x_{1}{\rm d}^{D}y_{1}{\rm e}^{-i(\ell_{1}\cdot x_{1})+i(-q+\ell_{1})\cdot y_{1}}(\partial_{x_{3}}^{+}G_{0}^{R})(x_{3},x_{1})(\partial_{y_{3}}^{+}G_{0}^{A})(y_{1},y_{3})\label{eq:FTGG}\\
	& =\int_{-\infty}^{x_{3}^{+}}{\rm d}x_{1}^{+}\int_{-\infty}^{y_{3}^{+}}{\rm d}y_{1}^{+}{\rm e}^{-i\ell_{1}^{-}x_{1}^{+}+i(-q^{-}+\ell_{1}^{-})y_{1}^{+}}\nonumber \\
	& \times\int\frac{{\rm d}k_{1}^{-}}{2\pi}\frac{{\rm d}k_{2}^{-}}{2\pi}\frac{{\rm e}^{-ik_{1}^{-}(x_{3}^{+}-x_{1}^{+})-ik_{2}^{-}(y_{1}^{+}-y_{3}^{+})-i\ell_{1}^{+}(x_{3}^{-}-y_{3}^{-})+i\boldsymbol{\ell}_{1}\cdot(\boldsymbol{x}_{3}-\boldsymbol{y}_{3})-iq^{+}y_{3}^{-}}}{4\left(k_{1}^{-}-\frac{\boldsymbol{\ell}_{1}^{2}-i0}{2\ell_{1}^{+}}\right)\left(k_{2}^{-}-\frac{\boldsymbol{\ell}_{1}^{2}-i0}{2(-q^{+}+\ell_{1}^{+})}\right)}\nonumber .
\end{align}
Using the pole integral \eqn{eq:Cauchy-int} leads to:
	\begin{align}
	& \frac{1}{4}\theta(\ell_{1}^{+})\theta(q^{+}-\ell_{1}^{+})\int_{-\infty}^{x_{3}^{+}}{\rm d}x_{1}^{+}\int_{-\infty}^{y_{3}^{+}}{\rm d}y_{1}^{+}{\rm e}^{-i\ell_{1}^{-}x_{1}^{+}+i(-q^{-}+\ell_{1}^{-})y_{1}^{+}}\label{eq:FTGGpole}\\
	& \times{\rm e}^{-i\frac{\boldsymbol{\ell}_{1}^{2}-i0}{2\ell_{1}^{+}}(x_{3}^{+}-x_{1}^{+})-i\frac{\boldsymbol{\ell}_{1}^{2}-i0}{2(-q^{+}+\ell_{1}^{+})}(y_{1}^{+}-y_{3}^{+})-i\ell_{1}^{+}(x_{3}^{-}-y_{3}^{-})+i\boldsymbol{\ell}_{1}\cdot(\boldsymbol{x}_{3}-\boldsymbol{y}_{3})-iq^{+}y_{3}^{-}}\nonumber .
\end{align}
In the full cross section, we have to integrate w.r.t. $\ell_1^-$ and $\ell_2^-$, keeping in mind that the hard wave functions do not depend on these variables. This integral will then set $x_1^+=y_1^+$ and the conditions on light cone times imposed by the fact that one propagator was retarded and the other one was advanced now sets $x_1^+<{\rm min}(x_3^+,y_3^+)$. One is then left with the integral
\begin{equation}
	\int_{-\infty}^{{\rm min}(x_{3}^{+},y_{3}^{+})}{\rm d}x_{1}^{+}{\rm e}^{i(E_{\ell_{1}}^{-}-E_{\ell_{1}-q}^{-}-E_{q}^{-}-i0)x_{1}^{+}}=\frac{{\rm e}^{i(E_{\ell_{1}}^{-}-E_{\ell_{1}-q}^{-}-E_{q}^{-}){\rm min}(x_{3}^{+},y_{3}^{+})}}{i(E_{\ell_{1}}^{-}-E_{\ell_{1}-q}^{-}-E_{q}^{-}-i0)}\label{eq:denom-with-ordering}
\end{equation}
which is the equivalent of \eqn{eq:denomint} with the additional condition that the photon splitting time $x_1^+$ has to occur before the first scattering time ${\rm min}(x_{3}^{+},y_{3}^{+})$. We now have an energy denominator, along with a phase which would not appear in the shock wave approximation.
In a similar fashion, the integration w.r.t. $x_2$ and $y_2$ involves the complex conjugate energy denominator:
\begin{equation}
	\int_{{\rm max}(x_{4}^{+},y_{4}^{+})}^{+\infty}{\rm d}x_{2}^{+}{\rm e}^{i(E_{q}^{-}+E_{\ell_{2}-q}^{-}-E_{\ell_{2}}^{-}+i0)x_{2}^{+}}=\frac{-{\rm e}^{i(E_{q}^{-}+E_{\ell_{2}-q}^{-}-E_{\ell_{2}}^{-}){\rm max}(x_{4}^{+},y_{4}^{+})}}{i(E_{q}^{-}+E_{\ell_{2}-q}^{-}-E_{\ell_{2}}^{-}+i0)}\label{eq:denom-with-antiordering},
\end{equation}
where the light cone time $x_2^+$ of the quark-antiquark pair merging into a photon now has to occur after the last scattering time ${\rm max}(y_4^+,x_4^+)$.
We finally have two relations:
\begin{align}
	& \int\frac{{\rm d}\ell_{1}^{-}}{2\pi}\int{\rm d}^{D}x_{1}{\rm d}^{D}y_{1}{\rm e}^{-i(\ell_{1}\cdot x_{1})+i(-q+\ell_{1})\cdot y_{1}}(\partial_{x_{3}}^{+}G_{0}^{R})(x_{3},x_{1})(\partial_{y_{3}}^{+}G_{0}^{A})(y_{1},y_{3})\nonumber \\
	& =\frac{1}{4i}\frac{\theta(\ell_{1}^{+})\theta(q^{+}-\ell_{1}^{+})}{\frac{q^{+}}{2\ell_{1}^{+}(q^{+}-\ell_{1}^{+})}\boldsymbol{\ell}_{1}^{2}-q^{-}-i0}{\rm e}^{-i\ell_{1}^{+}x_{3}^{-}+i\boldsymbol{\ell}_{1}\cdot(\boldsymbol{x}_{3}-\boldsymbol{y}_{3})-i(q^{+}-\ell_{1}^{+})y_{3}^{-}}\label{eq:x1y1int}\\
	& \times\left[\theta(x_{3}^{+}-y_{3}^{+}){\rm e}^{-iq^{-}y_{3}^{+}-i\frac{\boldsymbol{\ell}_{1}^{2}-i0}{2\ell_{1}^{+}}(x_{3}^{+}-y_{3}^{+})}+\theta(y_{3}^{+}-x_{3}^{+}){\rm e}^{-iq^{-}x_{3}^{+}-i\frac{\boldsymbol{\ell}_{1}^{2}-i0}{2(q^{+}-\ell_{1}^{+})}(y_{3}^{+}-x_{3}^{+})}\right]\nonumber ,
\end{align}
and
\begin{align}
	& \int\frac{{\rm d}\ell_{2}^{-}}{2\pi}\int{\rm d}^{D}x_{2}{\rm d}^{D}y_{2}{\rm e}^{i(\ell_{2}\cdot x_{2})-i(-q+\ell_{2})\cdot y_{2}}(\partial_{x_{4}}^{+}G_{0}^{R})(x_{2},x_{4})(\partial_{y_{4}}^{+}G_{0}^{A})(y_{4},y_{2})\nonumber \\
	& =\frac{1}{4i}{\rm e}^{i\ell_{2}^{+}x_{4}^{-}+i(q^{+}-\ell_{2}^{+})y_{4}^{-}-i\boldsymbol{\ell}_{2}\cdot(\boldsymbol{x}_{4}-\boldsymbol{y}_{4})}\frac{\theta(\ell_{2}^{+})\theta(q^{+}-\ell_{2}^{+})}{\frac{q^{+}}{2\ell_{2}^{+}(q^{+}-\ell_{2}^{+})}\boldsymbol{\ell}_{2}^{2}-q^{-}-i0}\label{eq:x2y2int}\\
	& \left[\theta(x_{4}^{+}-y_{4}^{+}){\rm e}^{iq^{-}x_{4}^{+}-i\frac{\boldsymbol{\ell}_{2}^{2}-i0}{2(q^{+}-\ell_{2}^{+})}(x_{4}^{+}-y_{4}^{+})}+\theta(y_{4}^{+}-x_{4}^{+}){\rm e}^{iq^{-}y_{4}^{+}-i\frac{\boldsymbol{\ell}_{2}^{2}-i0}{2\ell_{2}^{+}}(y_{4}^{+}-x_{4}^{+})}\right]\nonumber. 
\end{align}
Note that for $x_3^+\rightarrow y_3^+$ and $x_4^+ \rightarrow y_4^+$ as is set by the $\delta$ functions in \eqn{eq:GRGAfin}, the brackets in \eqn{eq:x1y1int} and in \eqn{eq:x2y2int} simply reduce to $\rme^{-iq^-x_3^+} $ and $\rme^{iq^-x_4^+}$, respectively.
Plugging these two relations into \eqn{eq:GRGAfin}, then renaming local variables with 3 (resp. 4) subscripts with 1 (resp. 2) subscripts for more readability, we obtain:
\begin{align}
	& \int\frac{{\rm d}\ell_{1}^{-}}{2\pi}\frac{{\rm d}\ell_{2}^{-}}{2\pi}G_{{\rm scal}}(\ell_{2},\ell_{1})G_{{\rm scal}}(-q+\ell_{1},-q+\ell_{2}) \nonumber \\
	& =g^{2}\int{\rm d}^{D}x_{1}\int{\rm d}^{D}x_{2}\int{\rm d}^{D}y_{1}\int{\rm d}^{D}y_{2}\delta(y_{1}^{+}-x_{1}^{+})\delta(x_{2}^{+}-y_{2}^{+}){\rm e}^{-iq^{-}(x_{1}^{+}-x_{2}^{+})}\nonumber \\
	& \times{\rm e}^{-i\ell_{1}^{+}x_{1}^{-}-i(q^{+}-\ell_{1}^{+})y_{1}^{-}+i\ell_{2}^{+}x_{2}^{-}+i(q^{+}-\ell_{2}^{+})y_{2}^{-}+i\boldsymbol{\ell}_{1}\cdot(\boldsymbol{x}_{1}-\boldsymbol{y}_{1})-i\boldsymbol{\ell}_{2}\cdot(\boldsymbol{x}_{2}-\boldsymbol{y}_{2})}\label{eq:GGfin} \\
	& \times{\rm tr}_{c}\left\{ G_{{\rm scal}}^{R}(x_{2},x_{1})\left[A^{-}(x_{1})-A^{-}(y_{1})\right]G_{{\rm scal}}^{A}(y_{1},y_{2})\left[A^{-}(y_{2})-A^{-}(x_{2})\right]\right\} \nonumber \\
	& \times\frac{\theta(\ell_{1}^{+})\theta(q^{+}-\ell_{1}^{+})}{\frac{q^{+}}{2\ell_{1}^{+}(q^{+}-\ell_{1}^{+})}\boldsymbol{\ell}_{1}^{2}-q^{-}-i0}\frac{\theta(\ell_{2}^{+})\theta(q^{+}-\ell_{2}^{+})}{\frac{q^{+}}{2\ell_{2}^{+}(q^{+}-\ell_{2}^{+})}\boldsymbol{\ell}_{2}^{2}-q^{-}-i0} \nonumber.
\end{align}
In the present case, we can write this equation in a more compact way, introducing back $z\equiv \ell_1^+ /q^+=\ell_2^+/q^+\equiv 1-z$, the notation $x_{12}\equiv x_1-x_2$, and using the fact that
\begin{equation}
	\frac{1}{\frac{q^{+}}{2\ell_{1,2}^{+}(q^{+}-\ell_{1,2}^{+})}\boldsymbol{\ell}_{1,2}^{2}-q^{-}-i0}=\frac{2z\bar{z}q^{+}}{\boldsymbol{\ell}_{1,2}^{2}+2z\bar{z}Q^{2}}\label{eq:denom-simp}.
\end{equation}
Then,
\begin{align} & \int\frac{{\rm d}\ell_{1}^{-}}{2\pi}\frac{{\rm d}\ell_{2}^{-}}{2\pi}G_{{\rm scal}}(\ell_{2},\ell_{1})G_{{\rm scal}}(-q+\ell_{1},-q+\ell_{2})\nonumber \\
	& =\int{\rm d}^{D}x_{1}\int{\rm d}^{D}x_{2}\int{\rm d}^{D}y_{1}\int{\rm d}^{D}y_{2}\delta(y_{1}^{+}-x_{1}^{+})\delta(x_{2}^{+}-y_{2}^{+})\nonumber \\
	& \times{\rm e}^{-izq^{+}x_{12}^{-}-i\bar{z}q^{+}y_{12}^{-}+i\boldsymbol{\ell}_{1}\cdot(\boldsymbol{x}_{1}-\boldsymbol{y}_{1})-i\boldsymbol{\ell}_{2}\cdot(\boldsymbol{x}_{2}-\boldsymbol{y}_{2})-iq^{-}x_{12}^{+}}\label{eq:GG-fin-2} \\
	& \times{\rm tr}_{c}\left\{ G_{{\rm scal}}^{R}(x_{2},x_{1})\left[A^{-}(x_{1})-A^{-}(y_{1})\right]G_{{\rm scal}}^{A}(y_{1},y_{2})\left[A^{-}(y_{2})-A^{-}(x_{2})\right]\right\}\nonumber  \\
	& \times\frac{\alpha_{s}}{\pi}\frac{z^{2}\bar{z}^{2}(q^{+})^{2}\theta(z)\theta(1-z)}{(\boldsymbol{\ell}_{1}^{2}+2z\bar{z}Q^{2})(\boldsymbol{\ell}_{2}^{2}+2z\bar{z}Q^{2})}.\nonumber 
\end{align}
Through a bit of involved operator algebra for the scalar propagator, we finally managed to accomplish two things. On the one hand, we extracted the so-called energy denominators, in the last line in Eq.~(\ref{eq:GGfin}). This will allow us to reconstruct the complete wave functions in the next Section. On the other hand, we have an explicit dependence on the time of the first scattering of the quark-antiquark pair with the target. Whereas in the shock wave framework (see Section~\ref{sec:shock-waves}) we would have entirely decoupled those times from the photon splitting time, we now have a naturally imposed ordering between splittings and scatterings.

\subsection{The hadronic tensor: full result}\label{sec:fulltensor}

We will now combine the wave functions from Section~\ref{sec:wf} with the energy denominators and operators from Section~\ref{sec:denom}. It is convenient to introduce the $(d+1)$-dimensional propagator from \eqn{eq:cG-def} in order to integrate out the $-$ components of the coordinates:
\begin{equation}
	\int{\rm d}x_{2}^{-}{\rm d}x_{1}^{-}{\rm e}^{ix_{2}^{-}\ell_{2}^{+}-ix_{1}^{-}\ell_{1}^{+}}G_{{\rm scal}}^{R}(x_{2},x_{1})=(2\pi)\delta(\ell_{2}^{+}-\ell_{1}^{+}) \frac{1}{2i \ell_1^+} (\boldsymbol{x}_{2}|{\cal G}_{\ell_{1}^{+}}(x_{2}^{+},x_{1}^{+})|\boldsymbol{x}_{1})\label{eq:intl-},
\end{equation}
with the Fourier transform
\begin{equation}
	(\boldsymbol{x}_{2}|{\cal G}_{\ell_{1}^{+}}(x_{2}^{+},x_{1}^{+})|\boldsymbol{x}_{1})\equiv\int\frac{{\rm d}^{d}\boldsymbol{\ell}_{2}}{(2\pi)^{d}}\frac{{\rm d}^{d}\boldsymbol{\ell}_{1}}{(2\pi)^{d}}{\rm e}^{i(\boldsymbol{\ell}_{2}\cdot\boldsymbol{x}_{2})-i(\boldsymbol{\ell}_{1}\cdot\boldsymbol{x}_{1})}(\boldsymbol{\ell}_{2}|{\cal G}_{\ell_{1}^{+}}(x_{2}^{+},x_{1}^{+})|\boldsymbol{\ell}_{1})\label{eq:curly-G-tf}.
\end{equation}
After a few trivial integrals, the amplitude reads:
\begin{align}
	{\cal A}_{T,L} & =e^{2}g^{2}(2\pi)\delta(0^{+})\sum_{f}q_{f}^{2}\int_{0}^{1}\frac{{\rm d}z}{2\pi}\int\frac{{\rm d}^{d}\boldsymbol{\ell}_{1}}{(2\pi)^{d}}\int\frac{{\rm d}^{d}\boldsymbol{\ell}_{2}}{(2\pi)^{d}}\nonumber \\
	& \times\int{\rm d}x_{1}^{+}\int{\rm d}x_{2}^{+}\int{\rm d}^{d}\boldsymbol{x}_{1}\int{\rm d}^{d}\boldsymbol{x}_{2}\int{\rm d}^{d}\boldsymbol{y}_{1}\int{\rm d}^{d}\boldsymbol{y}_{2}\label{eq:W-fin-mom}\\
	& \times{\rm e}^{-iq^{-}(x_{1}^{+}-x_{2}^{+})}{\rm e}^{i\boldsymbol{\ell}_{1}\cdot(\boldsymbol{x}_{1}-\boldsymbol{y}_{1})-i\boldsymbol{\ell}_{2}\cdot(\boldsymbol{x}_{2}-\boldsymbol{y}_{2})}\frac{z\bar{z}q^{+}{\cal T}_{T,L}\left(z,\boldsymbol{\ell}_{1},\boldsymbol{\ell}_{2}-\boldsymbol{\ell}_{1}\right)}{(\boldsymbol{\ell}_{1}^{2}+z\bar{z}Q^{2})(\boldsymbol{\ell}_{2}^{2}+z\bar{z}Q^{2})}\nonumber \\
	& \times{\rm tr}_{c}\left\{ (\boldsymbol{x}_{2}|{\cal G}_{zq^{+}}^{R}(x_{2}^{+},x_{1}^{+})|\boldsymbol{x}_{1})\left[A^{-}(x_{1}^{+},\boldsymbol{x}_{1})-A^{-}(x_{1}^{+},\boldsymbol{y}_{1})\right]\right.\nonumber \\
	& \left.\times(\boldsymbol{y}_{1}|{\cal G}_{-\bar{z}q^{+}}^{A}(x_{1}^{+},x_{2}^{+})|\boldsymbol{y}_{2})\left[A^{-}(x_{2}^{+},\boldsymbol{y}_{2})-A^{-}(x_{2}^{+},\boldsymbol{x}_{2})\right]\right\} \nonumber ,
\end{align}
The $\delta(0^+)$ term might seem worrisome at first glance, but the reason for its existence is that we are currently computing a forward amplitude in view of using the optical theorem as in \eqn{eq:opt-theo-1}. In said theorem, a $(2\pi)^D\delta^D(0)$ factor, to which the present $\delta(0^+)$ term contributes, is removed from the forward amplitude. 

With the energy denominators, it is possible to relate ${\cal T}_{T,L}$ to the well known photon wave functions. It is more conveniently seen in coordinate space. Using the standard integrals 
\begin{equation}
	\int\frac{\rmd^{d}\boldsymbol{k}}{(2\pi)^{d}}\frac{e^{i(\boldsymbol{k}\cdot\boldsymbol{r})}}{\boldsymbol{k}^{2}+z\bar{z}Q^{2}}=\frac{1}{(2\pi)^{\frac{d}{2}}}\left(\frac{Q\sqrt{z\bar{z}}}{\left|\boldsymbol{r}\right|}\right)^{\frac{d}{2}-1}K_{\frac{d}{2}-1}(Q\sqrt{z\bar{z}}|\boldsymbol{r}|)\label{eq:K0-int-d}
\end{equation}
and
\begin{equation}
	\int\frac{\rmd^{d}\boldsymbol{k}}{(2\pi)^{d}}\frac{e^{i(\boldsymbol{k}\cdot\boldsymbol{r})}}{\boldsymbol{k}^{2}+z\bar{z}Q^{2}}\boldsymbol{k}^{i}=\frac{i\boldsymbol{r}^{i}}{(2\pi)^{\frac{d}{2}}}\left(\frac{Q\sqrt{z\bar{z}}}{\left|\boldsymbol{r}\right|}\right)^{\frac{d}{2}}K_{\frac{d}{2}}(Q\sqrt{z\bar{z}}|\boldsymbol{r}|)\label{eq:K1-int-d},
\end{equation}
to perform the $\boldsymbol{\ell}_1$ and $\boldsymbol{\ell}_2$ integrations with explicit expressions for ${\cal T}_{L}$ and ${\cal T}_{T}$, we get:
\begin{align}
	& \Phi_{L}(z,\boldsymbol{x}_1-\boldsymbol{y}_1,\boldsymbol{x}_2-\boldsymbol{y}_2) \nn
	& \equiv  \int\frac{{\rm d}^{d}\boldsymbol{\ell}_{1}}{(2\pi)^{d}}\frac{{\rm d}^{d}\boldsymbol{\ell}_{2}}{(2\pi)^{d}}{\rm e}^{i\boldsymbol{\ell}_{1}(\boldsymbol{x}_{1}-\boldsymbol{y}_{1})-i\boldsymbol{\ell}_{2}\cdot(\boldsymbol{x}_{2}-\boldsymbol{y}_{2})}\frac{{\cal T}_{L}(z,\boldsymbol{\ell}_{1},\boldsymbol{\ell}_{2}-\boldsymbol{\ell}_{1})}{(\boldsymbol{\ell}_{1}^{2}+z\bar{z}Q^{2})(\boldsymbol{\ell}_{2}^{2}+z\bar{z}Q^{2})}\label{eq:TLL-coord}\\
	& =\frac{32z^{2}\bar{z}^{2}\left(q^{+}\right)^{2}Q^{2}}{(2\pi)^{d}}\left(\frac{z\bar{z}Q^{2}}{\left|\boldsymbol{x}_{1}-\boldsymbol{y}_{1}\right|\left|\boldsymbol{x}_{2}-\boldsymbol{y}_{2}\right|}\right)^{\frac{d}{2}-1}K_{\frac{d}{2}-1}(Q\sqrt{z\bar{z}}|\boldsymbol{x}_{1}-\boldsymbol{y}_{1}|)K_{\frac{d}{2}-1}(Q\sqrt{z\bar{z}}|\boldsymbol{x}_{2}-\boldsymbol{y}_{2}|),\nonumber 
\end{align}
and
\begin{align}
	& \Phi_{T}(z,\boldsymbol{x}_1-\boldsymbol{y}_1,\boldsymbol{x}_2-\boldsymbol{y}_2) \nn
	& \equiv \int\frac{{\rm d}^{d}\boldsymbol{\ell}_{1}}{(2\pi)^{d}}\frac{{\rm d}^{d}\boldsymbol{\ell}_{2}}{(2\pi)^{d}}{\rm e}^{i\boldsymbol{\ell}_{1}(\boldsymbol{x}_{1}-\boldsymbol{y}_{1})-i\boldsymbol{\ell}_{2}\cdot(\boldsymbol{x}_{2}-\boldsymbol{y}_{2})}\frac{{\cal T}_{T}(z,\boldsymbol{\ell}_{1},\boldsymbol{\ell}_{2}-\boldsymbol{\ell}_{1})}{(\boldsymbol{\ell}_{1}^{2}+z\bar{z}Q^{2})(\boldsymbol{\ell}_{2}^{2}+z\bar{z}Q^{2})}\label{eq:TTT-coord}\\
	& =\frac{8\left(q^{+}\right)^{2}}{(2\pi)^{d}}\left(1-\frac{4}{d}z\bar{z}\right)(\boldsymbol{x}_{1}-\boldsymbol{y}_{1})\cdot(\boldsymbol{x}_{2}-\boldsymbol{y}_{2})\left(\frac{z\bar{z}Q^{2}}{\left|\boldsymbol{x}_{1}-\boldsymbol{y}_{1}\right|\left|\boldsymbol{x}_{2}-\boldsymbol{y}_{2}\right|}\right)^{\frac{d}{2} } \nn 
	& \times K_{\frac{d}{2}}(Q\sqrt{z\bar{z}}|\boldsymbol{x}_{1}-\boldsymbol{y}_{1}|)K_{\frac{d}{2}}(Q\sqrt{z\bar{z}}|\boldsymbol{x}_{2}-\boldsymbol{y}_{2}|)\nonumber .
\end{align}
In $D=d+2=4$ dimensions, we would recognize photon wave function overlaps, see e.g.~\cite{Kowalski:2006hc}:
\begin{align}
	 \Phi_{L}(z,\boldsymbol{x}_1-\boldsymbol{y}_1,\boldsymbol{x}_2-\boldsymbol{y}_2) & =\frac{8\left(q^{+}\right)^{2}}{(2\pi)^{2}}\left\{ 4z^{2}\bar{z}^{2}Q^{2}K_{0}(Q\sqrt{z\bar{z}}|\boldsymbol{x}_{1}-\boldsymbol{y}_{1}|)K_{0}(Q\sqrt{z\bar{z}}|\boldsymbol{x}_{2}-\boldsymbol{y}_{2}|)\right\} \nn
	& \equiv \frac{4\left(q^{+}\right)^{2}}{ N_{c}q_{f}^{2}e^2}\Psi_{L}^{f}(\boldsymbol{x}_{1}-\boldsymbol{y}_{1},z,Q)\Psi_{L}^{f\ast}(\boldsymbol{x}_{2}-\boldsymbol{y}_{2},z,Q)\label{eq:TLL-wf} ,
\end{align}
and 
\begin{align}
	\Phi_{T}(z,\boldsymbol{x}_1-\boldsymbol{y}_1,\boldsymbol{x}_2-\boldsymbol{y}_2) & =\frac{8\left(q^{+}\right)^{2}}{(2\pi)^{2}}(z^{2}+\bar{z}^{2})\frac{(\boldsymbol{x}_{1}-\boldsymbol{y}_{1})\cdot(\boldsymbol{x}_{2}-\boldsymbol{y}_{2})}{\left|\boldsymbol{x}_{1}-\boldsymbol{y}_{1}\right|\left|\boldsymbol{x}_{2}-\boldsymbol{y}_{2}\right|} \nn & \times z\bar{z}Q^{2}K_{1}(Q\sqrt{z\bar{z}}|\boldsymbol{x}_{1}-\boldsymbol{y}_{1}|)K_{1}(Q\sqrt{z\bar{z}}|\boldsymbol{x}_{2}-\boldsymbol{y}_{2}|)\nonumber\\
	& \equiv \frac{4\left(q^{+}\right)^{2}}{ N_{c}q_{f}^{2}e^2}\Psi_{T}^{f}(\boldsymbol{x}_{1}-\boldsymbol{y}_{1},z,Q)\Psi_{T}^{f\ast}(\boldsymbol{x}_{2}-\boldsymbol{y}_{2},z,Q)\label{eq:TTT-wf}.
\end{align}
The physics of the amplitude in the present framework is particularly interesting in coordinate space as a comparison to the shock wave approximated result~\cite{Kowalski:2006hc}. In the latter, the wave functions are evaluated at the same dipole size in the wave function and in its complex conjugate. This dipole size is that of the quark-antiquark pair at light cone time 0, where the scattering occurs in its entirety. Meanwhile, the dipole sizes in the wave functions in \eqn{eq:TLL-wf} and \eqn{eq:TTT-wf} are actually distinct. In fact, it can be seen in \eqn{eq:W-fin-mom} that these dipole sizes correspond precisely to the transverse separation of the quark-antiquark pair  at the times of the first and last scatterings with the target.
The connection with QCD factorization is however more explicit in momentum space, to which we will stick from now on.

\begin{figure}
\begin{center}
\includegraphics[width=0.4\textwidth]{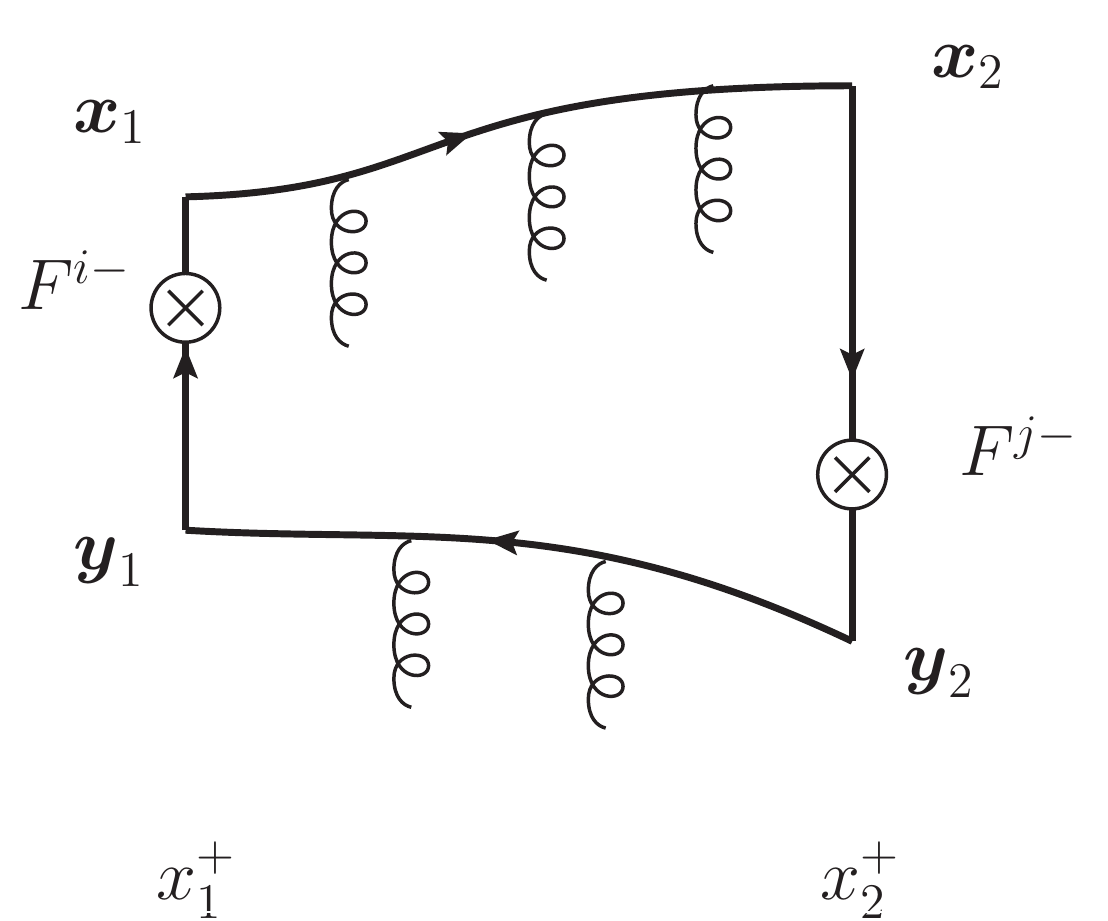} 
\end{center}
\caption{Illustration of the generalized dipole operator ${\cal U}^{ij}$.  }
\label{fig:long-DIS-gluon-dist-2}
\end{figure}
With full wave functions in the amplitude, the only remaining step is to introduce the proper non-perturbative target matrix elements. Using \eqn{eq:AtoF} to relate the differences of gluon fields to field strength tensors, we can identify the proper generalization of the dipole operator in momentum space as follows (see Figure~\ref{fig:long-DIS-gluon-dist-2} for an illustration):
\begin{align}
& {\cal U}^{ij}(z,q,\boldsymbol{\ell}_1,\boldsymbol{\ell}_2) \equiv \int \rmd x_1^+ \rmd x_2^+ \int \rmd^d \boldsymbol{x}_1\rmd^d \boldsymbol{x}_2\rmd^d \boldsymbol{y}_1\rmd^d \boldsymbol{y}_2 \int_0^1\rmd s \rmd s^\prime \rme^{-iq^-(x_1^+-x_2^+)+i \boldsymbol{\ell}_1\cdot(\boldsymbol{x}_1-\boldsymbol{y}_1)-i\boldsymbol{\ell}_2\cdot(\boldsymbol{x}_2-\boldsymbol{y}_2)} \nn
& \times {\rm tr}_c\left\{ (\boldsymbol{x}_2 | {\cal G}^R_{zq^+}(x_2^+,x_1^+) | \boldsymbol{x}_1)  F^{i-}(x_1^+,s \boldsymbol{x}_1+\bar{s}\boldsymbol{y}_1) (\boldsymbol{y}_1|{\cal G}^A_{-\bar{z}q^+}(x_1^+,x_2^+)|\boldsymbol{y}_2)F^{j-}(x_2^+,s^\prime\boldsymbol{x}_2+\bar{s}^\prime\boldsymbol{y}_2)\right\}, \label{eq:dip-op}
\end{align}
with $\bar{s}\equiv1-s$ and $\bar{s}^\prime = 1-s^\prime$. This distribution appears explicitly in the amplitude after the use of \eqn{eq:AtoF}. With an integration by parts involving the $(\boldsymbol{x}_1-\boldsymbol{y}_1)^i$ and $(\boldsymbol{x}_2-\boldsymbol{y}_2)^j$ prefactors from that equation, \eqn{eq:W-fin-mom} becomes:
\begin{align}
	{\cal A}_{T,L} & =-e^{2}g^{2}(2\pi)\delta(0^{+})\sum_{f}q_{f}^{2} \int_{0}^{1}\frac{{\rm d}z}{2\pi}\int\frac{{\rm d}^{d}\boldsymbol{\ell}_{1}}{(2\pi)^{d}}\int\frac{{\rm d}^{d}\boldsymbol{\ell}_{2}}{(2\pi)^{d}}\nonumber \\
	&\times {\cal U}^{ij}(z,q,\boldsymbol{\ell}_1,\boldsymbol{\ell}_2)  \, \frac{\partial}{\partial \boldsymbol{\ell}_1^i}\frac{\partial}{\partial \boldsymbol{\ell}_2^j} \frac{z\bar{z}q^{+}{\cal T}_{T,L}\left(z,\boldsymbol{\ell}_{1},\boldsymbol{\ell}_{2}-\boldsymbol{\ell}_{1}\right)}{(\boldsymbol{\ell}_{1}^{2}+z\bar{z}Q^{2})(\boldsymbol{\ell}_{2}^{2}+z\bar{z}Q^{2})} \label{eq:W-with-U}.
\end{align}
In what follows, we will further simplify the non-perturbative tensor ${\cal U}^{ij}$ by performing a partial twist expansion which will enable us to reduce the number of variables, leading to a new gluon distribution.

\section{The unintegrated gluon distribution from a partial twist expansion}\label{sec:uPDF}
In the previous Section, we have factorized out the hard subamplitude from the target matrix elements in full generality. In this Section, we will now study these matrix elements more closely, and we will extract a more useful form for them which will be valid up to $x_{\rm Bj}\Lambda_{\rm QCD}^2/Q^2$ corrections, hence at leading power accuracy both in the Regge limit and in the Bjorken limit.
The action $\langle P | {\cal U}^{ij} | P\rangle$ of the generalized dipole operator ${\cal U}^{ij}$ on diagonal target states is a function of momenta $q$ and $P$, transverse momenta $\boldsymbol{\ell}_1$ and $\boldsymbol{\ell}_2$, as well as the number $z$ which corresponds to the longitudinal momentum fraction of the quark in the photon wave function. From $q$ and $P$, neglecting the target mass corrections, we can build the following Lorentz scalars: $Q^2=-q^2$, $W = (q+P)^2$, or equivalently $x_{\rm Bj}$ and $Q$. $\boldsymbol{\ell}_1$ and $z$ are remnants of an incomplete factorization with the hard part and they will be reabsorbed into it later on. Finally, $\boldsymbol{\ell}_1-\boldsymbol{\ell}_2$ represents an intrinsic transverse momentum in the target, which means it is at most of order $Q_s$ in the saturated Regge limit.

\subsection{Partial twist expansion (PTE)}\label{sec:expansion}

\begin{figure}
\begin{center}
\includegraphics[width=0.5\textwidth]{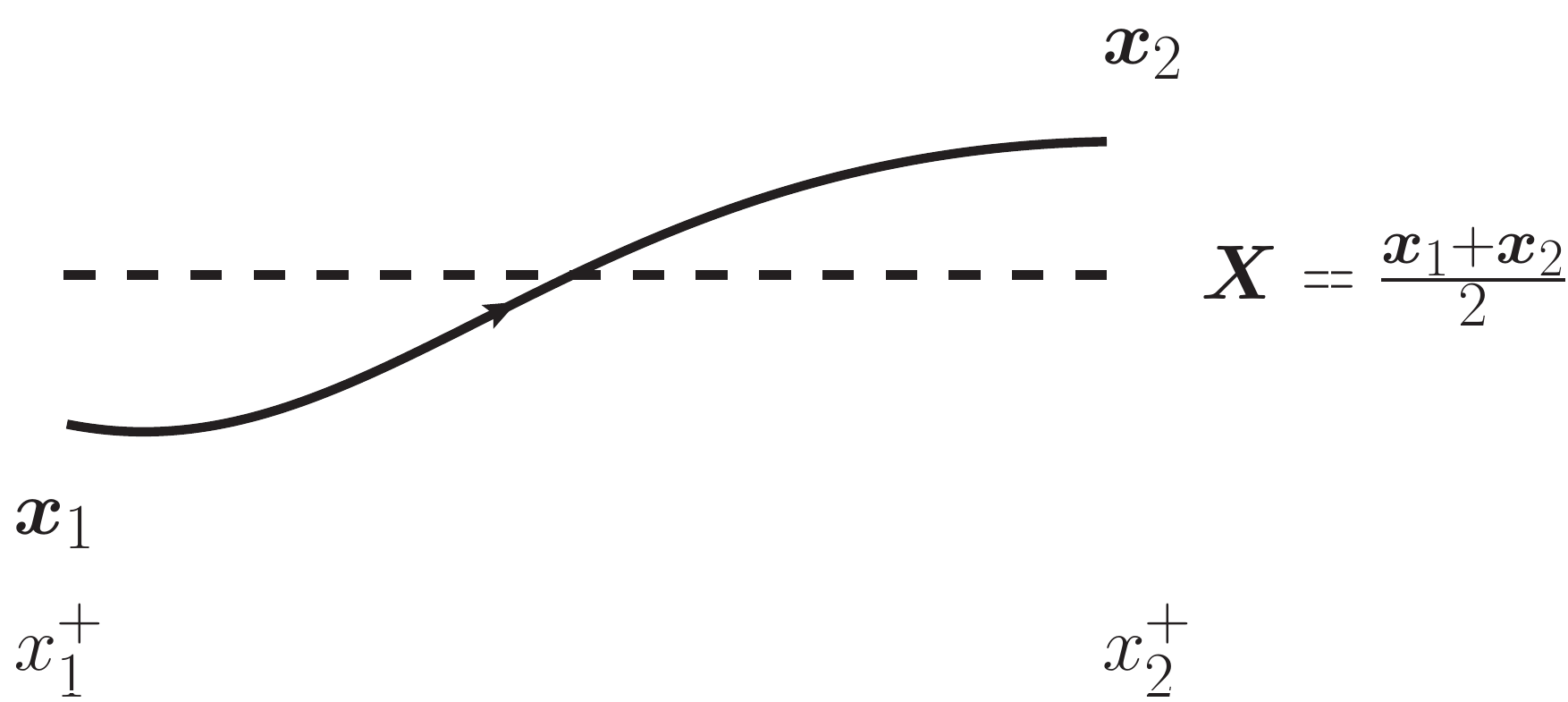} 
\end{center}
\caption{Approximation of the (non-eikonal) scalar propagator inside the target by a Wilson line (times a free propagator) in the PTE scheme evaluated at the mean transverse position $\X=(\x_2+\x_1)/2$. }
\label{fig:long-DIS-PTE}
\end{figure}

Let us consider a single propagator in the operator: $(\boldsymbol{x}_{2}|\mathcal{G}_{p^{+}}^{R}(x_{2}^{+},x_{1}^{+})|\boldsymbol{x}_{1})$ as illustrated in Figure~\ref{fig:long-DIS-PTE} and extract an extremal phase from it. For example, we may use
the integral form of the Schr\"{o}dinger equation:
\begin{align}
	(\boldsymbol{x}_{2}|\mathcal{G}_{p^{+}}^{R}(x_{2}^{+},x_{1}^{+})|\boldsymbol{x}_{1}) & =(\boldsymbol{x}_{2}|\mathcal{G}_{p^{+}}^{(0)R}(x_{2}^{+},x_{1}^{+})|\boldsymbol{x}_{1})\label{eq:intSc}\\
	& +ig\int_{x_{1}^{+}}^{x_{2}^{+}}\!{\rm d}x_{3}^{+}\int\!{\rm d}^{d}\boldsymbol{x}_{3}(\boldsymbol{x}_{2}|\mathcal{G}_{p^{+}}^{(0)R}(x_{2}^{+},x_{3}^{+})|\boldsymbol{x}_{3})A^{-}(x_{3})(\boldsymbol{x}_{3}|\mathcal{G}_{p^{+}}^{R}(x_{3}^{+},x_{1}^{+})|\boldsymbol{x}_{1}),\nonumber 
\end{align}
which is a direct consequence of the integrated Klein-Gordon equation~\eqn{eq:G-position}. The free propagators $(\boldsymbol{x}_{n}|\mathcal{G}_{p^{+}}^{(0)}(x_{n}^{+},x_{1}^{+})|\boldsymbol{x}_{1})$
are Gaussians in the transverse separation:
\begin{align}
	(\boldsymbol{x}_{2}|{\cal G}_{k^{+}}^{(0)R}(x_{2}^{+},x_{1}^{+})|\boldsymbol{x}_{1}) & =-i\left(\frac{-ik^{+}}{x_{2}^{+}-x_{1}^{+}}\right)^{\frac{d}{2}}\frac{{\rm e}^{i\frac{k^{+}}{2(x_{2}^{+}-x_{1}^{+})}\left[(\boldsymbol{x}_{2}-\boldsymbol{x}_{1})^{2}+i0\right]}}{2k^{+}(2\pi)^{\frac{d}{2}}}\theta(k^{+}).\label{eq:curvyG0}
\end{align}
These Gaussians describe quantum diffusion in the transverse plane and peak at $(\boldsymbol{x}_{2}-\boldsymbol{x}_{1})^{2}\sim2(x_{2}^{+}-x_{1}^{+})/k^{+}$.
Parametrically, $k^{+}\sim q^{+}$ and for the target matrix elements
of the operator we are studying now, $x_{2}^{+}-x_{1}^{+}\sim1/P^{-}$.
This means that during propagation, the change in the transverse position
of a parton is of order $1/(P^{-}q^{+})$. In other words,
\begin{equation}
	(\boldsymbol{x}_{2}-\boldsymbol{x}_{1})^{2}\sim x_{{\rm Bj}}/Q^{2}.\label{eq:recoil}
\end{equation}
This quantity is $x_{{\rm Bj}}$-suppressed in the eikonal limit,
and it is twist suppressed in the Bjorken limit. We can thus safely
expand all transverse positions around the average position 
\beq
\frac{\boldsymbol{x}_{1}+\boldsymbol{x}_{2}}{2} \equiv\boldsymbol{X}\,,
\eeq
without losing accuracy in either limit. Then in the leading approximation
one can factorize a Wilson line with phases using
the integrated Schr\"{o}dinger equation. We shall refer to this classical expansion as the ``partial twist expansion"  (PTE) in what follows. 

More explicitly, we consider first the second
term in the r.h.s. of Eq.~(\ref{eq:intSc}), expand it around $A^{-}(x_{3})\simeq A^{-}(x_{3}^{+},\boldsymbol{X})$
then use Eq.~(\ref{eq:intSc}) itself again:
\begin{align}
	& ig\int_{x_{2}^{+}}^{x_{1}^{+}}\!{\rm d}x_{3}^{+}\int\!{\rm d}^{d}\boldsymbol{x}_{3}(\boldsymbol{x}_{1}|\mathcal{G}_{p^{+}}^{(0)R}(x_{1}^{+},x_{3}^{+})|\boldsymbol{x}_{3})A^{-}(x_{3})(\boldsymbol{x}_{3}|\mathcal{G}_{p^{+}}^{R}(x_{3}^{+},x_{2}^{+})|\boldsymbol{x}_{2})\nonumber \\
	& \simeq ig\int_{x_{2}^{+}}^{x_{1}^{+}}\!{\rm d}x_{3}^{+}A^{-}(x_{3}^{+},\boldsymbol{X})\int\!{\rm d}^{d}\boldsymbol{x}_{3}(\boldsymbol{x}_{1}|\mathcal{G}_{p^{+}}^{(0)R}(x_{1}^{+},x_{3}^{+})|\boldsymbol{x}_{3})\label{eq:iteration}\\
	& \times\left[(\boldsymbol{x}_{3}|\mathcal{G}_{p^{+}}^{(0)R}(x_{3}^{+},x_{2}^{+})|\boldsymbol{x}_{2})\nonumber \right.\\
	& \left. +ig\int_{x_{2}^{+}}^{x_{3}^{+}}\!{\rm d}x_{4}^{+}\int\!{\rm d}^{d}\boldsymbol{x}_{4}(\boldsymbol{x}_{3}|\mathcal{G}_{p^{+}}^{(0)R}(x_{3}^{+},x_{4}^{+})|\boldsymbol{x}_{4})A^{-}(x_{4})(\boldsymbol{x}_{4}|\mathcal{G}_{p^{+}}^{R}(x_{4}^{+},x_{2}^{+})|\boldsymbol{x}_{2})\right].\nonumber 
\end{align}
The first term in the bracket can be simplified using the composition
law for the propagators:
\begin{equation}
	\int\!{\rm d}^{d}\boldsymbol{x}_{3}(\boldsymbol{x}_{1}|\mathcal{G}_{p^{+}}^{(0)R}(x_{1}^{+},x_{3}^{+})|\boldsymbol{x}_{3})(\boldsymbol{x}_{3}|\mathcal{G}_{p^{+}}^{(0)R}(x_{3}^{+},x_{2}^{+})|\boldsymbol{x}_{2})=(\boldsymbol{x}_{1}|\mathcal{G}_{p^{+}}^{(0)R}(x_{1}^{+},x_{2}^{+})|\boldsymbol{x}_{2}).
\end{equation}
Then the first iteration allows to identify the following term:
\begin{align}
	(\boldsymbol{x}_{1}|\mathcal{G}_{p^{+}}^{R}(x_{1}^{+},x_{2}^{+})|\boldsymbol{x}_{2}) & =(\boldsymbol{x}_{1}|\mathcal{G}_{p^{+}}^{(0)R}(x_{1}^{+},x_{2}^{+})|\boldsymbol{x}_{2})\left[1+ig\int_{x_{2}^{+}}^{x_{1}^{+}}\!{\rm d}x_{3}^{+}A^{-}(x_{3}^{+},\boldsymbol{X})\right]+...\label{eq:},
\end{align}
where the dots represent the second term in the bracket above. Here, one can recognize the perturbative expansion of a gauge link.
Step by step, the approximation where all gluon fields are evaluated
at $\boldsymbol{X}$ and the use of the integrated Schrödinger equation
allows to reconstruct the full link as illustrated in Figure \ref{fig:long-DIS-PTE}, and one eventually finds:
\begin{equation}
	(\boldsymbol{x}_{1}|\mathcal{G}_{p^{+}}^{R}(x_{1}^{+},x_{2}^{+})|\boldsymbol{x}_{2})\simeq \theta(p^+) \, (\boldsymbol{x}_{1}|\mathcal{G}_{p^{+}}^{(0)R}(x_{1}^{+},x_{2}^{+})|\boldsymbol{x}_{2})\left[x_{1}^{+},x_{2}^{+}\right]_{\frac{\boldsymbol{x}_{1}+\boldsymbol{x}_{2}}{2}}\,.\label{eq:classicalExpR}
\end{equation}
In a similar fashion, we obtain 
\begin{equation}
	(\boldsymbol{y}_{2}|\mathcal{G}_{p^{+}}^{A}(x_{2}^{+},x_{1}^{+})|\boldsymbol{y}_{1})\simeq\theta(-p^+)\, (\boldsymbol{y}_{2}|\mathcal{G}_{p^{+}}^{(0)A}(x_{2}^{+},x_{1}^{+})|\boldsymbol{y}_{1})\left[x_{2}^{+},x_{1}^{+}\right]_{\frac{\boldsymbol{y}_{1}+\boldsymbol{y}_{2}}{2}}\,.\label{eq:classicalExpA}
\end{equation}
Comparable so-called classical approximations were derived in previous works~\cite{subeikonal-tolga-1,subeikonal-tolga-2,subeikonal-tolga-3,subeikonal-tolga-4,subeikonal-tolga-5,subeikonal-tolga-6}, but in there the approximations were taken further by assuming $p^+$ is large and systematically expanding all factors including free propagators $G_0$ in order to 
compute corrections to the leading power in the Regge limit.
Because our purpose is to provide expressions which are correct in both Bjorken and Regge regimes, we will not take these additional steps which would only yield approximated results in the Bjorken limit.
Because the Gaussians in the propagators impose that $\boldsymbol{x}_1\simeq \boldsymbol{x}_2 \simeq (\boldsymbol{x}_1+\boldsymbol{x}_2)/2$ and the same condition for $\boldsymbol{y}_{1,2}$ positions, we can also 
approximate 
\begin{equation}
	F^{i-}(x_{1}^{+},s\boldsymbol{x}_{1}+\bar{s}\boldsymbol{y}_{1})\simeq F^{i-}\left(x_{1}^{+},s\frac{\boldsymbol{x}_{1}+\boldsymbol{x}_{2}}{2}+\bar{s}\frac{\boldsymbol{y}_{1}+\boldsymbol{y}_{2}}{2}\right),\label{eq:Fim}
\end{equation}
and
\begin{equation}
	F^{j-}(x_{2}^{+},s^{\prime}\boldsymbol{x}_{2}+\bar{s}^{\prime}\boldsymbol{y}_{2})\simeq F^{j-}\left(x_{2}^{+},s^{\prime}\frac{\boldsymbol{x}_{1}+\boldsymbol{x}_{2}}{2}+\bar{s}^{\prime}\frac{\boldsymbol{y}_{1}+\boldsymbol{y}_{2}}{2}\right)\label{eq:Fjm}.
\end{equation}
These approximations allow us to perform explicit integrals over some variables. Indeed, the purely non-perturbative elements comprised of field strength tensors and finite length Wilson lines depend only on $\boldsymbol{b}\equiv(\boldsymbol{x}_1+\boldsymbol{x}_2)/2$ and $\boldsymbol{b}^\prime \equiv (\boldsymbol{y}_1+\boldsymbol{y}_2)/2$ so we can integrate the free propagators w.r.t. $\boldsymbol{r}\equiv\boldsymbol{x}_1-\boldsymbol{x}_2$ and $\boldsymbol{r}^\prime \equiv \boldsymbol{y}_1-\boldsymbol{y}_2$. In terms of these local variables, the exponent which defines the generalized dipole operator are given by:
\begin{equation}
	i\boldsymbol{\ell}_{1}\cdot(\boldsymbol{x}_{1}-\boldsymbol{y}_{1})-i\boldsymbol{\ell}_{2}\cdot(\boldsymbol{x}_{2}-\boldsymbol{y}_{2})=i(\boldsymbol{\ell}_{1}-\boldsymbol{\ell}_{2})\cdot(\boldsymbol{b}-\boldsymbol{b}^{\prime})+i\frac{\boldsymbol{\ell}_{1}+\boldsymbol{\ell}_{2}}{2}\cdot(\boldsymbol{r}-\boldsymbol{r}^{\prime})\label{eq:exponents}.
\end{equation}
Using the integrals
\begin{equation}
	\int{\rm d}^{d}\boldsymbol{r}{\rm e}^{i\frac{\boldsymbol{\ell}_{1}+\boldsymbol{\ell}_{2}}{2}\cdot\boldsymbol{r}}(\boldsymbol{x}_{2}|{\cal G}_{zq^{+}}^{(0)R}(x_{2}^{+},x_{1}^{+})|\boldsymbol{x}_{1})=-i\frac{\theta(x_{2}^{+}-x_{1}^{+})}{2zq^{+}}{\rm e}^{-i\frac{(x_{2}^{+}-x_{1}^{+})}{2zq^{+}}\left(\frac{\boldsymbol{\ell}_{1}+\boldsymbol{\ell}_{2}}{2}\right)^{2}+i0},\label{eq:G0Rint}
\end{equation}
and
\begin{equation}
	\int{\rm d}^{d}\boldsymbol{r}^{\prime}{\rm e}^{-i\frac{\boldsymbol{\ell}_{1}+\boldsymbol{\ell}_{2}}{2}\cdot\boldsymbol{r}^{\prime}}(\boldsymbol{y}_{1}|{\cal G}_{-\bar{z}q^{+}}^{(0)A}(x_{1}^{+},x_{2}^{+})|\boldsymbol{y}_{2})=-i\frac{\theta(x_{2}^{+}-x_{1}^{+})}{2\bar{z}q^{+}}{\rm e}^{-i\frac{(x_{2}^{+}-x_{1}^{+})}{2\bar{z}q^{+}}\left(\frac{\boldsymbol{\ell}_{1}+\boldsymbol{\ell}_{2}}{2}\right)^{2}+i0}\label{eq:GoAint},
\end{equation}
we can finally approximate the generalized dipole operator given by \eqn{eq:dip-op}, up to $x_{\rm Bj}/Q^2$ corrections,
\begin{align}
	\langle P | {\cal U}^{ij}(z,q,\boldsymbol{\ell}_{1},\boldsymbol{\ell}_{2})|P\rangle & \simeq-\int{\rm d}x_{1}^{+}{\rm d}x_{2}^{+}\int \! {\rm d}^{d}\boldsymbol{b}\,{\rm d}^{d}\boldsymbol{b}^{\prime}\int_{0}^{1}\!{\rm d}s{\rm d}s^{\prime}{\rm e}^{i(\boldsymbol{\ell}_{1}-\boldsymbol{\ell}_{2})\cdot(\boldsymbol{b}-\boldsymbol{b}^{\prime})}\nonumber \\
	& \times \langle P | {\rm tr}_{c}\left\{ [x_{2}^{+},x_{1}^{+}]_{\boldsymbol{b}}F^{i-}(x_{1}^{+},s\boldsymbol{b}+\bar{s}\boldsymbol{b}^{\prime})[x_{1}^{+},x_{2}^{+}]_{\boldsymbol{b}^{\prime}}F^{j-}(x_{2}^{+},s^{\prime}\boldsymbol{b}+\bar{s}^{\prime}\boldsymbol{b}^{\prime})\right\}|P\rangle ,\nonumber \\
	& \times\frac{\theta(x_{2}^{+}-x_{1}^{+})}{4z\bar{z}(q^{+})^{2}}{\rm e}^{-i(x_{2}^{+}-x_{1}^{+})\left[\frac{1}{2z\bar{z}q^{+}}\left(\frac{\boldsymbol{\ell}_{1}+\boldsymbol{\ell}_{2}}{2}\right)^{2}-q^{-}-i0 \right]}\label{eq:Uij-int}.
\end{align}
Defining $\boldsymbol{v}\equiv\boldsymbol{b}^\prime-\boldsymbol{b}$ and $v^+ \equiv x_2^+ -x_1^+ $ then using the invariance of forward matrix elemets under translations in order to integrate out $(\boldsymbol{b}+\boldsymbol{b}^\prime)/2$ and $(x_1^++x_2^+)/2$ into $\delta$ functions, we get:

\begin{align}
	& \langle P|{\cal U}^{ij}(z,q,\boldsymbol{\ell}_{1},\boldsymbol{\ell}_{2})|P\rangle \nn & =-(2\pi)^{d+1}\delta(0^{-})\delta^{d}(\boldsymbol{0})\int{\rm d}v^{+}\!\!\int{\rm d}^{d}\boldsymbol{v}\int_{0}^{1}\!{\rm d}s\,{\rm d}s^{\prime}{\rm e}^{-i(\boldsymbol{\ell}_{1}-\boldsymbol{\ell}_{2})\cdot\boldsymbol{v}}\frac{\theta(v^{+})}{4z\bar{z}(q^{+})^{2}}{\rm e}^{-i\big[\frac{\left(\frac{\boldsymbol{\ell}_{1}+\boldsymbol{\ell}_{2}}{2}\right)^{2}}{2z\bar{z}q^{+}}-q^{-}-i0\big]v^{+}}\label{eq:Uij-int} \\
	& \times \big\langle P \big| {\rm tr}_{c}\big\{  \big[v^{+},0^+\big]_{\boldsymbol{0}}F^{i-}\big(0^+,s\boldsymbol{v}\big)\big[0^+,v^{+}\big]_{\boldsymbol{v}}F^{j-}\big(v^{+},s^{\prime}\boldsymbol{v}\big)\big\}\big|P \big\rangle\,. \nonumber
\end{align}
We now have the $\delta(0)$ functions to complete the one from \eqn{eq:W-fin-mom} into the full $\delta^D(0)$ which will be taken out by the optical theorem. It is worth noting that as expected, $\boldsymbol{\ell}_1-\boldsymbol{\ell}_2$ which is an intrinsic transverse momentum transferred from the target, decouples from the hard splitting momentum $(\boldsymbol{\ell}_1+\boldsymbol{\ell}_2)/2$. In practice, the latter will be reabsorbed into the hard sub-amplitude where it belongs.

\begin{figure}
\begin{center}
\includegraphics[width=0.45\textwidth]{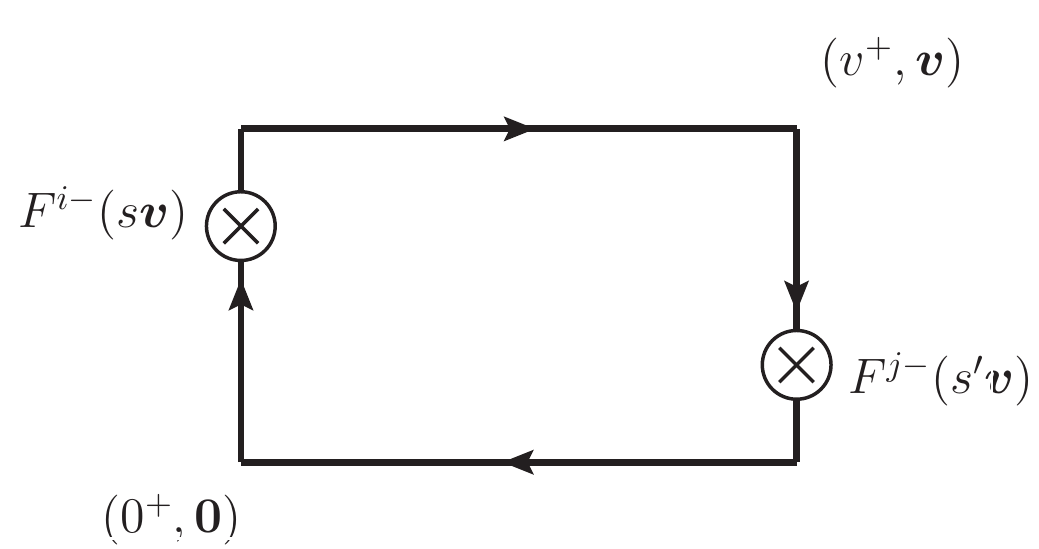} 
\end{center}
\caption{Diagrammatic representation of the $x$-dependent unintegrated gluon distribution (x-uPDF) defined in \eqn{eq:Gijdef}.}
\label{fig:long-DIS-gluon-dist}
\end{figure}

Parton distributions are usually defined without light cone time ordering. In order to get a more common form for our non-perturbative matrix element, let us get rid of the $\theta$ function using the following identity which is valid for any test function $f$:
\begin{equation}
	\int{\rm d}v^{+}\theta\left(v^{+}\right){\rm e}^{-iv^{+}\left(k^{-}-i0\right)}f(v^{+})=i\int\frac{{\rm d}x}{x-\frac{k^{-}}{P^{-}}+i0}\int_{-\infty}^{\infty}\frac{{\rm d}u^{+}}{2\pi}{\rm e}^{-ixP^{-}u^{+}}f(u^{+})\label{eq:trick}.
\end{equation}
Renaming the dummy variable $u^+$ back to $v^+$ again for more consistent notations, we finally obtain:
\begin{align}
	& \langle P|{\cal U}^{ij}(z,q,\boldsymbol{\ell}_{1},\boldsymbol{\ell}_{2})|P\rangle\nonumber \\
	& =-i(2\pi)^{d+1}\delta(0^{-})\delta^{d}(\boldsymbol{0})\frac{1}{4z\bar{z}(q^{+})^{2}}\int\frac{{\rm d}x}{x-\frac{\left(\frac{\boldsymbol{\ell}_{1}+\boldsymbol{\ell}_{2}}{2}\right)^{2}+z\bar{z}Q^{2}}{2z\bar{z}q^{+}P^{-}}+i0}\nonumber \\
	& \times\int_{-\infty}^{\infty}\frac{{\rm d}v^{+}}{2\pi}{\rm e}^{-ixP^{-}v^{+}}\int{\rm d}^{d}\boldsymbol{v}{\rm e}^{-i(\boldsymbol{\ell}_{1}-\boldsymbol{\ell}_{2})\cdot\boldsymbol{v}}\int_{0}^{1}{\rm d}s{\rm d}s^{\prime}\label{eq:Uij-not-ordered}\\
		& \times \big\langle P \big| {\rm tr}_{c}\big\{  \big[v^{+},0^+\big]_{\boldsymbol{0}}F^{i-}\big(0^+,s\boldsymbol{v}\big)\big[0^+,v^{+}\big]_{\boldsymbol{v}}F^{j-}\big(v^{+},s^{\prime}\boldsymbol{v}\big)\big\}\big|P \big\rangle \nonumber
\end{align}
The first line in the r.h.s. of \eqn{eq:Uij-not-ordered} will be absorbed  into the hard parts while the second line which depends only on intrinsic target variables $x$ and $\boldsymbol{\ell}_1 - \boldsymbol{\ell}_2$ defines the unintegrated gluon distribution, as illustrated in Figure \ref{fig:long-DIS-gluon-dist} \cite{Boussarie:2020fpb},
\begin{align}
	xG^{ij}(x,\boldsymbol{k}) & \equiv\frac{1}{P^{-}}\int\frac{{\rm d}v^{+}}{2\pi}{\rm e}^{ixP^{-}v^{+}}\int\frac{{\rm d}^{d}\boldsymbol{v}}{(2\pi)^{d}}{\rm e}^{-i(\boldsymbol{k}\cdot\boldsymbol{v})}\int_{0}^{1}{\rm d}s{\rm d}s^{\prime}\label{eq:Gijdef}\\
	& \times \big\langle P \big| {\rm tr}_{c}\big\{  \big[v^{+},0^+\big]_{\boldsymbol{0}}F^{i-}\big(0^+,s\boldsymbol{v}\big)\big[0^+,v^{+}\big]_{\boldsymbol{v}}F^{j-}\big(v^{+},s^{\prime}\boldsymbol{v}\big)\big\}\big|P \big\rangle\,, \nonumber
\end{align}
which is related to the generalized dipole operator (\ref{eq:dip-op}) by 
\begin{align}
	& \langle P|{\cal U}^{ij}(z,q,\boldsymbol{\ell}_{1},\boldsymbol{\ell}_{2})|P\rangle\nonumber \\
	& =-i(2\pi)^{d+1}\delta(0^{-})\delta^{d}(\boldsymbol{0})\frac{(2\pi)^{d}P^{-}}{4z\bar{z}(q^{+})^{2}}\int{\rm d}x\frac{xG^{ij}(x,\boldsymbol{\ell}_{2}-\boldsymbol{\ell}_{1})}{x- x_{\rm Bj} -\frac{\left(\frac{\boldsymbol{\ell}_{1}+\boldsymbol{\ell}_{2}}{2}\right)^{2}}{2z\bar{z}q^{+}P^{-}}+i0}+O(x_{\rm Bj}/Q^2).\label{eq:UijGij}
\end{align}
This concludes the last key step to our approach, where we have achieved a reduction of variables in the gluon operator that now depends only on intrinsic target variables $x$ and $\k=\boldsymbol{\ell}_{2}-\boldsymbol{\ell}_{1}$ while the hard loop variables $z$, $\bar z=1-z$ and $\bell=(\boldsymbol{\ell}_{2}+\boldsymbol{\ell}_{1})/2$ are factored out into a phase-space factor and a denominator that will be absorbed in the final hard matrix element.

To sum things up so far:
\begin{itemize}
	\item We first derived a generic cross section in the background field beyond the shock wave approximation, in terms of the scalar propagator. This gaves us the numerator of the $\gamma^\ast \rightarrow q\bar{q}$ wave functions.
	\item In a second step, we extracted the denominator of these wave functions through operator algebra without further approximation. At the same time, the times of the first and last scattering on the target appeared explicitly, thus yielding a dynamical longitudinal extent for our target. 
	\item Finally in this section, further analysis of the operators using a partial twist expansion provided a more suitable approximated expression for the non-perturbative matrix elements. We found a new distribution for the DIS cross section, to which we will refer as an $x$-dependent unintegrated gluon distribution. 
\end{itemize} 
As we shall see in later sections, the distribution which is defined in Eq.~(\ref{eq:Gijdef}) encodes both the so-called dipole operator and the collinear PDF. A few of its features may surprise the attentive reader, although they are expected in hindsight. 

To the expert of the Bjorken regime, it could be surprising that contrary to TMD distributions, the transverse momentum in our distribution is not Fourier conjugated to the transverse distance between the arguments of the field strength tensors. This property seems to break the partonic picture of the distribution. However, this property should not come as a surprise beyond the leading twist: in the Regge limit, the partonic interpretation is never the correct one. Be it Reggeons or Wilson line operators, Regge physics is always formulated in terms of object whose complexity is beyond the simple gluonic picture. Actually, as can be seen in Eq.~(\ref{eq:Reggeons}), the transverse distance $\boldsymbol{x}_1-\boldsymbol{x}_2$ in the dipole operator is not conjugated to the Reggeon momenta in the dilute limit either. Finally, it is worth mentioning here that in the presence of transverse momenta, the partonic picture is also broken in the Bjorken regime because of the non-cancellability of the gauge link structures. Among all possible gluon TMD distributions, one only gets to have a partonic interpretation within a given choice of gauge and subgauge conditions: the Weizs\"{a}cker-Williams TMD distribution~\cite{Kharzeev:2003wz}.

To the expert of the Regge limit, the gauge link structure of the distribution may feel awkward. Indeed, in the strict eikonal limit, observables such as the DIS cross section which involve the (fundamental) dipole operator in the CGC are described via the dipole-type TMD distribution $f^D(x,\boldsymbol{k})$ in its $x=0$ limit~\cite{Dominguez:2011wm}. The gauge link structure in this distribution involves links that extend to $x^+ = \pm \infty$, while our distribution has finite links. However the gauge link structure in the Bjorken regime, because one does not postulate a light cone time decoupling in that limit, is actually constrained by the physics of the process. In general, gauge links at $+\infty$ (resp. $-\infty$) light cone times are associated with multiple scatterings in the final (resp. initial) state~\cite{Belitsky:2002sm}. In inclusive DIS, initial and final states correspond to photons, hence no rescattering is expected at $\pm \infty$ light cone times. As a result, it should come as no surprise that the gauge link structure beyond the strict eikonal limit involves finite length gauge links even though one could naively expect to see $f^D(x\neq 0,\boldsymbol{k})$ instead. As we shall see in Section~\ref{sec:Regge}, we still recover the dipole operator and thus $f^D(x=0,\boldsymbol{k})$ and its infinite gauge links in the strict eikonal limit.

\subsection{DIS cross section }

Let us now summarize the results from Sections~\ref{sec:fulltensor} and~\ref{sec:expansion}. Using \eqn{eq:W-with-U} and \eqn{eq:UijGij}, we get:
\begin{align}
	{\rm Im} \frac{{\cal A}_{T,L}}{i(2\pi)^D\delta^D(0)} & =-e^{2}g^{2}\sum_{f}q_{f}^{2}\,{\rm Im}\int_{0}^{1}\frac{{\rm d}z}{2\pi}\int\frac{{\rm d}^{d}\boldsymbol{\ell}_{1}}{(2\pi)^{d}}\int\frac{{\rm d}^{d}\boldsymbol{\ell}_{2}}{(2\pi)^{d}}\nonumber \\
	&\times \frac{\partial}{\partial \boldsymbol{\ell}_1^i}\frac{\partial}{\partial \boldsymbol{\ell}_2^j} \frac{z\bar{z}q^{+}{\cal T}_{T,L}\left(z,\boldsymbol{\ell}_{1},\boldsymbol{\ell}_{2}-\boldsymbol{\ell}_{1}\right)}{(\boldsymbol{\ell}_{1}^{2}+z\bar{z}Q^{2})(\boldsymbol{\ell}_{2}^{2}+z\bar{z}Q^{2})}\nn 	
	&\times \frac{(2\pi)^{d}P^{-}}{4z\bar{z}(q^{+})^{2}}\int{\rm d}x\frac{xG^{ij}(x,\boldsymbol{\ell}_{2}-\boldsymbol{\ell}_{1})}{x-\frac{\left(\frac{\boldsymbol{\ell}_{1}+\boldsymbol{\ell}_{2}}{2}\right)^{2}+z\bar{z}Q^{2}}{2z\bar{z}q^{+}P^{-}}+i0}\,.
	\label{eq:W-Gij}
\end{align}
Let us note that $(G^{ij})^\ast=G^{ji}$. From the explicit expressions in Eqs.~(\ref{eq:traceLL}) and (\ref{eq:traceTT}), we can see that ${\cal T}_{L,T}$ has real values and that the derivatives are symmetric under $(i \leftrightarrow j, 1 \leftrightarrow 2)$ exchange. This leads to an additional simplification for these expressions, because the imaginary part we need is only embedded in a single denominator. We can indeed use:
\begin{align}
	{\rm Im}\frac{1}{x-\frac{\boldsymbol{\ell}^{2}+z\bar{z}Q^{2}}{2z\bar{z}q^{+}P^{-}}+i0} & =-\pi\delta\left(x-\frac{\boldsymbol{\ell}^{2}+z\bar{z}Q^{2}}{2z\bar{z}q^{+}P^{-}}\right)\label{eq:ipidelta}\\
	& =-\pi\delta\left(x-x_{{\rm Bj}}\bigl(1+\frac{\boldsymbol{\ell}^{2}}{z\bar{z}Q^{2}}\bigr)\right)\nonumber .
\end{align}
One has to keep in mind that this specific value for the Feynman $x$ variable is only valid for inclusive DIS. For more complicated processes such as exclusive Compton scattering, where both real and imaginary parts contribute to the amplitude, we would not set $x$ to this value. Concequently, taking the derivatives of the explicit expressions for the ${\cal T}_{T,L}$ tensors then adding the factors from~\eqn{eq:FLT} finally yields the DIS structure functions in the following form:
\begin{align}
	F_{T,L} & =g^{2}\sum_{f}q_{f}^{2}\int_{0}^{1}\frac{{\rm d}z}{2\pi}\int\frac{{\rm d}^{d}\boldsymbol{\ell}}{(2\pi)^{d}}\int{\rm d}^{d}\boldsymbol{k}\nonumber \\
	& \times(\partial^{i}\psi_{T,L})(z,\boldsymbol{\ell}-\boldsymbol{k}/2)(\partial^{j}\psi_{T,L}^{\ast})(z,\boldsymbol{\ell}+\boldsymbol{k}/2),\nonumber \\
		& \times\int{\rm d}x\, xG^{ij}(x,\boldsymbol{k}) \, \delta \!\left[x-x_{{\rm Bj}}\left(1+\frac{\boldsymbol{\ell}^{2}}{z\bar{z}Q^{2}}\right)\right]\label{eq:FTL-gen}
\end{align}
where $\psi_{T,L}$ are the photon wave functions in momentum space. In the transverse case, their overlap is implicitly summed over helicities. Explicitely, the structure functions then read:
\begin{align}
	F_{L}(x_{\rm Bj},Q^2) & = g^{2}Q^2\sum_{f}q_{f}^{2}\int_{0}^{1}\frac{{\rm d}z}{2\pi}\int\frac{{\rm d}^{d}\boldsymbol{\ell}}{(2\pi)^{d}}\int{\rm d}^{d}\boldsymbol{k}\nonumber \\
	& \times\int{\rm d}x\, \delta \!\left[x-x_{{\rm Bj}}\left(1+\frac{\boldsymbol{\ell}^{2}}{z\bar{z}Q^{2}}\right)\right] xG^{ij}(x,\boldsymbol{k})\label{eq:WLL}\\
	& \times\frac{16z^{2}\bar{z}^{2}Q^{2}\bigl(\boldsymbol{\ell}^i-\frac{\boldsymbol{k}^i}{2}\bigr)\bigl(\boldsymbol{\ell}^j+\frac{\boldsymbol{k}^j}{2}\bigr)}{\bigl((\boldsymbol{\ell}-\frac{\boldsymbol{k}}{2})^{2}+z\bar{z}Q^{2}\bigr)^{2}\bigl((\boldsymbol{\ell}+\frac{\boldsymbol{k}}{2})^{2}+z\bar{z}Q^{2}\bigr)^{2}},\nonumber 
\end{align}
for the longitudinal contribution, and 
\begin{align}
	F_{T}(x_{\rm Bj},Q^2) & =g^{2} Q^2 \sum_{f}q_{f}^{2}\int_{0}^{1}\frac{{\rm d}z}{2\pi}\int\frac{{\rm d}^{d}\boldsymbol{\ell}}{(2\pi)^{d}}\int{\rm d}^{d}\boldsymbol{k}\nonumber \\
	& \times\int{\rm d}x \, \delta \!\left[x-x_{{\rm Bj}}\left(1+\frac{\boldsymbol{\ell}^{2}}{z\bar{z}Q^{2}}\right)\right]xG^{ij}(x,\boldsymbol{k}) \left(1-\frac{4}{d}z\bar{z}\right)\label{eq:WTT}\\
	& \times\left[\frac{\delta^{ik}}{(\boldsymbol{\ell}-\frac{\boldsymbol{k}}{2})^{2}+z\bar{z}Q^{2}}-2\frac{(\boldsymbol{\ell}^{i}-\frac{\boldsymbol{k}^{i}}{2})(\boldsymbol{\ell}^{k}-\frac{\boldsymbol{k}^{k}}{2})}{\bigl((\boldsymbol{\ell}-\frac{\boldsymbol{k}}{2})^{2}+z\bar{z}Q^{2}\bigr)^{2}}\right]\nonumber \\
	& \times\left[\frac{\delta^{jk}}{(\boldsymbol{\ell}+\frac{\boldsymbol{k}}{2})^{2}+z\bar{z}Q^{2}}-2\frac{(\boldsymbol{\ell}^{j}+\frac{\boldsymbol{k}^{j}}{2})(\boldsymbol{\ell}^{k}+\frac{\boldsymbol{k}^{k}}{2})}{\bigl((\boldsymbol{\ell}+\frac{\boldsymbol{k}}{2})^{2}+z\bar{z}Q^{2}\bigr)^{2}}\right]\nonumber 
\end{align}
for the transverse contribution.

Eqs.~(\ref{eq:WLL}) and (\ref{eq:WTT}) together with \eqn{eq:Gijdef} are the core results of this study. They provide expressions which unify the gluon mediated DIS structure functions  in both known cases: the Bjorken limit and the Regge limit. In effect, these expressions are valid up to $x_{\rm Bj}/Q^2$ corrections. In the former, they are valid because of the $1/Q^2$ suppression, and in the latter they are valid thanks to the $x_{\rm Bj}$ suppression.
In the following sections, we will proceed to show that in the Bjorken limit the standard collinear result is recovered in its entirety including full collinear logarithms and splitting functions, then that in the Regge limit the well known expression involving the dipole-proton scattering matrix element is fully recovered as well.

\section{The Bjorken limit: gluon contribution to one loop DIS}\label{sec:Bjorken}

In the Bjorken limit, observables are factorized using QCD factorization. In the simplest inclusive cases such as the DIS cross section it is also known as \textit{collinear} factorization due to the fact that non-perturbative matrix elements involved then describe the physics of partons which are collinear to their mother hadron, carrying no transverse momentum. It is obtained by putting the separation in the bilocal correlator which defines the gluon distribution on the light cone. Equivalently, it can be obtained by neglecting the transfer of transverse momentum from the target in the hard subamplitude, assuming that any intrinsic transverse momentum to the hadron is negligible compared to the hard splitting momentum and the hard scale: $\boldsymbol{|k|}\ll |\boldsymbol{\ell}| \sim Q$ in~\eqn{eq:WLL} and~\eqn{eq:WTT}, or $\boldsymbol{v}\rightarrow \boldsymbol{0}$ in the second line of~ \eqn{eq:Gijdef}. Either way, the gluon PDF will appear from the $\boldsymbol{k}$ integral of the unintegrated distribution $G^{ij}(x,\boldsymbol{k})$. 
With 
\begin{align}
	xg(x) & \equiv\frac{\delta^{ij}}{P^{-}}\int\frac{{\rm d}v^{+}}{2\pi}{\rm e}^{ixP^{-}v^{+}} \langle P|F^{i-}(v^+)[v^+,0^+]F^{j-}(0^+)[0^+,v^+]|P\rangle\,,\label{eq:PDF-def} 
\end{align}
it is easy to check that:
\beq
\int \rmd^d \boldsymbol{k} \, \delta^{ij}G^{ij}(x,\boldsymbol{k})=g(x).\label{eq:Gtog}
\eeq
Let us focus on the hard subamplitude, to show how the $\delta^{ij}$ projection emerges. Neglecting $|\boldsymbol{k}|\ll |\boldsymbol{\ell}|$ in \eqn{eq:WLL} and \eqn{eq:WTT} yields:
\begin{align}
	F_L(x_{\rm Bj},Q^2\rightarrow\infty) & = g^{2}Q^2 \sum_{f}q_{f}^{2}\int{\rm d}x\int{\rm d}^{d}\boldsymbol{k}\, xG^{ij}(x,\boldsymbol{k}) \nonumber \\
	& \times \int_{0}^{1}\frac{{\rm d}z}{2\pi}\int\frac{{\rm d}^{d}\boldsymbol{\ell}}{(2\pi)^{d}} \delta\left(x-x_{\rm Bj}\frac{\boldsymbol{\ell}^{2}+z\bar{z}Q^{2}}{z\bar{z}Q^2}\right) \label{eq:WLL-col}\\
	& \times\frac{16z^{2}\bar{z}^{2}Q^{2}\boldsymbol{\ell}^i\boldsymbol{\ell}^j}{\bigl(\boldsymbol{\ell}^{2}+z\bar{z}Q^{2}\bigr)^{4}}\nonumber 
\end{align}
for the longitudinal case, and 
\begin{align}
	F_T(x_{\rm Bj},Q^2\rightarrow\infty) & =g^{2}Q^2\sum_{f}q_{f}^{2}\int{\rm d}x\int{\rm d}^{d}\boldsymbol{k}\, xG^{ij}(x,\boldsymbol{k}) \nonumber \\
	& \times\int_{0}^{1}\frac{{\rm d}z}{2\pi}\int\frac{{\rm d}^{d}\boldsymbol{\ell}}{(2\pi)^{d}} \delta \left(x-x_{\rm Bj}\frac{\boldsymbol{\ell}^{2}+z\bar{z}Q^{2}}{z\bar{z}Q^2}\right) \left(1-\frac{4}{d}z\bar{z}\right)\label{eq:WTT-col}\\
	& \times\left(\frac{\delta^{ij}}{(\boldsymbol{\ell}^{2}+z\bar{z}Q^{2})^{2}}-4\frac{\boldsymbol{\ell}^{i}\boldsymbol{\ell}^{j}}{(\boldsymbol{\ell}^{2}+z\bar{z}Q^{2})^{3}}+4\frac{\boldsymbol{\ell}^{i}\boldsymbol{\ell}^{j}\boldsymbol{\ell}^{2}}{(\boldsymbol{\ell}^{2}+z\bar{z}Q^{2})^{4}}\right) \nonumber
\end{align}
for the transverse case. The $\delta^{ij}$ projection appears from the usual substitution
\beq
\boldsymbol{\ell}^i \boldsymbol{\ell}^j \rightarrow \frac{1}{d}\delta^{ij}\boldsymbol{\ell}^2,
\eeq
which is valid as long as the transverse integral does not involve other vectors, as it is the case here. It is actually possible to decouple the integrals w.r.t. $z$ and $\boldsymbol{\ell}$ via the simple change of variable $\boldsymbol{\ell}\rightarrow \boldsymbol{p}\equiv\boldsymbol{\ell}/{\sqrt{z\bar{z}}}$. Then we find:
\begin{align}
	F_L(x_{\rm Bj},Q^2\rightarrow\infty) & =g^{2}Q^2\sum_{f}q_{f}^{2}\int\frac{{\rm d}x}{2\pi}\int{\rm d}^{d}\boldsymbol{k}x\delta^{ij}G^{ij}(x,\boldsymbol{k})\nonumber \\
	& \times\int_{0}^{1}{\rm d}z\left(z\bar{z}\right)^{\frac{d}{2}-1}\int\frac{{\rm d}^{d}\boldsymbol{p}}{(2\pi)^{d}}\delta\left(x-x_{{\rm Bj}}-x_{{\rm Bj}}\frac{\boldsymbol{p}^{2}}{Q^{2}}\right)\nonumber \\
	& \times\frac{1}{d}\frac{16Q^{2}\boldsymbol{p}^{2}}{\bigl(\boldsymbol{p}^{2}+Q^{2}\bigr)^{4}}\label{eq:WLL-col-dec}
\end{align}

and
\begin{align}
	F_T(x_{\rm Bj},Q^2\rightarrow\infty) & =g^{2}Q^2\sum_{f}q_{f}^{2}\int\frac{{\rm d}x}{2\pi}\int{\rm d}^{d}\boldsymbol{k}x\delta^{ij}G^{ij}(x,\boldsymbol{k})\nonumber \\
	& \times\int_{0}^{1}{\rm d}z\left[\left(z\bar{z}\right)^{\frac{d}{2}-2}-\frac{4}{d}\left(z\bar{z}\right)^{\frac{d}{2}-1}\right]\int\frac{{\rm d}^{d}\boldsymbol{p}}{(2\pi)^{d}}\delta\left(x-x_{{\rm Bj}}-x_{{\rm Bj}}\frac{\boldsymbol{p}^{2}}{Q^{2}}\right)\nonumber \\
	& \times\left(\frac{1}{(\boldsymbol{p}^{2}+Q^{2})^{2}}-\frac{4}{d}\frac{\boldsymbol{p}^{2}}{(\boldsymbol{p}^{2}+Q^{2})^{3}}+\frac{4}{d}\frac{\boldsymbol{p}^{4}}{(\boldsymbol{p}^{2}+Q^{2})^{4}}\right)\,.\label{eq:WTT-col-dec}
\end{align}
The $z$ integral will be taken using the standard integral
\begin{equation}
	\int_{0}^{1}{\rm d}z\bar{z}^{p}z^{q}=\frac{\Gamma\left(p+1\right)\Gamma\left(q+1\right)}{\Gamma\left(p+q+2\right)}\label{eq:intz}.
\end{equation}
For the $\boldsymbol{p}$ integral, it is convenient to note that the integrand is a function of $\boldsymbol{p}^2$ only. We can thus integrate out the angular dependence into the volume of the hypersphere: for any function $\varphi(\boldsymbol{p}^2)$, we have
\begin{equation}
	\int\frac{{\rm d}^{d}\boldsymbol{p}}{(2\pi)^{d}}\varphi(\boldsymbol{p}^{2})=\frac{\pi^{\frac{d}{2}}}{\Gamma\left(\frac{d}{2}\right)}\int_{0}^{\infty}\frac{{\rm d}\boldsymbol{p}^{2}}{(2\pi)^{d}}\left(\boldsymbol{p}^{2}\right)^{\frac{d}{2}-1}\varphi(\boldsymbol{p}^{2})\label{eq:ang-int}.
\end{equation}
In fact, we can actually go one step further. For any function $\varphi(\boldsymbol{p}^2)$, we have:
\begin{align}
	& \int\frac{{\rm d}^{d}\boldsymbol{p}}{(2\pi)^{d}}\delta\left(x-x_{{\rm Bj}}-x_{{\rm Bj}}\frac{\boldsymbol{p}^{2}}{Q^{2}}\right)\varphi(\boldsymbol{p}^{2})\label{eq:int-ang-delta}\\
	& =\frac{\theta(x-x_{{\rm Bj}})}{(4\pi)^{\frac{d}{2}}\Gamma\left(\frac{d}{2}\right)}\frac{Q^{2}}{x_{{\rm Bj}}}\left[Q^{2}\left(\frac{x-x_{{\rm Bj}}}{x_{{\rm Bj}}}\right)\right]^{\frac{d}{2}-1}\varphi\left[Q^{2}\left(\frac{x-x_{{\rm Bj}}}{x_{{\rm Bj}}}\right)\right].\nonumber 
\end{align}
Putting everything together by using~\eqn{eq:intz}, \eqn{eq:Gtog} and \eqn{eq:int-ang-delta} in~\eqn{eq:WLL-col-dec} and~\eqn{eq:WTT-col-dec}, we get the final results before the expansion around 4 dimensions:
\begin{align}
	F_{L}(x_{{\rm Bj}},Q^{2}\rightarrow\infty) & =8\frac{g^{2}}{\pi(4\pi)^{\frac{d}{2}}}\sum_{f}q_{f}^{2}\int{\rm d}x\theta(x-x_{{\rm Bj}})xg(x)\label{eq:WLL-fin-gen}\\
	& \times\frac{\Gamma\left(\frac{d}{2}\right)}{\Gamma(d+1)}\frac{x_{{\rm Bj}}^{2}(x-x_{{\rm Bj}})}{x^{4}}\left[Q^{2}\left(\frac{x-x_{{\rm Bj}}}{x_{{\rm Bj}}}\right)\right]^{\frac{d}{2}-1}\nonumber 
\end{align}
and
\begin{align}
	F_{T}(x_{{\rm Bj}},Q^{2}\rightarrow\infty) & =2\frac{g^{2}}{\pi(4\pi)^{\frac{d}{2}}}\sum_{f}q_{f}^{2}\int{\rm d}x\theta(x-x_{{\rm Bj}})xg(x)\label{eq:WTT-fin-gen}\\
	& \times\frac{\Gamma\left(\frac{d}{2}\right)}{\Gamma\left(d\right)}\left(\frac{d-1}{d-2}-\frac{1}{d}\right)x_{{\rm Bj}}\frac{x^{2}-\frac{4}{d}x_{{\rm Bj}}\left(x-x_{{\rm Bj}}\right)}{x^{4}}\left[Q^{2}\left(\frac{x-x_{{\rm Bj}}}{x_{{\rm Bj}}}\right)\right]^{\frac{d}{2}-1}.\nonumber 
\end{align}
In $d=2+2\epsilon$ dimensions, introducing $y\equiv x_{
\rm Bj}/x$, these become:
\begin{align}
	F_{L}(x_{{\rm Bj}},Q^{2}\rightarrow\infty) & =\frac{\alpha_s}{\pi}\sum_{f}q_{f}^{2}\int_{x_{{\rm Bj}}}^{1}{\rm d}y\left[xg(x)\right]_{x=x_{{\rm Bj}}/y}4y(1-y)\label{eq:FL-col-dim4}, 
\end{align}
and
\begin{align}
	 &F_{T}(x_{{\rm Bj}},Q^{2}\rightarrow\infty)  =\frac{\alpha_s}{\pi}\sum_{f}q_{f}^{2}\int_{x_{{\rm Bj}}}^{1}{\rm d}y\left[xg(x)\right]_{x=x_{{\rm Bj}}/y}\label{eq:FT-col-dim4}\\
	& \times\left\{ \frac{1}{\epsilon}\left(\frac{{\rm e}^{\gamma_{E}}}{4\pi}\right)^{\epsilon}\left[(1-y)^{2}+y^{2}\right]+\left[(1-y)^{2}+y^{2}\right]\ln\left[\frac{Q^{2}(1-y)}{\mu^{2}y}\right]-1+4y\left(1-y\right)\right\} .\nonumber 
\end{align}
The $qg$ splitting function ${\cal P}^{qg}(y)\equiv (1-y)^2+y^2$ can immediately be recognized in the first, divergent, term in the brackets in~\eqn{eq:FT-col-dim4}. In fact, this term is exactly canceled by the Dokshitzer, Gribov, Lipatov, Altarelli, Parisi (DGLAP) evolution equation~\cite{dglap} for the gluon distribution in the $\overline{MS}$ renormalization scheme. It corresponds to the collinear divergence in the $t$ channel, when the quark or the antiquark is collinear to the target. Using the evolution of the distribution simply amounts to removing this term and replacing the dimensional regularization parameter $\mu$ by the factorization scale $\mu_F$:
\begin{align}
	F_{T}(x_{{\rm Bj}},Q^{2}\rightarrow\infty) & =\frac{g^{2}}{4\pi^{2}}\sum_{f}q_{f}^{2}\int_{x_{{\rm Bj}}}^{1}{\rm d}y\left[xg(x)\right]_{x=x_{{\rm Bj}}/y}\label{eq:FT-col-dim4-ren}\\
	& \times\left\{ \left[(1-y)^{2}+y^{2}\right]\ln\left[\frac{Q^{2}(1-y)}{\mu_F^{2}y}\right]-1+4y\left(1-y\right)\right\} .\nonumber 
\end{align}
We finally recovered the well known result for the gluon contribution to DIS at one loop accuracy~\cite{Bardeen:1978yd}. Note that we found it in its entirety, despite the similarities our framework shares with standard CGC. The small $x_{\rm Bj}$ limit of these structure functions will be discussed in Section~\ref{sec:log-discussion}.

\section{The Regge limit: the dipole cross section}\label{sec:Regge}
In the Regge limit, the most efficient factorization scheme for dilute-dense scattering is the semi-classical formalism one recovers by taking the shock wave approximation in the present framework. In the intermediate steps of the computation, it is tantamount to neglecting the phases of type $i\boldsymbol{k}^2 x^+/(2k^+)$ since all scatterings are then assumed to occur at light cone times $x^+$ close to $0$, and $1/k^+ \sim 1/\sqrt{s} \simeq 0$ in the Regge power counting, and to extending Wilson line operators to infinite times. See Section~\ref{sec:shock-waves}. However, it is more interesting to study how the shock wave limit is recovered from the final factorized results~\eqn{eq:WLL} and~\eqn{eq:WTT}. The $\delta$ function in these equations sets
\beq
x=x_{\rm Bj}\left(1 + \frac{\boldsymbol{\ell}^2}{z\bar{z}Q^2}\right)\label{x-exp}.
\eeq
Provided that the quark loop variable $\boldsymbol{\ell}^2/(z\bar{z})$ does not diverge fast enough to compensate for the smallness of $x_{\rm Bj}$, taking the Regge limit thus comes down to taking the $x \simeq x_{\rm Bj}\simeq 0$ approximation. It is most conveniently taken in coordinate space, which is often considered to be the suitable space for semi-classical small $x$ effective theories. It is also more standard to revert back to before the use of translation invariance in the definition of the gluon distribution. Instead of~\eqn{eq:Gijdef}, we will use:
\begin{align}
	& \delta(0^{-})\delta^{d}(\boldsymbol{0})xG^{ij}(x,\boldsymbol{k})\label{eq:deltaGij}\\
	& =\frac{1}{P^{-}}\int\frac{{\rm d}v_{1}^{+}}{2\pi}\int\frac{{\rm d}v_{2}^{+}}{2\pi}{\rm e}^{ixP^{-}(v_{1}^{+}-v_{2}^{+})}\int\frac{{\rm d}^{d}\boldsymbol{v}_{1}}{(2\pi)^{d}}\int\frac{{\rm d}^{d}\boldsymbol{v}_{2}}{(2\pi)^{d}}{\rm e}^{-i\boldsymbol{k}\cdot(\boldsymbol{v}_{1}-\boldsymbol{v}_{2})}\int_{0}^{1}{\rm d}s{\rm d}s^{\prime}\nonumber \\
	& \times\bigl\langle P\bigl|{\rm tr}_{c}\bigl\{\bigl[v_{2}^{+},v_{1}^{+}\bigr]_{\boldsymbol{v}_{1}}F^{i-}\bigl(v_{1}^{+},s\boldsymbol{v}_{1}+\bar{s}\boldsymbol{v}_{2}\bigr)\bigl[v_{1}^{+},v_{2}^{+}\bigr]_{\boldsymbol{v}_{2}}F^{j-}\bigl(v_{2}^{+},s^{\prime}\boldsymbol{v}_{1}+\bar{s}^{\prime}\boldsymbol{v}_{2}\bigr)\bigr\}\bigr|P\bigr\rangle.\nonumber
\end{align}
Using the Fourier transform of the hard parts, as computed explicitely in~\eqn{eq:K0-int-d} and~\eqn{eq:K1-int-d},
\begin{align*}
	\frac{{\cal T}_{T,L}(z,\boldsymbol{\ell}_{1},\boldsymbol{\ell}_{2}-\boldsymbol{\ell}_{1})}{(\boldsymbol{\ell}_{1}^{2}+z\bar{z}Q^{2})(\boldsymbol{\ell}_{2}^{2}+z\bar{z}Q^{2})} & =\int{\rm d}^{d}\boldsymbol{r}_{1}{\rm d}^{d}\boldsymbol{r}_{2}{\rm e}^{-i(\boldsymbol{\ell}_{1}\cdot\boldsymbol{r}_{1})+i(\boldsymbol{\ell}_{2}\cdot\boldsymbol{r}_{2})}\Phi_{T,L}(z,\boldsymbol{r}_{1},\boldsymbol{r}_{2}),
\end{align*}
and neglecting $x$ based on the discussion above, we find:
\begin{align}
	\lim_{x_{\rm Bj}\rightarrow 0}{\cal A}_{T,L} & = i\pi e^{2}g^{2}(2\pi)^{D}\delta(0^{+})\sum_{f}q_{f}^{2}\int_{0}^{1}\frac{{\rm d}z}{2\pi}\int{\rm d}^{d}\boldsymbol{r}_{1}{\rm d}^{d}\boldsymbol{r}_{2}\frac{(2\pi)^{d}P^{-}}{4q^{+}}\Phi_{T,L}(z,\boldsymbol{r}_{1},\boldsymbol{r}_{2})\label{eq:W-coord-x0}\\
	& \times\int\frac{{\rm d}^{d}\boldsymbol{\ell}_{1}}{(2\pi)^{d}}\int\frac{{\rm d}^{d}\boldsymbol{\ell}_{2}}{(2\pi)^{d}}{\rm e}^{-i(\boldsymbol{\ell}_{1}\cdot\boldsymbol{r}_{1})+i(\boldsymbol{\ell}_{2}\cdot\boldsymbol{r}_{2})}\delta(0^{-})\delta^{d}(\boldsymbol{0})\boldsymbol{r}_{1}^{i}\boldsymbol{r}_{2}^{j}\left[xG^{ij}(x,\boldsymbol{\ell}_{2}-\boldsymbol{\ell}_{1})\right]_{x=0}.\nonumber 
\end{align}
The second line, studied separately along with~\eqn{eq:deltaGij}, yields :
\begin{align}
	& \int\frac{{\rm d}^{d}\boldsymbol{\ell}_{1}}{(2\pi)^{d}}\int\frac{{\rm d}^{d}\boldsymbol{\ell}_{2}}{(2\pi)^{d}}{\rm e}^{-i(\boldsymbol{\ell}_{1}\cdot\boldsymbol{r}_{1})+i(\boldsymbol{\ell}_{2}\cdot\boldsymbol{r}_{2})}\delta(0^{-})\delta^{d}(\boldsymbol{0})\boldsymbol{r}_{1}^{i}\boldsymbol{r}_{2}^{j}\left[xG^{ij}(x,\boldsymbol{\ell}_{2}-\boldsymbol{\ell}_{1})\right]_{x=0}\nonumber \\
	& =\frac{\delta^{d}(\boldsymbol{r}_{1}-\boldsymbol{r}_{2})}{P^{-}(2\pi)^d}\int\frac{{\rm d}v_{1}^{+}}{2\pi}\int\frac{{\rm d}v_{2}^{+}}{2\pi}\int\frac{{\rm d}^{d}\boldsymbol{v}_{2}}{(2\pi)^{d}}\int_{0}^{1}{\rm d}s\int_{0}^{1}{\rm d}s^{\prime}\label{eq:int-G-x0}\\
	& \times\bigl\langle P\bigl|{\rm tr}_{c}\bigl\{\bigl[v_{2}^{+},v_{1}^{+}\bigr]_{\boldsymbol{v}_{2}+\boldsymbol{r}_{1}}\boldsymbol{r}_{1}^{i}F^{i-}\bigl(v_{1}^{+},\boldsymbol{v}_{2}+s\boldsymbol{r}_{1}\bigr)\bigl[v_{1}^{+},v_{2}^{+}\bigr]_{\boldsymbol{v}_{2}}\boldsymbol{r}_{1}^{j}F^{j-}\bigl(v_{2}^{+},\boldsymbol{v}_{2}+s^{\prime}\boldsymbol{r}_{1}\bigr)\bigr\}\bigr|P\bigr\rangle.\nonumber 
\end{align}
Note that the presence of a $\delta$ function which sets the dipole sizes to be equal in the hard part is entirely due to the fact that the distribution is evaluated at null values of the Feynman $x$ variable, because the $x$ variable is dependent on the transverse momenta which integrate into the $\delta$. Any dependence on $x$ in the distribution would force dipole sizes to be distinct.
Using \eqn{eq:AtoF} in order to relate the $s$ integral of field strength tensors to $A^-$ fields, then~\eqn{eq:WLder} in order to relate differences of $A^-$ fields with Wilson lines to the derivatives of Wilson lines leads to the relation
\begin{align}
	& \int\frac{{\rm d}v_{1}^{+}}{2\pi}\bigl[v_{2}^{+},v_{1}^{+}\bigr]_{\boldsymbol{v}_{2}+\boldsymbol{r}_{1}}\int_{0}^{1}{\rm d}s\boldsymbol{r}_{1}^{i}F^{i-}\bigl(v_{1}^{+},\boldsymbol{v}_{2}+s\boldsymbol{r}_{1}\bigr)\bigl[v_{1}^{+},v_{2}^{+}\bigr]_{\boldsymbol{v}_{2}}\nonumber \\
	& =-\int\frac{{\rm d}v_{1}^{+}}{2\pi}\bigl[v_{2}^{+},v_{1}^{+}\bigr]_{\boldsymbol{v}_{2}+\boldsymbol{r}_{1}}\bigl[A^{-}(v_{1}^{+},\boldsymbol{v}_{2}+\boldsymbol{r}_{1})-A^{-}(v_{1}^{+},\boldsymbol{v}_{2})\bigr]\bigl[v_{1}^{+},v_{2}^{+}\bigr]_{\boldsymbol{v}_{2}}\nonumber \\
	& =\frac{-i}{2\pi g}\int{\rm d}v_{1}^{+}\frac{\partial}{\partial v_{1}^{+}}\left(\bigl[v_{2}^{+},v_{1}^{+}\bigr]_{\boldsymbol{v}_{2}+\boldsymbol{r}_{1}}\bigl[v_{1}^{+},v_{2}^{+}\bigr]_{\boldsymbol{v}_{2}}\right)\label{eq:v1-int}\\
	& =\frac{-i}{2\pi g}\left\{ \bigl[v_{2}^{+},\infty^+\bigr]_{\boldsymbol{v}_{2}+\boldsymbol{r}_{1}}\bigl[\infty^+,v_{2}^{+}\bigr]_{\boldsymbol{v}_{2}}-\bigl[v_{2}^{+},-\infty^+\bigr]_{\boldsymbol{v}_{2}+\boldsymbol{r}_{1}}\bigl[-\infty^+,v_{2}^{+}\bigr]_{\boldsymbol{v}_{2}}\right\}. \nonumber 
\end{align}
Using the same steps for the other field strength tensors, for $t^+=\pm \infty$:
\begin{align}
	& \int\frac{{\rm d}v_{2}^{+}}{2\pi}\bigl[t^{+},v_{2}^{+}\bigr]_{\boldsymbol{v}_{2}}\int_{0}^{1}{\rm d}s^{\prime}\boldsymbol{r}_{1}^{j}F^{j-}\bigl(v_{2}^{+},\boldsymbol{v}_{2}+s^{\prime}\boldsymbol{r}_{1}\bigr)\bigl[v_{2}^{+},t^{+}\bigr]_{\boldsymbol{v}_{2}+\boldsymbol{r}_{1}}\label{eq:v2int}\\
	& =\frac{i}{2\pi g}\left\{ \bigl[t^{+},\infty^{+}\bigr]_{\boldsymbol{v}_{2}}\bigl[\infty^{+},t^{+}\bigr]_{\boldsymbol{v}_{2}+\boldsymbol{r}_{1}}-\bigl[t^{+},-\infty^{+}\bigr]_{\boldsymbol{v}_{2}}\bigl[-\infty^{+},t^{+}\bigr]_{\boldsymbol{v}_{2}+\boldsymbol{r}_{1}}\right\}. \nonumber 
\end{align}
Eventually, we find the usual infinite Wilson lines:
\begin{align}
	& \int\frac{{\rm d}^{d}\boldsymbol{\ell}_{1}}{(2\pi)^{d}}\int\frac{{\rm d}^{d}\boldsymbol{\ell}_{2}}{(2\pi)^{d}}{\rm e}^{-i(\boldsymbol{\ell}_{1}\cdot\boldsymbol{r}_{1})+i(\boldsymbol{\ell}_{2}\cdot\boldsymbol{r}_{2})}\delta(0^{-})\delta^{d}(\boldsymbol{0})\boldsymbol{r}_{1}^{i}\boldsymbol{r}_{2}^{j}\left[xG^{ij}(x,\boldsymbol{\ell}_{2}-\boldsymbol{\ell}_{1})\right]_{x=0}\nonumber \\
	& =\frac{\delta^{d}(\boldsymbol{r}_{1}-\boldsymbol{r}_{2})}{(2\pi)^{D}g^{2}P^{-}}\int\frac{{\rm d}^{d}\boldsymbol{v}_{2}}{(2\pi)^{d}}\bigl\langle P\bigl|2N_{c}-{\rm tr}_{c}\bigl(U^{\phantom{\dagger}}_{\boldsymbol{v}_{2}+\boldsymbol{r}_{1}}U_{\boldsymbol{v}_{2}}^{\dagger}\bigr)-{\rm tr}_{c}\bigl(U_{\boldsymbol{v}_{2}+\boldsymbol{r}_{1}}^{\dagger}U^{\phantom{\dagger}}_{\boldsymbol{v}_{2}}\bigr)\bigr\}\bigr|P\bigr\rangle\label{eq:GintoU},
\end{align}
where we used the fact that any operator of the form $[t^+,t^+]_{\boldsymbol{t}}$ is actually unity.
We recognize the real part of the standard dipole-proton scattering matrix element: the second line involves
\begin{equation}
	D(\boldsymbol{x}-\boldsymbol{y})\equiv2{\rm Re}\bigl\langle P\bigl|1-\frac{1}{N_{c}}{\rm tr}(U^{\phantom{\dagger}}_{\boldsymbol{x}}U_{\boldsymbol{y}}^{\dagger})\bigr|P\bigr\rangle\label{eq:dipole-def}.
\end{equation}
Note that because of the invariance of forward matrix elements under translation, it is a function of the dipole size only. In~\eqn{eq:GintoU}, there is an implicit $\delta^d(\boldsymbol{0})$ factor from the $\boldsymbol{v}_2$ integral. In general, this factor is compensated by the use of normalized matrix elements $D(\boldsymbol{x}-\boldsymbol{y})/\langle P|P \rangle$ with $\langle P^\prime | P \rangle = 2P^- (2\pi)^{d+1} \delta(P^{\prime -}-P^-)\delta^d(\boldsymbol{P}^\prime-\boldsymbol{P})$.
Let us plug \eqn{eq:GintoU} back into~\eqn{eq:W-coord-x0}, while renaming $\boldsymbol{v}=\boldsymbol{b}$ and $\boldsymbol{r}_1=\boldsymbol{r}_2=\boldsymbol{r}$. We can recreate the complete $\delta^D(0)$ function to be removed in the optical theorem by compensating with the proton normalization factor. With $\langle ... \rangle_P \equiv \langle P | ... | P \rangle/\langle P|P \rangle $,
\begin{align}
	\lim_{x_{\rm Bj}\rightarrow 0} {\cal A}_{T,L} & = 2i\pi e^{2}(2\pi)^{D}\delta^{D}(0)\sum_{f}q_{f}^{2}\int_{0}^{1}{\rm d}z\int{\rm d}^{d}\boldsymbol{r}\frac{(2\pi)^{d-2}P^{-}\Phi_{T,L}(z,\boldsymbol{r},\boldsymbol{r})}{4q^{+}}\nonumber \\
	& \times2{\rm Re}\int \frac{{\rm d}^{d}\boldsymbol{b}}{(2\pi)^d}\langle N_{c}-{\rm tr}_{c}\bigl(U^{\phantom{\dagger}}_{\boldsymbol{b}+\boldsymbol{r}}U_{\boldsymbol{b}}^{\dagger}\bigr)\rangle_P.\label{eq:Wmunu-x0}
\end{align}
With the explicit hard parts, given by~\eqn{eq:K0-int-d} and~\eqn{eq:K1-int-d}, we can get the final expression for the Regge limit of the $LL$ transition
\begin{align}
	F_{L}(x_{{\rm Bj}}\rightarrow 0,Q^{2}) & =\frac{2}{\pi^{2}}\sum_{f}q_{f}^{2}\int_{0}^{1}{\rm d}z\int{\rm d}^{d}\boldsymbol{r}\label{eq:WLL-eik}\\
	& \times\left(\frac{z\bar{z}Q^{2}}{\boldsymbol{r}^{2}}\right)^{\frac{d}{2}-1}z^{2}\bar{z}^{2}Q^{4}\bigl[K_{\frac{d}{2}-1}(Q\sqrt{z\bar{z}}|\boldsymbol{r}|)\bigr]^{2}\nonumber \\
	& \times2{\rm Re}\int\frac{{\rm d}^{d}\boldsymbol{b}}{(2\pi)^d}\langle N_{c}-{\rm tr}_{c}\bigl(U_{\boldsymbol{b}+\boldsymbol{r}}^{\phantom{\dagger}}U_{\boldsymbol{b}}^{\dagger}\bigr)\rangle_{P},\nonumber 
\end{align}
and the $TT$ transition
\begin{align}
	F_{T}(x_{{\rm Bj}}\rightarrow 0,Q^{2}) & =\frac{Q^{2}}{2\pi^{2}}\sum_{f}q_{f}^{2}\int_{0}^{1}{\rm d}z\int{\rm d}^{d}\boldsymbol{r}\label{eq:WTT-eik}\\
	& \times\left(1-\frac{4}{d}z\bar{z}\right)\boldsymbol{r}^{2}\left(\frac{z\bar{z}Q^{2}}{\boldsymbol{r}^{2}}\right)^{\frac{d}{2}}\bigl[K_{\frac{d}{2}}(Q\sqrt{z\bar{z}}|\boldsymbol{r}|)\bigr]^{2}\nonumber \\
	& \times2{\rm Re}\int\frac{{\rm d}^{d}\boldsymbol{b}}{(2\pi)^d}\langle N_{c}-{\rm tr}_{c}\bigl(U_{\boldsymbol{b}+\boldsymbol{r}}^{\phantom{\dagger}}U_{\boldsymbol{b}}^{\dagger}\bigr)\rangle_{P}.\nonumber 
\end{align}

In this limit, we thus managed to recover the standard results in their entirety. Let us insist once more on the fact that these results are obtained by taking the strict $x=0$ limit in the non-perturbative distribution. The fact that the target matrix element is simply the dipole-target scattering element relies on this approximation, because otherwise the matrix elements would be evaluated at different positions $(\boldsymbol{b},\boldsymbol{r},\boldsymbol{b}^\prime,\boldsymbol{r}^\prime) $ and one would find the more involved operator
\begin{equation}
	N_{c}-{\rm tr}(U_{\boldsymbol{b}+\boldsymbol{r}}^{\phantom{\dagger}}U_{\boldsymbol{b}}^{\dagger})-{\rm tr}(U_{\boldsymbol{b}^{\prime}}^{\phantom{\dagger}}U_{\boldsymbol{b}^{\prime}+\boldsymbol{r}^{\prime}}^{\dagger})+{\rm tr}(U_{\boldsymbol{b}+\boldsymbol{r}}^{\phantom{\dagger}}U_{\boldsymbol{b}}^{\dagger}U_{\boldsymbol{b}^{\prime}}^{\phantom{\dagger}}U_{\boldsymbol{b}^{\prime}+\boldsymbol{r}^{\prime}}^{\dagger})\label{eq:quadrupole}
\end{equation}
in lieu of real part of the dipole. We will now elaborate on this statement in the following section.

\section{Collinear logs at small $x_{\rm Bj}$: Towards a top down approach}\label{sec:log-discussion}

It is particularly informative to study the leading power of $1/Q$ in the Regge limit of the structure functions. It can be recovered thanks to the observation that the Bessel functions in the hard sub-amplitude peak around $\boldsymbol{r}^2 \propto 1/Q^2$, along with the Taylor expansion of the dipole-target matrix element~\cite{Dominguez:2011wm}:
\begin{align}
	& \int\frac{{\rm d}^{d}\boldsymbol{b}}{(2\pi)^d}\bigl\langle N_{c}-{\rm tr}_{c}\bigl(U^{\phantom{\dagger}}_{\boldsymbol{b}+\boldsymbol{r}}U_{\boldsymbol{b}}^{\dagger}\bigr)\bigr\rangle_{P} \nonumber \\
	& =\frac{\boldsymbol{r}^{i}\boldsymbol{r}^{j}}{2}\int\frac{{\rm d}^{d}\boldsymbol{b}}{(2\pi)^d}\bigl\langle{\rm tr}_{c}(\partial^{i}U_{\boldsymbol{b}})(\partial^{j}U_{\boldsymbol{b}}^{\dagger})\bigr\rangle_{P}+O(|\boldsymbol{r}|^{3})\label{eq:UUtoFF}\\
	& =\frac{\boldsymbol{r}^{i}\boldsymbol{r}^{j}g^2}{4P^{-}(2\pi)^d}\int{\rm d}v^{+}\bigl\langle P\bigl|[0^{+},v^{+}]F^{i-}(v^{+})[v^{+},0^{+}]F^{j-}(0^{+})\bigr|P\bigr\rangle+O(|\boldsymbol{r}|^{3})\nonumber .
\end{align}
Given that the hard parts only depend on $\boldsymbol{r}^2$, we can perform the substitution $\boldsymbol{r}^i \boldsymbol{r}^j \rightarrow \boldsymbol{r}^2 \delta^{ij}/d$. Finally using the definition~\eqn{eq:PDF-def} we get the relation:
\begin{align}
	\int\frac{{\rm d}^{d}\boldsymbol{b}}{(2\pi)^d}\bigl\langle N_{c}-{\rm tr}_{c}\bigl(U^{\phantom{\dagger}}_{\boldsymbol{b}+\boldsymbol{r}}U_{\boldsymbol{b}}^{\dagger}\bigr)\bigr\rangle_{P} & =(2\pi)^{2-d}\frac{\alpha_{s}\boldsymbol{r}^{2}}{2d}\left[xg(x)\right]_{x=0}+O(|\boldsymbol{r}|^{3})\label{eq:UUtoPDF}.
\end{align} 
A bit of algebra, detailed in Appendix~\ref{sec:double-log-details}, leads to an intermediate result which is reminiscent of~\eqn{eq:WLL-col}:
\begin{align}
	\lim_{Q^{2}\rightarrow\infty}F_{L}(x_{{\rm Bj}}\rightarrow0,Q^{2}) & =g^{2}Q^{2}\sum_{f}q_{f}^{2}\int{\rm d}x\left[xg(x)\right]_{x=0}\label{eq:WLL-eik-LT-2}\\
	& \times\int_{0}^{1}\frac{{\rm d}z}{2\pi}\int\frac{{\rm d}^{d}\boldsymbol{\ell}}{(2\pi)^{d}}\frac{1}{d}\frac{16z^{2}\bar{z}^{2}Q^{2}\boldsymbol{\ell}^{2}}{\bigl(\boldsymbol{\ell}^{2}+z\bar{z}Q^{2}\bigr)^{4}}\delta(x),\nonumber 
\end{align}
and a similar one for $F_T$. Comparing this to its equivalent in the Bjorken limit from Eq.~(\ref{eq:WLL-col}) reveals the proper way to obtain the small $x$ limit of that result: by taking $x_{\rm Bj}=0$ in the $\delta$ function which sets $x$ as a function of $x_{\rm Bj}$. This procedure can lead to issues in the case where the proportionality factor between the two, which is comprised of loop variables, diverges and compensates for the smallness of $x_{\rm Bj}$. This factor, $\boldsymbol{\ell}^2/{z\bar{z}Q^2}$, actually diverges in the $z\in \{0,1\}$ limit where the quark or the antiquark is collinear to the target, which is precisely the limit where the DGLAP evolution equation arises. This will eventually prove to be the reason why the collinear sector of small-$x$ observables is incorrectly dealt with, as we will see later in this section.
Eventually, the leading twist limit of the Regge limit of $F_L$ and $F_T$ read respectively:
\begin{equation}
	\lim_{Q^{2}\rightarrow\infty}F_{L}(x_{{\rm Bj}}\rightarrow0,Q^{2}) =\frac{2}{3}\frac{\alpha_{s}}{\pi}\sum_{f}q_{f}^{2}\left[xg(x)\right]_{x=0},\label{eq:FL-R-B-fin}
\end{equation}
and
\begin{align}
	& \lim_{Q^{2}\rightarrow\infty}F_{T}(x_{{\rm Bj}}\rightarrow0,Q^{2})\label{eq:FT-R-B-fin}\\
	& =\frac{\alpha_{s}}{\pi}\sum_{f}q_{f}^{2}\left[xg(x)\right]_{x=0}\int_{0}^{1}{\rm d}y\Biggl\{\frac{1}{\epsilon}\left(\frac{{\rm e}^{\gamma_{E}}}{4\pi}\right)^{\epsilon}\left[(1-y)^{2}+y^{2}\right]\nonumber\\
	& +\left[(1-y)^{2}+y^{2}\right]\ln\left[\frac{Q^{2}(1-y)}{\mu^{2}y}\right]-1+4y(1-y)\Biggr\}\nonumber.
\end{align}
In~\eqn{eq:FT-R-B-fin}, we recognize the DGLAP $qg$ splitting function ${\cal P}^{qg}(y)=(1-y)^2+y^2$ once again. However, the way it appears in that equation is different from the way it appeared in~\eqn{eq:FT-col-dim4}. Indeed, rather than the DGLAP convolution of the parton distribution with the splitting function, we find the integral of the splitting function taken independently from the distribution:
\begin{equation}
	\left.\lim_{Q^{2}\rightarrow\infty}F_{T}(x_{{\rm Bj}}\rightarrow0,Q^{2})\right|_{{\rm div}}=\frac{\alpha_{s}}{\pi}\sum_{f}q_{f}^{2}\frac{1}{\epsilon}\left(\frac{{\rm e}^{\gamma_{E}}}{4\pi}\right)^{\epsilon}\left[xg(x)\right]_{x=0}\int_{0}^{1}{\rm d}y{\cal P}^{qg}(y)\label{eq:FT-R-B-div}
\end{equation}
The collinear pole can thus be compensated by a collinear logarithm of the physical factorization scale by renormalizing the gluon distribution if and only if:
\begin{equation}
	\lim_{x_{{\rm Bj}}\rightarrow0}\int_{x_{{\rm Bj}}}^{1}{\rm d}y\left[xg(x)\right]_{x=x_{{\rm Bj}}/y}{\cal P}^{qg}(y)=\left[xg(x)\right]_{x=0}\int_{0}^{1}{\rm d}y{\cal P}^{qg}(y).\label{eq:DGLAP-assumption-1}
\end{equation}
In other words, in order to cancel out the collinear divergence from the semi-classical Regge limit result it is necessary to assume that, in the $x=0$ limit, the parton distributions are constant and that the splitting functions are integrable on $[0,1]$. Note that similar observations have been made in~\cite{Hatta:2017cte} in the context of Deeply Virtual Compton Scattering, where the distribution is explicitly assumed to be a constant for the leading twist limit matching discussed in the appendix. This leads to potential issues.

First of all, this equation is obviously only correct if the PDF is a constant at $x=0$. It can be justified, incidentally, at leading logarithmic accuracy provided that the PDF is a power $x^{-\gamma}$ at small values of $x$ and that $\gamma=O(\alpha_{s})$ has a perturbative expansion\footnote{In other words, it works if the PDF is a constant at $x=0$ up to higher loop corrections.}. The fact that the distribution is evaluated exactly at $x=0$ is the fundamental origin of the problems of small $x_{{\rm Bj}}$ observables in the collinear corner of the phase space: the equality above is not correct if there is any dependence on $x$ in the parton distribution.

It is commonly accepted in CGC phenomenology that the BK evolution
equation allows us to restore some form of dependence on $x$ in the
parton distribution, which would then mean that said distribution
is not actually evaluated at $x=0$ in practice. We will now build an argument against this idea, or at least against some interpretations of it. Mathematically, several notions of $x$ variables must be distinguished, even though we tend to consider them to be equivalent in the Regge approximation where they are all close to $0$. First, the $(x_{i})_{i=1...n}$ variables in a gauge invariant distribution with $n$ partons represent the intrinsic longitudinal momentum fractions inside the target hadron. They are Fourier conjugated to the light cone distances between the partonic fields inside the target and are intrinsic properties of the distribution, independently from any observable they might couple to. Second, a process dependent $x_{F}$ variable represents the total fraction of longitudinal momentum transferred from the target, as set by the hard partonic subpart of the process to which the distribution couples. Lastly, the process dependent Bjorken variable $x_{{\rm Bj}}$ is given as a function of the hard scale of the process and of its center-of-mass energy. It is in principle the variable whose large logarithms can compensate for the smallness of the coupling constant $\alpha_{s}$ in the Regge
limit. The basic idea of semi-classical small-$x$ physics is that the BK evolution equation allows to resum all powers of $\alpha_{s}\ln(1/x_{{\rm Bj}})$: it yields the logarithmic dependence on a rapidity regulator $\eta$~\footnote{It is common to denote this variable as an $x$, but to avoid any confusion with the $x$ variables that exist in the Bjorken limit we will refrain from using this notation.} which is then set to $\eta=\ln1/x_{{\rm Bj}}$. This procedure is the Regge analog of the logarithmic dependence on a scale $\mu_{F}$ introduced by QCD factorization and the DGLAP evolution equation where eventually one sets $\mu_{F}=Q$.

With the current example of the inclusive DIS cross section in mind,
let us study for a while how the dependence on each of the notions
of ``$x$'' variables goes.

The most straightforward one is the $x$ variable from the definition
of parton distribution functions:
\begin{equation}
xg(x)=\frac{\delta^{ij}}{P^{-}}\int\frac{{\rm d}v^{+}}{2\pi}{\rm e}^{ixP^{-}v^{+}}\langle P|F^{i-}(v^{+})[v^{+},0^{+}]F^{j-}(0^{+})[0^{+},v^{+}]|P\rangle.
\end{equation}
The fact that one encounters infinite length Wilson line operators in the shock wave approximation is inextricable from the fact that parton distributions are evaluated at $x=0$: the Wilson line is the
leading power in the $x$ expansion~\cite{Dominguez:2011wm}. Since $x$ is Fourier conjugated to a longitudinal distance in the target and said distance is taken to be infinite in the shock wave approximation, it is actually natural for $x$ to be null in that limit. In fact, any action of Wilson line operators on target states can be rewritten into parton distributions with all $(x_{i})_{i}$ variables set to 0~\cite{Altinoluk:2019wyu, Boussarie:2020vzf}. This $x$ variable in the distribution describes an intrinsic property of the target which does not depend on a given process, and in full generality there is no straightforward relation between the scalar $x$ and the energy $\sqrt{s}$. 

The second notion of ``$x$'' variables is the process-dependent $x_{F}$ variable which is the total fraction of longitudinal momentum transferred from the target as determined by the hard subprocess. At leading genuine twist\footnote{Beyond the leading genuine twist where only one collinear gluon is
extracted, we would have several $x_{i}$ variables in the
distribution, and the hard part would only set $\sum_i x_i=x_{F}$.} , and unless one is considering an exclusive process with non-zero skewedness\footnote{In an exclusive process, $x_{F}$ is related to the skewedness parameter
	$\xi$, and the relation between $x$ and $\xi$ obtained from computing
	the hard part is not always as simple as $x=\xi=x_{F}$.}, the hard part imposes that $x=x_{F}$, which is why these two variables tend to be used interchangeably. For example in the inclusive
DIS cross section considered in this article, $x_{F}$ is given by
$x_{F}=x_{{\rm Bj}}(1+\frac{\boldsymbol{\ell}^{2}}{z\bar{z}Q^{2}})$
and the hard part provides a $\delta(x-x_{F})$ function. This could 
allow to mimick an $x$ dependence in the parton distribution by introducing instead an $x_{F}$ dependence through the evolution equation. We will now show that the shock wave description of the DIS cross section is also incompatible with any dependence on this $x_{F}$ variable in the distribution, be it from BK evolution or not.

Let us consider the $\gamma^{\ast}P\rightarrow q\bar{q}X$ amplitude
in the shock wave approximation. It reads~\cite{Boussarie:2021lkb}:
\begin{align}
	{\cal A}_{L/T} & =z\bar{z}e_{q}Q\delta(p_{q}^{+}+p_{\bar{q}}^{+}-q^{+})\int{\rm d}^{2}\boldsymbol{x}_{1}{\rm d}^{2}\boldsymbol{x}_{2}{\rm e}^{-i\left(\boldsymbol{p}_{q}\cdot\boldsymbol{x}_{1}\right)-i\left(\boldsymbol{p}_{q}\cdot\boldsymbol{x}_{2}\right)}\nonumber \\
	& \times(1-U_{\boldsymbol{x}_{1}}^{\phantom{\dagger}}U_{\boldsymbol{x}_{2}}^{\dagger})\varphi_{L/T}(z,\boldsymbol{x}_{1}-\boldsymbol{x}_{2}),
\end{align}
with 
\begin{equation}
	\varphi_{L}(z,\boldsymbol{x}_{1}-\boldsymbol{x}_{2})=2K_{0}(\sqrt{z\bar{z}Q^{2}\boldsymbol{r}^{2}})(\bar{u}_{p_{q}}\gamma^{+}v_{p_{\bar{q}}})
\end{equation}
and 
\begin{equation}
	\varphi_{T}(z,\boldsymbol{x}_{1}-\boldsymbol{x}_{2})=-i\frac{\varepsilon_{q\perp\mu}r_{\perp\nu}}{\left|\boldsymbol{r}\right|\sqrt{z\bar{z}}}K_{1}(\sqrt{z\bar{z}Q^{2}\boldsymbol{r}^{2}})\bar{u}_{p_{q}}\left[z(\gamma_{\perp}^{\mu}\gamma_{\perp}^{\nu})-\bar{z}(\gamma_{\perp}^{\nu}\gamma_{\perp}^{\mu})\right]\gamma^{+}v_{p_{\bar{q}}}.
\end{equation}
The $\gamma^{\ast}P\rightarrow q\bar{q}X$ cross section is thus given
by:
\begin{align}
	{\rm d}\sigma_{L/T;L/T}^{\gamma^{\ast}P\rightarrow q\bar{q}X} & =\frac{{\rm d}p_{q}^{+}{\rm d}^{2}\boldsymbol{p}_{q}{\rm d}p_{\bar{q}}^{+}{\rm d}^{2}\boldsymbol{p}_{\bar{q}}}{4(q^{+})^{3}(2\pi)^{6}}\delta(p_{q}^{+}+p_{\bar{q}}^{+}-q^{+})\\
	& \times z\bar{z}q_{f}^{2}\alpha_{{\rm em}}Q^{2}\int{\rm d}^{2}\boldsymbol{x}_{1}{\rm d}^{2}\boldsymbol{x}_{2}{\rm d}^{2}\boldsymbol{x}_{1}^{\prime}{\rm d}^{2}\boldsymbol{x}_{2}^{\prime}{\rm e}^{i\boldsymbol{p}_{q}\cdot\left(\boldsymbol{x}_{1}^{\prime}-\boldsymbol{x}_{1}\right)+i\boldsymbol{p}_{q}\cdot\left(\boldsymbol{x}_{2}^{\prime}-\boldsymbol{x}_{2}\right)}\nonumber \\
	& \times\langle{\rm tr}_{c}(1-U_{\boldsymbol{x}_{1}}^{\phantom{\dagger}}U_{\boldsymbol{x}_{2}}^{\dagger})(1-U_{\boldsymbol{x}_{2}^{\prime}}^{\phantom{\dagger}}U_{\boldsymbol{x}_{1}^{\prime}}^{\dagger})\rangle_{P}\varphi_{L/T}(z,\boldsymbol{x}_{1}-\boldsymbol{x}_{2})\varphi_{L/T}^{\ast}(z,\boldsymbol{x}_{1}^{\prime}-\boldsymbol{x}_{2}^{\prime}).\nonumber 
\end{align}
We can now recover the leading perturbative power of the DIS cross
section without using the optical theorem, simply by integrating
w.r.t. the quark and antiquark momenta and summing over the flavors
in the loop:
\begin{align}
	{\rm d}\sigma_{L/T;L/T}^{\gamma^{\ast}P\rightarrow X} & =\sum_{f}\frac{q_{f}^{2}\alpha_{{\rm em}}Q^{2}}{4(q^{+})^{2}(2\pi)^{2}}\int{\rm d}^{2}\boldsymbol{x}_{1}{\rm d}^{2}\boldsymbol{x}_{2}\langle{\rm tr}_{c}(1-U_{\boldsymbol{x}_{1}}^{\phantom{\dagger}}U_{\boldsymbol{x}_{2}}^{\dagger})(1-U_{\boldsymbol{x}_{2}}^{\phantom{\dagger}}U_{\boldsymbol{x}_{1}}^{\dagger})\rangle_{P}\nonumber \\
	& \times\int{\rm d}z(z\bar{z})\varphi_{L/T}(z,\boldsymbol{x}_{1}-\boldsymbol{x}_{2})\varphi_{L/T}^{\ast}(z,\boldsymbol{x}_{1}-\boldsymbol{x}_{2}).
\end{align}
The most crucial step here is that the integral w.r.t. $\boldsymbol{p}_{q}$
and $\boldsymbol{p}_{\bar{q}}$ have set $\boldsymbol{x}_{1}=\boldsymbol{x}_{1}^{\prime}$
and $\boldsymbol{x}_{2}=\boldsymbol{x}_{2}^{\prime}$. Then the target
matrix element becomes 
\begin{equation}
	\langle{\rm tr}_{c}(1-U_{\boldsymbol{x}_{1}}^{\phantom{\dagger}}U_{\boldsymbol{x}_{2}}^{\dagger})(1-U_{\boldsymbol{x}_{2}}^{\phantom{\dagger}}U_{\boldsymbol{x}_{1}}^{\dagger})\rangle_{P}=2{\rm Re}\langle N_{c}-{\rm tr}_{c}(U_{\boldsymbol{x}_{1}}^{\phantom{\dagger}}U_{\boldsymbol{x}_{2}}^{\dagger})\rangle_{P}.
\end{equation}
Suppose we had introduced a dependence on $x_{F}$ in the distribution.
Given that $x_{F}=p_{q}^{-}+p_{\bar{q}}^{-}-q^{-}=\frac{\bar{z}\boldsymbol{p}_{q}^{2}+z\boldsymbol{p}_{\bar{q}}^{2}+z\bar{z}Q^{2}}{2z\bar{z}q^{+}}$
depends on the transverse momenta, we would not have been able to
set $\boldsymbol{x}_{1}=\boldsymbol{x}_{1}^{\prime}$ and $\boldsymbol{x}_{2}=\boldsymbol{x}_{2}^{\prime}$
using the momentum integrations. Instead of having a dipole matrix
element for the target, we would have found: 
\begin{equation}
\langle{\rm tr}_{c}(1-U_{\boldsymbol{x}_{2}^{\prime}}^{\phantom{\dagger}}U_{\boldsymbol{x}_{1}^{\prime}}^{\dagger}-U_{\boldsymbol{x}_{1}}^{\phantom{\dagger}}U_{\boldsymbol{x}_{2}}^{\dagger}+U_{\boldsymbol{x}_{1}}^{\phantom{\dagger}}U_{\boldsymbol{x}_{2}}^{\dagger}U_{\boldsymbol{x}_{2}^{\prime}}^{\phantom{\dagger}}U_{\boldsymbol{x}_{1}^{\prime}}^{\dagger})\rangle_{P}.
\end{equation}
In other words, using the dipole operator to describe the inclusive
DIS cross section is inherently incompatible with having any $x_{F}$
dependence in the non-perturbative matrix element. This was previously observed in a similar context in~\cite{Bialas:2000xs}, in which the authors found an incompatibility between the dipole model (the large $N_c$ limit of the CGC) and a so-called \textit{exact kinematics} small-$x$ factorization scheme based on a perturbative ansatz for the target where a dependence on $x$ is added by hand. This observation was then confirmed with a numerical anaysis of this scheme~\cite{updf}.
To summarize, for inclusive DIS gluon distributions are evaluated strictly at $x=x_{F}=0$ in the Regge limit.
Indeed, $x$ is $0$ because we have infinite Wilson line operators, and $x_{F}$ is $0$ because we have a dipole operator in particular.

It is thus impossible to correct for \eqn{eq:DGLAP-assumption-1} with an $x$ dependence in the distribution: the DGLAP convolution is irremediably undone by the shock wave approximation.

Finally, the last ``$x$'' variable is the Bjorken $x_{{\rm Bj}}$ variable. This is the variable whose logarithms the small-$x_{{\rm Bj}}$ evolution equations were initially designed to resum. In all known non-exclusive cases $x_{F}$ and $x_{{\rm Bj}}$ are proportional to each other, but the proportionality factor is not suppressed in either
the Bjorken regime or the Regge limit: in our inclusive DIS example,
$x_{F}=x_{{\rm Bj}}(1+\frac{\boldsymbol{\ell}^{2}}{z\bar{z}Q^{2}})$
and $\boldsymbol{\ell}^{2}/Q^{2}$ is leading power in twist counting\footnote{Indeed $\ell$ is a hard loop momentum so it can be of order $Q$.}
and in eikonal counting. However, the logarithms of $x_{F}$ and $x_{{\rm Bj}}$
are the same at leading logarithm of $s$ accuracy:
\begin{equation}
\ln(x_{F})=-\ln(s)+\ln(Q^{2}+\frac{\boldsymbol{\ell}^{2}}{z\bar{z}})\sim-\ln(s)\sim\ln(x_{{\rm Bj}}).
\end{equation}
This is finally why if we know the $x$ dependence in the PDF, we
know the $x_{{\rm Bj}}$ dependence of the inclusive DIS cross section:
first the hard part sets $x=x_{F}$, then the leading logarithm of
$x_{F}$ is noted to be equivalent to that of $x_{{\rm Bj}}$ at leading
logarithmic accuracy, which means any power dependence $x^{-\gamma}$
in the distribution will lead in practice to an $x_{{\rm Bj}}^{-\gamma}$
dependence in the observable, up to higher logarithmic corrections.
The converse is not necessarily true: knowing the $x_{\rm Bj}$ dependence of an observable does not mean we know the $x$ dependence in the involved distributions without further assumptions. We can, maybe incidentally, justify the disentanglement of the DGLAP convolution using an $x_{{\rm Bj}}$ dependence: if $xg(x)\sim_{x=0}x^{-\gamma}$,
we have
\begin{equation}
	\lim_{x_{{\rm Bj}}\rightarrow0}\int_{x_{{\rm Bj}}}^{1}{\rm d}y\left[xg(x)\right]_{x=x_{{\rm Bj}}/y}{\cal P}^{qg}(y)=x_{{\rm Bj}}^{-\gamma}\int_{0}^{1}{\rm d}y[y^{\gamma}{\cal P}^{qg}(y)].\label{eq:DGLAP-assumption-1-1}
\end{equation}
The power of $x_{{\rm Bj}}$ obtained via BK evolution yields the
$x_{{\rm Bj}}^{-\gamma}$ factor one would have obtained using the
more rigorous distribution $\left[xg(x)\right]_{x=x_{{\rm Bj}}}$
which, let us insist again, cannot be computed using infinite Wilson
line operators. For a fully consistent scheme, is it then absolutely
necessary to check order by order that the $\gamma$-th moments of
the splitting functions appear in each observable. It is also worth
noting that the ${\cal P}^{gg}$ splitting function whose integral
is divergent will yield serious issues with disentangling the DGLAP
convolution that this consideration cannot fix.

The negative cross section problem for small-$x$ physics was initially noticed by studies of observables at NLL accuracy. In practice, the BK resummation schemes mimic a dependence on $x$ in order to catch the leading collinear divergences by imposing by hand the entanglement of longitudinal and transverse variables: one can add additional \textit{ad hoc} kinematic constraints~\cite{collinear-logs-Beuf} in the evolution equation or directly include an additional resummation of logarithms~\cite{collinear-logs-Edmond-1,collinear-logs-Edmond-2,collinear-logs-Edmond-3,collinear-logs-Edmond-4,collinear-logs-Edmond-5,collinear-logs-zhongbo}. Such \textit{ad hoc} bottom-up modifications of the resummation scheme, albeit not providing a fully consistent picture of the collinear phase space at small $x$, have proven to be successful at postponing the negativity issue of small $x_{{\rm Bj}}$ cross sections to larger values of the hard scales at a given perturbative order.

Most of such bottom-up approaches to collinear logarithm resummation at small $x$ rely on imposing either one of two equivalent orderings: $k^-$ ordering in order to reconstruct the DGLAP ladder structure, or light cone time ordering along said ladder. In effect, such corrections allow to correct how one treats the kinematical phase space for the emission of gluons whose transverse momenta are large enough to break longitudinal momentum ordering along the gluon ladder: 
if $\boldsymbol{k}^2$ is large enough, even with $k^+ \gg \ell^+$, one may have $k^- \sim \boldsymbol{k}^2/(2k^+) \gg \ell^- \sim \boldsymbol{\ell}^2/(2\ell^+)$ which violates the $-$ ordering between $k$ and $\ell$.
In fact, it is possible to argue that the problem with this part of the phase space could already have been diagnosed at LL accuracy with only quarks. Indeed at leading twist where $|\boldsymbol{\ell}| \gg |\boldsymbol{k}|$, we can study the exact same corner of phase space for transverse momenta in the $z \rightarrow 0$ limit. This kinematics correspond precisely to the collinear (to the target) limit where one would derive the DGLAP kernel for quarks.

As already discussed earlier, there are two potential problems to the r.h.s. of Eq.~(\ref{eq:DGLAP-assumption-1}) that corresponds to the semi-classical small-$x$ scheme. First, it is correct only if the distribution is a constant at $x=0$. Second, it diverges if the splitting function is not integrable on $[0,1]$. Given the fact that this corner of phase space is the one that causes problems for gluons, and that as we proved earlier the actual $x$ variable appearing in DGLAP is automatically set to be strictly null by CGC schemes, we can finally see where the issue originates from. Indeed, the two problematic conditions are met with gluons: the gluon distribution potentially has a non-trivial dependence on $x$ around $x=0$ if saturation effects do not completely cancel out the intercept, and in a more obvious fashion the ${\cal P}^{gg}$ splitting function has poles at the end points $y=0,1$. This is the reason why difficulties were encountered for NLL small-$x$ studies.

Our approach has two advantages. First and foremost, the fact that our distribution is not evaluated in the strict $x=0$ approximation means that the DGLAP kernel is fully reproduced at least within LL accuracy, as we proved in Section~\ref{sec:Bjorken}. Second, the ordering in light cone time is built into the framework because we did not allow the light cone time decoupling which is the starting point of CGC computations. See the discussions in Section~\ref{sec:beyond-shock-waves} and around Eqs.~(\ref{eq:denom-with-ordering}) and (\ref{eq:denom-with-antiordering}). Applying our scheme to the evolution equation of our distribution will naturally incorporate the desired ordering. 
We traded the dipole operator which is defined only in the strict $x=0$ limit for the $x$-dependent gluon distribution~\eqn{eq:Gijdef} and we thus have one more variable when compared to standard semi-classical small-$x$ schemes. The distribution we found depends on a transverse momentum and on the rapidity factorization scale which allows to separate out classical target fields in $+$ momentum space using the BK equation, as is the case in these schemes, but also on the physical DGLAP $-$ momentum variable $x$ that is missing from them:
\beq
\varphi_{_G}(\k;\Lambda^+)\equiv \frac{1}{N_c}\int_\r \,\rme^{-i\r\cdot \k}\, \langle \tr \, U_\0 U^\dag_\r  \rangle_{\Lambda^+} \quad \to \quad G_{\Lambda^+} ^{ij}(x,\k)
\eeq
One of the main strategies adopted in previous studies to solve the problem of small-$x$ evolution in the collinear corner of phase-space has been to perform \textit{ad hoc} modifications to the  BK evolution equation by forcing a relation between the cutoff variable $\Lambda^+ > k^+$ and  momenta along the other light cone direction $k^-=P^-$ roughly as follows $2\Lambda^+ k^- \sim \k^2$, throughout the evolution. In our case, the $-$ light cone direction is fully accounted for by the presence of $x$.

What we conclude from the present discussion is
that although \textit{ad hoc} considerations allow us to extend the validity of the description of observables in the shock wave approximation, we argued that we should actually fix what is being evolved instead of fixing how it is evolved. As long as we have infinite Wilson lines, we have $x=0$. In semi-inclusive cases that involve jets $x_F$ is not integrated out at tree level so one can set a relation between $\Lambda^+$ and $x_{F}$, but it does not fix the collinear corner of phase space since DGLAP involves
$x$ rather than $x_{F}$. Beyond NLL accuracy, or in any other semi-inclusive case where $x_{F}$ is a convolution variable because of the presence of fragmentation functions, the relation between $\Lambda^+$ and $x_{F}$ cannot be imposed by hand anymore.
Changing how we evolve the dipole operator or the scale at which we evaluated it does not solve the $x=x_{F}=0$ problem which is due to the very fact that we are studying the dipole operator itself. Of course to complete the picture we need to investigate quantum evolution of the $x$-dependent unintegrated gluon distribution and compare to the aforementioned schemes, but the distribution found in this article does not present the fundamental flaw of being set to $x=0$ from the get-go.

\section{Conclusion and outlook }\label{sec:conclusion}

In this article, we provided a semi-classical scheme inspired by small-$x_{\rm Bj}$ frameworks, in view of a first principle interpolation between both major limits of perturbative QCD. We applied this scheme to the inclusive DIS cross section and obtained the leading power of $x_{\rm Bj}/Q^2$ for this observable. Our scheme is based on a partial twist expansion (PTE) that resums subclasses of powers of $s$ and $Q^2$ to all orders. We have derived the leading contribution in the PTE and showed that it encompasses both Bjorken and Regge regimes, and thus provides the desired interpolation.
Our computation yields a new, unexpected, form for an unintegrated gluon distribution with correct dependences on both transverse and longitudinal gluon momentum components. 
From our interpolating formula, we perform a top-down analysis of the collinear limit of Regge descriptions of our observable. We find that the shock wave approximation on which semi-classical small-$x_{\rm Bj}$ framework rely is inherently incompatible with a proper account of the DGLAP equation because of its treatment of the $x$ variable.
Much progress has been made towards including all subleading power corrections to the shock wave approximation, so-called subeikonal corrections~\cite{subeikonal-tolga-1,subeikonal-tolga-2,subeikonal-tolga-3,subeikonal-tolga-4,subeikonal-tolga-5,subeikonal-tolga-6, subeikonal-yuri-1,subeikonal-yuri-2,subeikonal-yuri-3,subeikonal-yuri-4,subeikonal-yuri-5,subeikonal-yuri-6,subeikonal-yuri-7,subeikonal-yoshitaka,subeikonal-giovanni-1,subeikonal-giovanni-2,subeikonal-jamal-1,subeikonal-jamal-2,subeikonal-jamal-3}. In such approaches, some of the subleading terms correspond to the expansion in the Feynman $x$ variables of the leading twist distribution. This would yield a correction to the assumption which is made in Eq.~(\ref{eq:DGLAP-assumption-1}): along with $[xg(x)]_{x=0}$, the second term in the Taylor expansion $[\frac{\rm d}{{\rm d} x}xg(x)]_{x=0}$ hides in the expressions one would find in the collinear limit of subeikonal results. Although this does not make Eq.~(\ref{eq:DGLAP-assumption-1}) valid in full generality, it would be insightful to study how such corrections would compare to an expansion of DGLAP from our top-down approach and whether or not they could allow to reconstruct the full equation and correct for the dangerous hypothesis in a bottom-up approach. 

In this article, we have focused on the leading order diagram in the small $x$ limit. However, since we encountered a quark loop at this order, we have partially tackled the question of quantum evolution by providing a prescription for dealing with the longitudinal structure of the target. Nevertheless, a complete picture would require the derivation of the analog of the BK equation for the unintegrated gluon distribution which we leave for a following work.

\section*{Acknowledgements} 
The work of Y. M.-T. is supported by the U.S. Department of Energy, Office of Science, Office of Nuclear Physics, under contract No. DE- SC0012704. Y. M.-T. acknowledges support from the RHIC Physics Fellow Program of the RIKEN BNL Research Center. 


\appendix

\section{On the inclusion of non-pure gauge transverse gluon fields}\label{sec:transverse-gluons}

This appendix aims at showing that although they contribute as gauge invariance fixing counterterms to $\partial^i A^-$ in $F^{i-}$, transverse gluon fields are not necessary to compute the cross section beyond a consistency check.
In the Regge regime, one always assumes that one can gauge away transverse gluon fields. In the Bjorken regime, it is not a valid assumption. However, for inclusive DIS, the leading (collinear) distributions do not involve a transverse
separation. As a result, the only contribution from $\boldsymbol{A}^{i}$
fields at our perturbative order is contained in the $F^{i-}$ tensors
that define the PDF. Let us write the DIS cross section in the factorized
form
\begin{equation}
	{\rm d}\sigma=\int{\rm d}x{\cal H}(x){\rm e}^{ixP^{-}z^{+}}\langle P|{\rm tr}F^{i-}(z^{+})[z^{+},0^{+}]F^{i-}(0^{+})[0^{+},z^{+}]|P\rangle,\label{eq:gen-fac}
\end{equation}
where ${\cal H}(x)$ is the collinear hard part. Let us note that
for any $y^{+},u^{+}$:
\begin{align*}
	& {\rm tr}[y^{+},u^{+}]F^{i-}(u^{+})[u^{+},y^{+}]F^{i-}(y^{+})\\
	& ={\rm tr}(\partial^{i}A^{-})[u^{+},y^{+}](\partial^{i}A^{-})(y^{+})[y^{+},u^{+}]\\
	& +\frac{\partial}{\partial u^{+}}\frac{\partial}{\partial y^{+}}{\rm tr}[y^{+},u^{+}]A^{i}(u^{+})[u^{+},y^{+}]A^{i}(y^{+})\\
	& -\frac{\partial}{\partial y^{+}}{\rm tr}(\partial^{i}A^{-})(u^{+})[u^{+},y^{+}]A^{i}(y^{+})[y^{+},u^{+}]\\
	& -\frac{\partial}{\partial u^{+}}{\rm tr}[u^{+},y^{+}](\partial^{i}A^{-})(y^{+})[y^{+},u^{+}]A^{i}(u^{+}).
\end{align*}
Once plugged into the convolution, integrations by parts w.r.t.
$y^{+}$ in the penultimate term and w.r.t. $x^{+}$ in the last term allow to cancel the last two lines. One is left with 
\begin{align}
	{\rm d}\sigma & =\int{\rm d}x{\cal H}(x){\rm e}^{ixP^{-}z^{+}}\langle P|{\rm tr}(\partial^{i}A^{-})(z^{+})[z^{+},0^{+}](\partial^{i}A^{-})(0^{+})[0^{+},z^{+}]|P\rangle\nonumber \\
	& +\int{\rm d}x{\cal H}(x){\rm e}^{ixP^{-}z^{+}}x^{2}(P^{-})^{2}\langle P|{\rm tr}A^{i}(z^{+})[z^{+},0^{+}]A^{i}(0^{+})[0^{+},z^{+}]|P\rangle.\label{eq:gen-ds}
\end{align}
The first line of this equation can be computed without any scattering with transverse gluon fields. We thus have an explicit expression to reconstruct the full gauge invariant cross section while only computing the hard part with pure
gauge transverse gluons: the term with non-pure gauge transverse gluons is obtained from the Ward-Takahashi identities and reads $x^{2}(P^{-})^{2}{\cal H}(x)$.
In a nutshell, it is possible to compute the full cross section while only including pure gauge transverse gluons in the Bjorken limit as
well.

Our purpose is to have exact results in both limits, but we are not
so much interested in the interpolating region. We can therefore conclude that unless one is trying to get an explicit proof of QCD gauge invariance for the considered observable, it is not necessary at the considered precision to explicitly include transverse gluons which are not pure gauge in the computation.

\section{Proof of \eqn{eq:conv-prop}\label{sec:convolution}}
The convolution property for scalar propagators \eqn{eq:conv-prop}  follows from the Klein-Gordon equations. Let us start by multiplying the two equations 
\beq
\left(-\Box_z + 2ig A^-(z) \del_z^+ \right) G_\scal(z,x_0) =\delta^D(z-x_0)  
\eeq
and 
\beq
G_\scal(x,z)  \left( \overleftarrow{\Box}_{z} + 2ig \overleftarrow{\del^+_{z}}A^-(z)  \right) =-\delta^D(z-x) \,, 
\eeq
respectively on the left by $G_\scal(x,z) $ and on the right by $G_\scal(z,x_0) $. For a given $u^+$, let us add them up and integrate over $z$ while imposing $z^+>  u^+$. 
This yields 
\begin{align}
	& \int\limits_{z^+>u^+} \!\!\!\! \rmd^D z \, G_{{\rm scal}}(x,z)\left(\overleftarrow{\Box}_{z}-\Box_{z}+2ig\overleftarrow{\partial_{z}^{+}}A^{-}(z)+2igA^{-}(z)\partial_{z}^{+}\right)G_{{\rm scal}}(z,x_{0})\nonumber \\
	& = \int\limits_{z^+>u^+}\!\!\!\!\rmd^D z  \left[\delta^{D}(z-x_{0})-\delta^{D}(x-z)\right]G_{{\rm scal}}(x,x_{0}).
\end{align}
The third and last terms in the brackets cancel each other after integration by part over $z^-$, neglecting boundary terms at infinity. The same holds for the transverse part of the d'Alembertian $\Box \equiv 2\del^+\del^--\boldsymbol{\del}^2$, which cancel out between the first and second terms. 
We are thus left with the following:
\beq
\left[\theta(y^+-u^+)-\theta(x^+-u^+)\right]\, G_\scal (x,y) & =&\!\!\!\!\int\limits_{z^+>u^+} \!\!\!\! \rmd^D z\, G_\scal (x,z)\left(-2 \overset{\rightarrow}{\del^+}\overset{\rightarrow}{\del^-}   +2 \overset{\leftarrow}{\del^+}  \overset{\leftarrow}{\del^-}   \right) G_\scal (z,y)\,\nn
&=& - \!\!\!\! \int\limits_{z^+>u^+} \!\!\!\! \rmd^D z\, G_\scal (x,z)\left(2 \overset{\rightarrow}{\del^+}\overset{\rightarrow}{\del^-}   +2 \overset{\rightarrow}{\del^+}  \overset{\leftarrow}{\del^-}   \right) G_\scal (z,y)\,\nn
&=& -  2\!\!\!\!  \int\limits_{z^+>u^+} \!\!\!\! \rmd^D z\,  \frac{\del}{\del z^+ } \left( G_\scal (x,z) \del_z^+  G_\scal (z,y) \right)\,\nn
&=& -  2 \int_{z^+=u^+} \!\!\!\! \rmd^d \z \int \rmd z^-\,  G_\scal (x,z) \del_z^+ G_\scal (z,y)\,, \nn
 \eeq
where we have repeatedly neglected boundary terms at $z^-=\pm \infty$, $z^+=+\infty$. Finally, for $x^+>y^+$ only the $x^+>u^+>y^+$ ordering yields a non-vanishing l.h.s: 
\beq
 G_\scal (x,y) =  2 \int_{z^+} \rmd \z \int \rmd z^-\,  G_\scal (x,z) \del_z^+  G_\scal (z,y)\, \,.
\eeq
Similarly, for $y^+>u^+>x^+$ we have
\beq
 G_\scal (x,y) = - 2 \int_{z^+} \rmd \z \int \rmd z^-\,  G_\scal (x,z) \del_z^+  G_\scal (z,y)\, \,.
\eeq
\section{Notes on QED gauge invariance\label{sec:QEDinv}}
Let us establish a few useful relations as a corollary for the Ward-Takahashi
identities. First of all, the Klein-Gordon equations in momentum space
read:
\begin{equation}
p^{2}G_{\mathrm{scal}}(p,p_{0}) =(2\pi)^{D}\delta^{D}(p-p_{0})-2gp^{+}\!\int\!\frac{{\rm d}^{D}k}{(2\pi)^{D}}A^{-}(k)G_{\mathrm{scal}}(p-k,p_{0})\label{eq:KGmomL}
\end{equation}
and
\begin{equation}
G_{\mathrm{scal}}(p,p_{0}) p_{0}^{2}=(2\pi)^{4}\delta^{4}(p-p_{0})-2gp^{+}\!\int\!\frac{{\rm d}^{4}k}{(2\pi)^{4}}G_{\mathrm{scal}}(p,p_{0}+k)A^{-}(k).\label{eq:KGmomR}
\end{equation}
As a direct consequence of these, one has:
\begin{equation}
\slashed{p}D_{F}(p,p_{0}) =i(2\pi)^{D}\delta^{D}(p-p_{0})-ig\gamma^{+}\slashed{p}_{0}\!\int\!\frac{{\rm d}^{D}k}{(2\pi)^{D}}A^{-}(k)G_{\mathrm{scal}}(p-k,p_{0}),\label{eq:pD}
\end{equation}
and 
\begin{equation}
D_{F}(p,p_{0})\slashed{p}_{0} =i(2\pi)^{D}\delta^{D}(p-p_{0})-ig\slashed{p}\gamma^{+}\!\int\!\frac{{\rm d}^{D}k}{(2\pi)^{D}}G_{\mathrm{scal}}(p,p_{0}+k)A^{-}(k).\label{eq:Dp}
\end{equation}
Let us check the Ward-Takahashi (WT) identity for the initial photon
by computing $q_\mu W^{\mu\nu}$. By writing
$\slashed{q}=\slashed{\ell}-(\slashed{\ell}-\slashed{q})$, then with the
help of Eq.~(\ref{eq:Dp}) for the first term and Eq.~(\ref{eq:pD})
for the second term, one gets:
\begin{align}
	& q_{\mu}W^{\mu\nu}  =-ie^{2}\sum_{f}q_{f}^{2}\int\!\frac{{\rm d}^{D}\ell}{(2\pi)^{D}}\int\!\frac{{\rm d}^{D}k}{(2\pi)^{D}}\,\label{eq:WT1}\\
	& \: \times \Bigl\{(2\pi)^{D}\delta^{D}(k)\langle P|{\rm tr}[\gamma^{\nu}D_{F}(-q+\ell,-q+\ell+k)-\gamma^{\nu}D_{F}(\ell+k,\ell)]|P\rangle\nonumber \\
	& -g\!\int\!\frac{{\rm d}^{D}k_{0}}{(2\pi)^{D}}\langle P|{\rm tr}[G_{{\rm scal}}(\ell+k,\ell+k_{0})A^{-}(k_{0})\gamma^{\nu}(\slashed{\ell}+\slashed{k})\gamma^{+}D_{F}(-q+\ell,-q+\ell+k)]|P\rangle\nonumber \\
	& +g\!\int\!\frac{{\rm d}^{D}k_{0}}{(2\pi)^{D}}\langle P|{\rm tr}[\gamma^{\nu}D_{F}(\ell+k,\ell)\gamma^{+}(-\slashed{q}+\slashed{\ell}+\slashed{k})A^{-}(k_{0})G_{{\rm scal}}(-q+\ell-k_{0},-q+\ell+k)]|P\rangle\Bigr\}\nonumber 
\end{align}
A simple change of variables allows to compensate the first term in the brackets with the second term. We will now use the explicit expression for the Dirac propagators in terms of scalar propagators in the remaining terms. Note that it is greatly simplified by the presence of $\gamma^+$ matrices in the trace:
\begin{align}
	q_{\mu}W^{\mu\nu} & =ige^{2}\sum_{f}q_{f}^{2}\int\!\frac{{\rm d}^{D}\ell}{(2\pi)^{D}}\int\!\frac{{\rm d}^{D}k}{(2\pi)^{D}}\,\!\int\!\frac{{\rm d}^{D}k_{0}}{(2\pi)^{D}}\nonumber \\
	& \times{\rm tr}_{s}[\gamma^{\nu}(\slashed{\ell}+\hat{k})\gamma^{+}(-\slashed{q}+\slashed{\ell}+\slashed{k})]\label{eq:WT2}\\
	& \times \langle P| \Bigl\{{\rm tr}_{c}[G_{{\rm scal}}(\ell+k,\ell+k_{0})A^{-}(k_{0})G_{{\rm scal}}(-q+\ell,-q+\ell+k)]\nonumber \\
	& -{\rm tr}_{c}[G_{{\rm scal}}(\ell+k,\ell)A^{-}(k_{0})G_{{\rm scal}}(-q+\ell-k_{0},-q+\ell+k)]\Bigr\}|P\rangle\nonumber . 
\end{align}
Finally, we can see that taking the changes of variables $\ell \rightarrow \ell+k_0$ and $ k \rightarrow k-k_0 $ in the second term casts it in a form where it explicitely cancels the first term. One is left with $q_\mu W^{\mu \nu}$, and the Ward-Takahashi identity thus holds. The proof for the outgoing photon is identical, and QED gauge invariance is finally confirmed.

\section{Cancellation of the monopoles\label{sec:monopoles}}

Let us now prove that the instantaneous terms from the quark propagators do not contribute to the amplitude. Let us start from the definition of the hadronic tensor
\begin{align}
	W^{\mu\nu} & =e^{2}\sum_{f}q_{f}^{2}\int\!\frac{{\rm d}^{D}\ell}{(2\pi)^{D}}\int\!\frac{{\rm d}^{D}k}{(2\pi)^{D}}\label{eq:Wmunu-rappel}\\
	& \times\langle P|{\rm tr}\bigl[\gamma^{\nu}D_{F}(\ell+k,\ell)\gamma^{\mu}D_{F}(-q+\ell,-q+\ell+k)\bigr]|P\rangle\nonumber .
\end{align}
With the help of the Ward-Takahashi identity we established in Appendix~\ref{sec:QEDinv}, we can subsitute $\gamma^{\mu,\nu}\rightarrow\gamma^{\mu,\nu}-n_2^{\mu,\nu}\slashed{q}/q^{+}$ and restrict ourselves to computing 
\begin{align}
	W^{\mu\nu} & =e^{2}\sum_{f}q_{f}^{2}\int\!\frac{{\rm d}^{D}\ell}{(2\pi)^{D}}\int\!\frac{{\rm d}^{D}k}{(2\pi)^{D}}\label{eq:Wmunu-WT-trick}\\
	& \times\langle P|{\rm tr}\bigl[\bigl(\gamma^{\nu}-n_2^{\nu}\frac{\slashed{q}}{q^{+}}\bigr)D_{F}(\ell+k,\ell)\bigl(\gamma^{\mu}-n_2^{\mu}\frac{\slashed{q}}{q^{+}}\bigr)D_{F}(-q+\ell,-q+\ell+k)\bigr]|P\rangle\nonumber 
\end{align}
instead. The instaneaneous contribution from the first propagator $D_F(\ell+k,\ell)$ reads:
\begin{align}
	W_{{\rm inst}(1)}^{\mu\nu} & =e^{2}\sum_{f}q_{f}^{2}\,{\rm Re}\int\!\frac{{\rm d}^{D}\ell}{(2\pi)^{D}}\frac{i}{2\ell^{+}}\label{eq:Winst1}\\
	& \times\langle P|{\rm tr}\bigl[\bigl(\gamma^{\nu}-n_2^{\nu}\frac{\slashed{q}}{q^{+}}\bigr)\gamma^{+}\bigl(\gamma^{\mu}-n_2^{\mu}\frac{\slashed{q}}{q^{+}}\bigr)D_{F}(-q+\ell,-q+\ell)\bigr]|P\rangle\nonumber.
\end{align}
When computing the spinor trace, it is worth noticing that the first and third brackets both have null components along the $+$ direction. This means that in the final result, the open $+$ index has to be contracted with the $D_F$ contribution. In other words, only the $\gamma^-$ component of the second $D_F$ propagator contributes. Since there is no momentum transfer in that propagator, it is easy to identify it:
\begin{align}
	W_{{\rm inst}(1)}^{\mu\nu} & =e^{2}\sum_{f}q_{f}^{2}\int\!\frac{{\rm d}^{D}\ell}{(2\pi)^{D}}\frac{i}{2\ell^{+}}\langle P|{\rm tr}_{c}G_{{\rm scal}}(-q+\ell,-q+\ell)|P\rangle\label{eq:Winst1-2}\\
	& \times{\rm tr}_{s}\bigl[\bigl(\gamma^{\nu}-n_2^{\nu}\frac{\slashed{q}}{q^{+}}\bigr)\gamma^{+}\bigl(\gamma^{\mu}-n_2^{\mu}\frac{\slashed{q}}{q^{+}}\bigr)(-q^{+}+\ell^{+})\gamma^{-}\bigr]\nonumber 
\end{align}
With the spinor trace 
\begin{align}
	{\rm tr}_{s}\bigl[\bigl(\gamma^{\nu}-n_2^{\nu}\frac{\slashed{q}}{q^{+}}\bigr)\gamma^{+}\bigl(\gamma^{\mu}-n_2^{\mu}\frac{\slashed{q}}{q^{+}}\bigr)(-q^{+}+\ell^{+})\gamma^{-} \bigr] & =4(q^{+}-\ell^{+})g_{\perp}^{\mu\nu},\label{eq:trace1}
\end{align}
we finally get:
\begin{align}
	W_{{\rm inst}(1)}^{\mu\nu} & =e^{2}g_{\perp}^{\mu\nu}\sum_{f}q_{f}^{2}\int\!\frac{{\rm d}^{D}\ell}{(2\pi)^{D}}\frac{2i(q^{+}-\ell^{+})}{\ell^{+}}\langle P|{\rm tr}_{c}G_{{\rm scal}}(-q+\ell,-q+\ell)|P\rangle\label{eq:Winst1-fin}
\end{align}
Similar steps lead to the following result for the contribution of the instantaneous term in the second propagator $D_F(-q+\ell,-q+\ell+k)$:
\begin{align}
	W_{{\rm inst}(2)}^{\mu\nu} & =e^{2}g_{\perp}^{\mu\nu}\sum_{f}q_{f}^{2}\int\!\frac{{\rm d}^{D}\ell}{(2\pi)^{D}}\frac{2i\ell^{+}}{(q^{+}-\ell^{+})}\langle P|{\rm tr}_{c}G_{{\rm scal}}(\ell,\ell)|P\rangle\label{eq:Winft2-fin}.
\end{align}
We finally need to cancel quantities of the form
\begin{equation}
	\int \frac{\rm d \ell^- }{2\pi} \langle P|{\rm tr}_{c}G_{{\rm scal}}(\ell,\ell)|P\rangle.
\end{equation}
In coordinate space, the integral sets the propagator to be between two identical light cone times. The physical argument for the cancellation of the monopoles thus relies on the fact that there is not a big enough light cone time window for scatterings with the target to occur, which means monopoles only contribute to disconnected diagrams. This can be seen mathematically from the integrated version of the Schr\"{o}dinger equation:
\begin{align}
	& G_{{\rm scal}}(x^{\prime},x)-G_{0}(x^{\prime}-x)\label{eq:}\\
	& =2g\int\!{\rm d}^{D}y\int\!\frac{{\rm d}^{D}k}{(2\pi)^{D}}{\rm e}^{-i(k\cdot y)} \int\frac{{\rm d}^{D}\ell}{(2\pi)^{D}}{\rm e}^{-i\ell\cdot(x^{\prime}-y)}G_{0}(\ell)\,(k\cdot A)(y)\,G_{{\rm scal}}(k,x).  \nonumber
\end{align}
Using the standard Cauchy integral from Eq.~(\ref{eq:Cauchy-int}), the integral over $\ell^-$ sets a strict ordering between $x^\prime$ and $y$. Similar considerations show a strict ordering for $x$ as well, and one can easily see that light cone times for the scatterings are strictly ordered. If the initial and final times are equal, the r.h.s. cancels due to the absence of such an ordering and thus $G = G_0$. This concludes our proof: in the case of monopoles, the integral over $\ell^-$ sets equal light cone times for the unique scalar propagator, which sets this propagator to the free propagator. Disconnected contributions without scatterings with the target being subtracted, monopoles do not contribute. 

\section{Derivation of the leading $1/Q$ power of the Regge limit}\label{sec:double-log-details}

The twist expansion of the dipole scattering matrix element in Eq.~(\ref{eq:UUtoPDF}) allows us to get the leading $1/Q$ power of~\eqn{eq:WLL-eik} and~\eqn{eq:WTT-eik}:
\begin{align}
	\lim_{Q^{2}\rightarrow\infty}F_{L}(x_{{\rm Bj}}\rightarrow0,Q^{2}) & =8\sum_{f}q_{f}^{2}\int_{0}^{1}{\rm d}z\int{\rm d}^{d}\boldsymbol{r}\frac{\alpha_{s}\boldsymbol{r}^{2}}{d(2\pi)^d}\left[xg(x)\right]_{x=0} \label{eq:WLL-eik-LT}\\
	& \times\left(\frac{z\bar{z}Q^{2}}{\boldsymbol{r}^{2}}\right)^{\frac{d}{2}-1}z^{2}\bar{z}^{2}Q^{4}\bigl[K_{\frac{d}{2}-1}(Q\sqrt{z\bar{z}}|\boldsymbol{r}|)\bigr]^{2},\nonumber
\end{align}
and
\begin{align}
	\lim_{Q^{2}\rightarrow\infty}F_{T}(x_{{\rm Bj}}\rightarrow0,Q^{2}) & =2Q^{2}\sum_{f}q_{f}^{2}\int_{0}^{1}{\rm d}z\int{\rm d}^{d}\boldsymbol{r} \frac{\alpha_{s}\boldsymbol{r}^{2}}{d(2\pi)^d}\left[xg(x)\right]_{x=0} \label{eq:WTT-eik-LT}\\
	& \times\left(1-\frac{4}{d}z\bar{z}\right)\boldsymbol{r}^{2}\left(\frac{z\bar{z}Q^{2}}{\boldsymbol{r}^{2}}\right)^{\frac{d}{2}}\bigl[K_{\frac{d}{2}}(Q\sqrt{z\bar{z}}|\boldsymbol{r}|)\bigr]^{2}\nonumber .
\end{align}
It is not too difficult to take the $\boldsymbol{r}$ and $z$ integrals at this point. However, it is more instructive to work in momentum space with the appropriate variables. Let us use~\eqn{eq:K0-int-d} and~\eqn{eq:K1-int-d}, then, after an integration by parts and a few trivial integrals we finally find a form and~\eqn{eq:WTT-col}:
\begin{align}
	\lim_{Q^{2}\rightarrow\infty}F_{L}(x_{{\rm Bj}}\rightarrow0,Q^{2}) & =g^{2}Q^{2}\sum_{f}q_{f}^{2}\int{\rm d}x\left[xg(x)\right]_{x=0}\label{eq:WLL-eik-LT-2}\\
	& \times\int_{0}^{1}\frac{{\rm d}z}{2\pi}\int\frac{{\rm d}^{d}\boldsymbol{\ell}}{(2\pi)^{d}}\frac{1}{d}\frac{16z^{2}\bar{z}^{2}Q^{2}\boldsymbol{\ell}^{2}}{\bigl(\boldsymbol{\ell}^{2}+z\bar{z}Q^{2}\bigr)^{4}}\delta(x),\nonumber 
\end{align}
and
\begin{align}
	\lim_{Q^{2}\rightarrow\infty}F_{T}(x_{{\rm Bj}}\rightarrow0,Q^{2}) & =g^{2}Q^{2}\sum_{f}q_{f}^{2}\int{\rm d}x\left[xg(x)\right]_{x=0}\label{eq:WTT-eik-LT-2}\\
	& \times\int_{0}^{1}\frac{{\rm d}z}{2\pi}\int\frac{d^{d}\boldsymbol{\ell}}{(2\pi)^{d}}\delta(x)\left(1-\frac{4}{d}z\bar{z}\right)\nonumber \\
	& \times\frac{1}{d}\left(\frac{d}{\bigl(\boldsymbol{\ell}^{2}+z\bar{z}Q^{2}\bigr)^{2}}-4\frac{\boldsymbol{\ell}^{2}}{\bigl(\boldsymbol{\ell}^{2}+z\bar{z}Q^{2}\bigr)^{3}}+4\frac{\boldsymbol{\ell}^{4}}{\bigl(\boldsymbol{\ell}^{2}+z\bar{z}Q^{2}\bigr)^{4}}\right).\nonumber 
\end{align}
In the Bjorken regime, we defined the $y\equiv x_{\rm Bj}/x$ variable. Because of the additional condition that was relating the Feynman $x$ variable to $x_{\rm Bj}$ and the loop variables, we could have equivalently defined it via
\begin{equation}
	\boldsymbol{\ell}^{2}=z\bar{z}\left(\frac{1-y}{y}\right)Q^{2}.\label{eq:ell-to-y}
\end{equation}
Let us use the definition above in~\eqn{eq:WLL-eik-LT-2} and~\eqn{eq:WTT-eik-LT-2}. Given the absence of angular dependence, we can perform the angular integral for $\boldsymbol{\ell}$ in a trivial way, by rewriting the integration measure as follows:
\begin{align}
	\int_{0}^{1}{\rm d}z\int{\rm d}^{d}\boldsymbol{\ell} & \rightarrow\frac{\pi^{\frac{d}{2}}}{\Gamma\left(\frac{d}{2}\right)}\int_{0}^{1}{\rm d}z\int_{0}^{\infty}{\rm d}\boldsymbol{\ell}^{2}\left(\boldsymbol{\ell}^{2}\right)^{\frac{d}{2}-1}\label{eq:dzdell}\\
	& \rightarrow \frac{\pi^{\frac{d}{2}}}{\Gamma\left(\frac{d}{2}\right)}\int_{0}^{1}{\rm d}z\int_{0}^{1}\frac{{\rm d}y}{y^{2}}\left(z\bar{z}Q^{2}\right)^{\frac{d}{2}}\left(\frac{1-y}{y}\right)^{\frac{d}{2}-1}\nonumber .
\end{align}
We find:
\begin{align}
	\lim_{Q^{2}\rightarrow\infty}F_{L}(x_{{\rm Bj}}\rightarrow0,Q^{2}) & =\frac{\alpha_{s}}{2\pi}\sum_{f}q_{f}^{2}\int{\rm d}x\int_{0}^{1}{\rm d}y\int_{0}^{1}{\rm d}z\left[xg(x)\right]_{x=0}\nonumber \\
	& \times\frac{1}{\Gamma\left(\frac{d}{2}\right)}\left[\frac{Q^{2}(1-y)}{4\pi\mu^{2}y}\right]^{\frac{d}{2}-1}(z\bar{z})^{\frac{d}{2}-1}\delta(x)\nonumber \\
	& \times\frac{16}{d}y(1-y),\label{eq:FL-R-B}
\end{align}
and
\begin{align}
	\lim_{Q^{2}\rightarrow\infty}F_{T}(x_{{\rm Bj}}\rightarrow0,Q^{2}) & =\frac{\alpha_{s}}{2\pi}\sum_{f}q_{f}^{2}\int{\rm d}x\int_{0}^{1}{\rm d}y\int_{0}^{1}{\rm d}z\left[xg(x)\right]_{x=0}\label{eq:FT-R-B}\\
	& \times\frac{1}{\Gamma\left(\frac{d}{2}\right)}\left[\frac{Q^{2}(1-y)}{4\pi\mu^{2}y}\right]^{\frac{d}{2}-1}\left[(z\bar{z})^{\frac{d}{2}-2}-\frac{4}{d}(z\bar{z})^{\frac{d}{2}-1}\right]\delta(x)\nonumber \\
	& \times\left[1-\frac{4}{d}y(1-y)\right].\nonumber 
\end{align}
Note the dimensional regularization parameter $\mu$ whose presence was implicit until now. Finally taking the $z$ integral and expanding around $d=2+2\epsilon \simeq 2$, we obtain:
\begin{align}
	\lim_{Q^{2}\rightarrow\infty}F_{L}(x_{{\rm Bj}}\rightarrow0,Q^{2}) & =4\frac{\alpha_{s}}{\pi}\sum_{f}q_{f}^{2}\left[xg(x)\right]_{x=0}\int_{0}^{1}{\rm d}yy(1-y)\nonumber \\
	& =\frac{2}{3}\frac{\alpha_{s}}{\pi}\sum_{f}q_{f}^{2}\left[xg(x)\right]_{x=0},
\end{align}
and
\begin{align}
	& \lim_{Q^{2}\rightarrow\infty}F_{T}(x_{{\rm Bj}}\rightarrow0,Q^{2})\\
	& =\frac{\alpha_{s}}{\pi}\sum_{f}q_{f}^{2}\left[xg(x)\right]_{x=0}\int_{0}^{1}{\rm d}y\nonumber \\
	& \times\left\{ \frac{1}{\epsilon}\left(\frac{{\rm e}^{\gamma_{E}}}{4\pi}\right)^{\epsilon}\left[(1-y)^{2}+y^{2}\right]+\left[(1-y)^{2}+y^{2}\right]\ln\left[\frac{Q^{2}(1-y)}{\mu^{2}y}\right]-1+4y(1-y)\right\} \nonumber \\
	& =\frac{\alpha_{s}}{3\pi}\sum_{f}q_{f}^{2}\left[xg(x)\right]_{x=0}\left[\frac{2}{\epsilon}\left(\frac{{\rm e}^{\gamma_{E}}}{4\pi}\right)^{\epsilon}+2\ln\left(\frac{Q^{2}}{\mu^{2}}\right)-1\right].\nonumber 
\end{align}

\bibliographystyle{apsrev4-1}

\end{document}